\documentclass[12pt,reqno,a4paper]{amsart}    %A4-format

\usepackage[totalwidth=440pt, totalheight=710pt]{geometry}
\usepackage{amsmath, amssymb}%, theorem}

\usepackage{graphicx}                      % for graphics
\newcommand{\color}[2][{}]{}        % figure-x.pstex_t file contains this
\usepackage{bbm}         % blackboard (see: /usr/share/texmf/mytex/bbm.dvi)

\theoremstyle{plain}            % body italics
\newtheorem{theorem}{Theorem}[section]

\newtheorem{lemma}[theorem]{Lemma}
\newtheorem{corollary}[theorem]{Corollary}

\theoremstyle{definition}       % body roman
\newtheorem{definition}[theorem]{Definition}
\newtheorem{assumption}[theorem]{Assumption}
\newtheorem{example}[theorem]{Example}

\theoremstyle{remark}
\newtheorem{remark}[theorem]{Remark}

\newcommand{\Sec}[1]{Section~\ref{sec:#1}}
\newcommand{\Secs}[2]{Sections~\ref{sec:#1} and~\ref{sec:#2}}
\newcommand{\App}[1]{Appendix~\ref{app:#1}}

\newcommand{\Ex}[1]{Example~\ref{ex:#1}}
\newcommand{\Exenum}[2]{Example~\ref{ex:#1}~(\ref{#2})}

\newcommand{\Thm}[1]{Theorem~\ref{thm:#1}}

\newcommand{\Thms}[2]{Thms.~\ref{thm:#1}--\ref{thm:#2}}
\newcommand{\Lem}[1]{Lemma~\ref{lem:#1}}
\newcommand{\Lemenum}[2]{Lemma~\ref{lem:#1}~(\ref{#2})}
\newcommand{\Lems}[2]{Lemmas~\ref{lem:#1}--\ref{lem:#2}}
\newcommand{\Cor}[1]{Corollary~\ref{cor:#1}}
\newcommand{\Cors}[2]{Corollaries~\ref{cor:#1}--\ref{cor:#2}}
\newcommand{\Rem}[1]{Remark~\ref{rem:#1}}
\newcommand{\Remenum}[2]{Remark~\ref{rem:#1}~(\ref{#2})}
\newcommand{\Def}[1]{Definition~\ref{def:#1}}
\newcommand{\Ass}[1]{Assumption~\ref{ass:#1}}
\newcommand{\Assenum}[2]{Assumption~\ref{ass:#1}~(\ref{#2})}

\newcommand{\Fig}[1]{Figure~\ref{fig:#1}}
\newcommand{\Figenum}[2]{Figure~\ref{fig:#1}~(\ref{#2})}

\newcommand{\Footnote}[1]{Footnote~\ref{fn:#1}}

\numberwithin{equation}{section}

\DeclareMathOperator{\dist}   {dist}
\DeclareMathOperator{\dom}    {dom}
\DeclareMathOperator{\ran}    {ran}

\DeclareMathOperator{\supp}   {supp}

\DeclareMathOperator{\tr} {tr} % trace
\DeclareMathOperator{\Ln} {Ln} % complex logarithm
\DeclareMathOperator{\leb}    {{\boldsymbol \lambda}}%{l}  % Lebegue measure
\newcommand{\weak}            {\mathrm w}    % suffix weak
\newcommand{\strong}          {\mathrm s}    % suffix strong
\DeclareMathOperator{\slim}   {\strong-lim}  % trace
\newcommand{\wint}     {\weak \operatorname*{-\!\!\int}} % weak integral

\renewcommand{\Re}     {\mathrm {Re}\,}
\renewcommand{\Im}     {\mathrm {Im}\,}

\newcommand{\spec}[2][{}]   {\sigma_{\mathrm{#1}}(#2)}

\newcommand{\SL}{\mathrm{SL}}

\newcommand{\R}{\mathbb{R}} % symbol for real numbers
\newcommand{\C}{\mathbb{C}} % symbol for complex numbers
\newcommand{\N}{\mathbb{N}} % symbol for natural numbers
\newcommand{\Z}{\mathbb{Z}} % symbol for integers
\newcommand{\eps}{\varepsilon} % shortcut
\renewcommand{\phi}{\varphi}   % shortcut
\newcommand{\e}{\mathrm e}  %Euler number
\newcommand{\im}{\mathrm i} % complex unit
\newcommand{\dd}{\,\mathrm d} % symbol for integration d

\newcommand{\wt}{\widetilde}           % shortcut
\newcommand {\qf}[1]{\mathfrak{#1}}    % font for quadratic forms

\newcommand {\Exp}[2][{}]{\mathbb{E}{#1}{(#2)}} % Expectation value
\newcommand {\Prob}   {{\mathbb P}}    % Prob. measure

\newcommand{\HS}{\mathcal H}           % symbol for Hilbert space
\newcommand{\Bsymb} {\mathcal B}       % symbol for bounded linear operators
\newcommand{\TRsymb}{\mathcal \Bsymb_1}% symbol for trace class
\newcommand{\HSsymb}{\mathcal \Bsymb_2}% symbol for Hilbert-Schmidt class

\newcommand{\Sobsymb} {\mathsf H}      % symbol for Sobolev space
\newcommand{\Contsymb} {\mathsf C}     % symbol for cont. space
\newcommand{\Lsymb}    {\mathsf L}     % symbol for int L-spaces
\newcommand{\lsymb}    {\ell}          % symbol for int l-spaces

\newcommand{\Ci} [2][{}]{\Contsymb^\infty_{#1} ({#2})}
\newcommand{\Cci}[1]{\Ci[\mathrm c]{#1}} % C_c^\infty(#1)-sp
\newcommand{\Cont}[2][{}]{\Contsymb_{#1}({#2})}
\newcommand{\Contn}[1]{\Cont[\circ]{#1}} % space of C^k fct. wit compact supp

\newcommand{\Lp}[2][p]{\Lsymb_{#1}({#2})} % L_#1(#2)-spaces

\newcommand{\Lsqr}[2][{}]{\Lsymb_2^{#1}({#2})} % L_2(#1)-spaces
\newcommand{\lsqr}[2][{}]{\lsymb_2^{#1}({#2})}   % l_2(#1)-spaces
\newcommand{\Linfty}[2][{}]{\Lsymb_\infty^{#1}({#2})} % L_2(#1)-spaces
\newcommand{\Sob}[2][1]{\Sobsymb^{#1}({#2})} % Sobolev spaces
\newcommand{\Sobn}[2][1]{\Sobsymb_\circ^{#1}({#2})}

\newcommand{\Sobc}[2][1]{\Sobsymb_\mathrm{c}^{#1}({#2})}

\newcommand{\norm}[2][{}]{\|{#2}\|_{{#1}}}    % norm
\newcommand{\normsqr}[2][{}]{\|{#2}\|^2_{#1}} % norm squared
\newcommand{\iprod}[3][{}]{\langle{#2},{#3}\rangle_{#1}}  % inner product
\newcommand{\bigiprod}[3][{}]{\bigl\langle{#2},{#3}\bigr\rangle_{#1}}

\newcommand{\set}[2]{\{ \, #1 \, | \, #2 \, \} } % set { #1 | #2 }
\newcommand{\bigset}[2]{\bigl\{ \, #1 \, \bigl|\bigr. \, #2 \, \bigr\} }
\newcommand{\Bigset}[2]{\Bigl\{ \, #1 \, \Bigl|\Bigr. \, #2 \, \Bigr\} }

\newcommand{\map}[3]{{#1}\colon{#2}\longrightarrow{#3}} % maps

\newcommand{\bd}  {\partial}                % symbol for boundary of a set
\newcommand{\clo}[1]{\overline{{#1}}}       % symbol for closure

\newcommand{\compl}[1]{#1^{\mathrm c}}       % complement of a set
\newcommand{\dcup}{\mathrel{\uplus}}         % symbol for disjoint union
\newcommand{\bigdcup}{\operatorname*{\biguplus}}
\newcommand{\restr}[1]{{\restriction}_{#1}} % symbol for map restriction

\newcommand{\conj}[1]{\overline {{#1}}}     % symbol for complex conjugation
\newcommand{\1}{\mathbbm 1}                    % blackboard 1
\newcommand{\Dir}{{\mathrm D}}              % symbol for Dirichlet bd cond
\newcommand{\laplacian}[2][{}]{\Delta_{{#2}}^{{#1}}}
\newcommand{\laplacianD}[1]{\laplacian[\Dir]{#1}} % symb f Dir-Laplacian
\newcommand{\hiddenfootnote}[1]{}
\newcommand{\Proj}[1]{\mathrm P({#1})}  % symbol for proj space
\newcommand{\proj}[1]{[#1]}  % symbol for proj element
\newcommand{\bigproj}[1]{\bigl[#1\bigr]}  % symbol for proj element
\DeclareMathOperator{\gen}{gen}
\DeclareMathOperator{\rad}{rad}
\newcommand{\Lsplit}{{\mathrm{split}}}

\begin{document}
\title[Localization for random quantum tree graphs]{Anderson
  Localization for radial
  tree-like random quantum graphs}

\author{Peter D.\ Hislop}
\thanks{PDH partially supported by NSF grant DMS-0503784 and OP partially
supported by DFG grant Po 1034/1-1.}
\address{Peter D.\ Hislop,
  Department of Mathematics\\
  University of Kentucky,
  753 Patterson Office Tower,
  Lexington, KY 40506-0027, USA}
\email{hislop@ms.uky.edu}
\author{Olaf Post}
\address{Olaf Post,
  Department of Mathematics\\
  University of Kentucky,
  751 Patterson Office Tower,
  Lexington, KY 40506-0027, USA\\
   \emph{Present Address}: Department of Mathematics\\
    Humboldt University, Rudower Chaussee~25, 12489 Berlin, Germany}
\email{post@math.hu-berlin.de}
\date{\today}

%------------------------------------------------------------
% Subject classifications
%------------------------------------------------------------
%Subjclass contains the Classification of the paper following the 1991
%Mathematics Subject Classifications

%\subjclass{35P20, 58G18, 47F05}
%\keywords{Eigenvalues, spectral gap, perturbation of periodic structures}

%------------------------------------------------------------
% Abstract.
%------------------------------------------------------------

\begin{abstract}
  We prove that certain random models associated with radial,
  tree-like, rooted quantum graphs exhibit Anderson localization at
  all energies. The two main examples are the random length model
  (RLM) and the random Kirchhoff model (RKM). In the RLM, the lengths
  of each generation of edges form a family of independent,
  identically distributed random variables (iid).  For the RKM, the
  iid random variables are associated with each generation of vertices
  and moderate the current flow through the vertex.  We consider
  extensions to various families of decorated graphs and prove
  stability of localization with respect to decoration. In particular,
  we prove Anderson localization for the random necklace model.
\end{abstract}

\maketitle

%%%%%%%%%%%%%%%%%%%%%%%%%%%%%%%%%%%%%%%%%%%%%%%%%%%%%%%%%%%%%%%%%%%%%%%
%----------------------------------------------------------------------
%
\section{Introduction}
\label{sec:intro}
%
%----------------------------------------------------------------------
%%%%%%%%%%%%%%%%%%%%%%%%%%%%%%%%%%%%%%%%%%%%%%%%%%%%%%%%%%%%%%%%%%%%%%%

Quantum mechanics on metric graphs is a subject with a long history
which can be traced back to the paper of Ruedenberg and
Scherr~\cite{ruedenberg-scherr:53} on spectra of aromatic carbohydrate
molecules elaborating an idea of L.~Pauling. A new impetus came in the
eighties from the need to describe semiconductor graph-type
structures, cf.~\cite{exner-seba:89}, and the interest to these
problems driven both by mathematical curiosity and practical
applications e.g.\ nano-technology, network theory, optics, chemistry
and medicine is steadily growing.

Mathematically, many of these problems can be described by suitably
definded Laplace operators on graphs. For example, relevant
information of the corresponding model like transport properties of
the medium may be infered by the spectrum of the Laplacian.  There are
basically two classes of operators on graphs: On a
\emph{combinatorial} or \emph{discrete} graph, the Laplacian or
Schr\"odinger operator is defined as a \emph{difference} operator on
function on the \emph{vertices}. The edges here only play the role of
an incidence relation.  In contrast, on a \emph{metric} graph, the
basic operator acts on each edge as a one-dimensional
Schr\"odinger-type operator with certain boundary conditions at each
vertex assuring that the global operator is self-adjoint. The metric
graph together with a self-adjoint differential operator is usually
called a \emph{quantum graph}.  It is almost impossible to give a
complete account to all relevant literature here. Instead, we refer to
the introductive surveys~\cite{kuchment:pre08,kuchment:05,kuchment:04} as
well as to the proceedings~\cite{efkk:pre08,bcfk:06} and the
references therein.

Since quantum graphs are supposed to model various real graph-like
structures with the transverse size which is small but non-zero, one
has to ask naturally how close are such system to an ``ideal'' graph
in the limit of zero thickness. For anwers to this question we refer
to the papers~\cite{kuchment-zeng:01, rubinstein-schatzman:01,
  post:06,exner-post:pre07b} and the references therein.

In this paper, we study families of infinite quantum graphs with some
inherent randomness and prove that the spectra of the associated
Schr\"odinger-type operators are almost surely pure point.  In this
manner, the radial random quantum graphs act as one-dimensional random
Schr\"odinger operators exhibiting Anderson localization at all
energies.

We consider quantum graphs consisting of a rooted infinite metric tree
that are \emph{radial}.  A radial quantum graph is one for which all
variables, such as the branching number, edge length, and vertex
boundary conditions, depend only on the generation. The generation of
a vertex is determined by the distance from the root vertex.  A common
example of a rooted infinite metric tree is the rooted Bethe lattice.

We study two main models of random quantum graphs for which the
randomness is introduced in two ways.  The \emph{Random Length Model}
(RLM) is a quantum graph for which the edge length $\ell_e$ is given,
for example, by $\ell_e ( \omega_e ) = \ell_0 e^{\omega_e}$, where $\{
\omega_e \}$ is a family of independent, identically distributed
(\emph{iid}) random variables.  In a \emph{radial RLM}, the family of
\emph{iid} random variables $\{ \omega_e \}$ depends only on the
generation, not on the individual edge.  The \emph{Random Kirchhoff
  Model} (RKM) is a quantum graph and a family $\{ q(v) \}$
\emph{iid} random variables associated with each vertex and entering
into the Kirchhoff boundary conditions at each vertex.  Roughly
speaking, if $E_v$ is the set of edges entering the vertex $v$, the
Kirchhoff boundary condition is
\begin{equation}
  \label{kirch1}
  \sum_{ e \in E_v} f_e' (v) = q(v) f(v) .
\end{equation}
where the precise formulation is given in
\eqref{eq:bd.cond1}--\eqref{eq:bd.cond2}.  Physically, the current
flow through the vertex is determined by the random coupling
$q(v)=\omega_v$. A \emph{radial RKM} is one for which the \emph{iid}
random variables $\{ \omega_v \}$ depend only on the generation of the
vertex.  Under some conditions, we prove that the almost sure spectrum
of both of these models is pure point with exponentially decaying
eigenfunctions.

Random quantum graphs have been studied more extensively only in the
last years. There are works concerning the existence and continuity
properties of the integrated density of states (IDS) of various random
graph models, see
e.g.~\cite{kostrykin-schrader:04,glv:pre07,glv:07,helm-veselic:07}.
Localization has been proved e.g.\
in~\cite{hkk:05,ehs:07,klopp-pankrashkin:08} where the considered
models resemble the RKM or random potential model on the edges but
where different methods are used.

There is one major article that we are aware of on random length
models. An important contribution and the basis for our work on the
\emph{nonradial} RLM is given by Aizenman, Sims, and
Warzel~\cite{asw:pre05b}. These authors consider the nonradial RLM in
the weak disorder limit. As for the radial RLM, the edge lengths
$\ell_e$ are given by $\ell_e ( \omega_e ) = \ell_0 e^{\tau
  \omega_e}$, where $\{ \omega_e \}$ is a family of independent,
identically distributed random variables and $\tau$ is a measure of
the disorder.  They prove that as the disorder parameter $\tau \to 0$,
there is some absolutely continuous spectrum near the absolutely
continuous spectrum of the unperturbed model with probability one.  As
we prove that the radial RLM always exhibits only localization for any
nonzero disorder, this shows that the assumption that the graph is
radial is a strong one.  One might expect that in the nonradial case
and for moderate disorder there are localized states near the band
edges of the unperturbed quantum graph, but the proof of this requires
different methods.  Proving localization for the radial case is a
first step.

As other applications of the methods developed here, we examine the
random necklace model of Kostrykin and Schrader
\cite{kostrykin-schrader:04} (see~\Sec{necklace}), and various
families of decorated graphs. The random necklace model consists of
loops with perimeters given by iid random variables and joined by
straight line segments of length one. Kostrykin and Schrader studied
the integrated density of states and proved the positivity of the
Lyapunov exponent for these models.  We complete this study by proving
Anderson localization for the random necklace model in
\Thm{loc.necklace2}.  Graph decorations have been studied as a
mechanism for introducing spectral gaps in the combinatorial
\cite{aizenman-schenker:00} and quantum \cite{kuchment:05} case.  We
consider decorated graphs obtained from the RLM or the RKM by
adjoining compact graphs at each generation. We prove that such
decorations preserve localization, although there is a discrete set of
exceptional energies determined by the Dirichlet Laplacian on the
compact decoration graphs.

The contents of this paper are as follows.  In \Sec{rad.tree.gr}, we
describe the basic family of radial metric trees and the corresponding
operators. We refer to a tree plus the corresponding differential
operator as a quantum graph. Using a symmetry reduction emphasized by
Solomyak \cite{solomyak:04}, we reduce the problem on rooted radial
trees to an effective half-line problem with certain singularities at
the vertices. We present a generalized version of this symmetry
reduction in \App{reduction} for completeness (cf.\ 
\cite{sobolev-solomyak:02} for the standard case).  Transfer matrix
methods can now be used to describe solutions to the generalized
eigenvalue problem on the effective half-line.  We conclude by
computing the spectrum of the periodic problems and the deterministic
spectrum of the random models.  \Sec{ran.tg.loc} is devoted to the
proof of localization for the RLM and the RKM (cf.~\Thm{main}). The
proof relies on the positivity of the Lyapunov exponent
\cite{ishii:73, kotani:86} and an extension of Kotani's spectral
averaging method \cite{kotani:86}. The spectral averaging technique
employed here is new as one must deal with \emph{complex} matrices in
$\SL_2(\C)$ instead of real ones in $\SL_2(\R)$. We consider general
decorated graphs in \Sec{gen.gr}. We define the permissible decoration
graphs and construct radial tree-like quantum graphs corresponding to
the RKM and RLM. By the symmetry reduction procedure, we obtain
line-like quantum graphs in analogy to the reduction of the RKM and
RLM, and construct their transfer matrices. In \Sec{ran.graphs}, we
extend the arguments of \Sec{ran.tg.loc} to these families of
decorated graphs and prove localization (cf.~\Thm{kotani2},
\Thms{loc.end.point}{loc.kirchhoff}).  We show how to prove
localization for the random necklace model by extending the methods
used here to the line, following the general arguments in Kotani and
Simon \cite{kotani-simon:87}.

There are many works on quantum graphs, cf.\ volume {\bf 14} of {\it
  Waves in Random Media} and two review papers of Kuchment
\cite{kuchment:04, kuchment:05}. Much of these works emphasize compact
quantum graphs or compact quantum graphs with leads extending to
infinity.  Both of these classes of quantum graphs are different from
those considered here.  There are many results on unbounded quantum
graphs that might be well-known to the experts but whose proofs we
could not find in the literature. In the appendices, we systematically
present these results.  \App{reduction} present the proof of the
symmetry reduction for generalized radial tree graphs. In \App{bd.ef},
we extend many known results concerning generalized eigenfunctions to
quantum graphs.  We apply these results to establish a functional
calculus using the generalized eigenfunctions.  \App{ll.gr.gen.ef} is
devoted to the extension of these results to the line-like graphs
obtained from decorated radial graphs by the symmetry reduction.
\App{tm.weyl} presents an application of the material on generalized
eigenfunctions to the transfer matrices and Weyl-Titchmarsh functions
associated with quantum graphs.  The Dirichlet-to-Neumann map for
quantum graphs is introduced and used to study the transfer matrix.
The Dirichlet-to-Neumann map is particularly useful in the analysis of
decorated graphs. The last \App{sp.av} is devoted to the extension of
spectral averaging needed in the proofs of localization.

%----------------------------------------------------------------------
\subsection*{Acknowlegdements}
%----------------------------------------------------------------------
PDH thanks Simone Warzel for discussions on random quantum graphs. OP
thanks Peter Kuchment for the invitation at TAMU and for suggesting
the approach to the monodromy matrix via the Dirichlet-to-Neumann map
(cf.~\Sec{trans.mat}). OP would also like to thank G\"unter Stolz for
the invitation to UAB and general remarks on localization.

%%%%%%%%%%%%%%%%%%%%%%%%%%%%%%%%%%%%%%%%%%%%%%%%%%%%%%%%%%%%%%%%%%%%%%%
%----------------------------------------------------------------------
%
\section{Radial quantum tree graphs and their reduction}
\label{sec:rad.tree.gr}
%
%----------------------------------------------------------------------
%%%%%%%%%%%%%%%%%%%%%%%%%%%%%%%%%%%%%%%%%%%%%%%%%%%%%%%%%%%%%%%%%%%%%%%
In this section, we define the basic concept of quantum tree graphs.
We specialize to the family of radial quantum tree graphs and state a
theorem on the reduction of the full graph Hamiltonian to a countable
family of half-line Hamiltonians, with singularities at a discrete set
of points. In the ergodic case, such as the RKM and RLM, these
half-line Hamiltonians are unitarily equivalent. Finally, we introduce
the transfer matrices on the half-line models.  Transfer matrices will
play an important role in the spectral theory of the random models.

%%%%%%%%%%%%%%%%%%%%%%%%%%%%%%%%%%%%%%%%%%%%%%%%%%%%%%%%%%%%%%%%%%%%%%%
%----------------------------------------------------------------------
\subsection{Tree graphs}
\label{sec:tree.gr}
%----------------------------------------------------------------------
A \emph{discrete graph} $T$ is given by a triple $T \equiv (V,E,\bd)$,
where $V=V(T)$ denotes the set of vertices, $E=E(T)$ the set of
(directed) edges and the map $\map \bd E {V \times V}$ maps an edge
$e$ onto its start/end point $\bd e=(\bd_-e, \bd_+e)$.  For two
vertices $v,w \in V$ such that there is an edge $e \in E$ with $\bd e
= (v,w)$ or $\bd e=(w,v)$ we write $v \sim w$.  For each vertex $v \in
V$ we set
\begin{equation}
  \label{eq:ed.vx}
  E_v^\pm := \set {e \in E} {\bd_\pm e = v} \qquad \text{and} \qquad
  E_v := E_v(T) := E_v^+ \dcup E_v^-,
\end{equation}
i.e., $E_v^\pm=E_v^\pm(T)$ consists of all edges starting ($-$),
respectively, ending ($+$) at $v$, and $E_v$ their \emph{disjoint}
union. Note that the \emph{disjoint} union is necessary in order to
allow self-loops, i.e., edges having the same starting and ending
point so that the edge occurs in both $E_v^+$ and $E_v^-$, whereas we
only want it to occur once in $E_v$.  The \emph{degree} $\deg v$ of a
vertex $v$ is given by the number of edges emanating from $v$, i.e.,
$\deg v:= |E_v|$.

A \emph{path of length $n$} from a vertex $v$ to a vertex $w$ is a
sequence of vertices $v_0=v, \dots, v_n=w$ such that $v_i \sim
v_{i+1}$. The \emph{discrete distance} $\delta(v,w)$ of $v$ and $w$ is
the shortest length of a simple path joining $v$ and $w$.

A \emph{tree graph} is a graph $T$ without (nontrivial) closed paths
(i.e., every closed path has length $0$). If we fix a vertex $o \in
V(T)$ (the \emph{root vertex}) we say that $T$ is \emph{rooted at
  $o$}. \emph{We will always assume that our tree graphs are rooted}.

On a rooted tree graph we can define the notion of the
\emph{generation} $\gen v$: Every vertex with $\delta(o,v)=n$ is said
to be in \emph{generation $n$}. All edges are supposed to be directed
\emph{away} from the root $o$, i.e.  $\bd_-e =w$ and $\bd_+e=v$ where
$\gen w = n-1 < \gen v = n$. The generation of an edge $e$ is then the
generation of the subsequent vertex, i.e., $\gen e := \gen \bd_+e=n$.
The \emph{branching number} of a vertex $v$ is the number of
\emph{succeeding} edges, i.e., $b(v):=\deg v - 1$.
%----------------------------------------------------------------------
\begin{figure}[h]
  \centering
%----------------------------------------------------------------------
%  \input{loc-graph-fig1.pstex_t}
\begin{picture}(0,0)%
\includegraphics{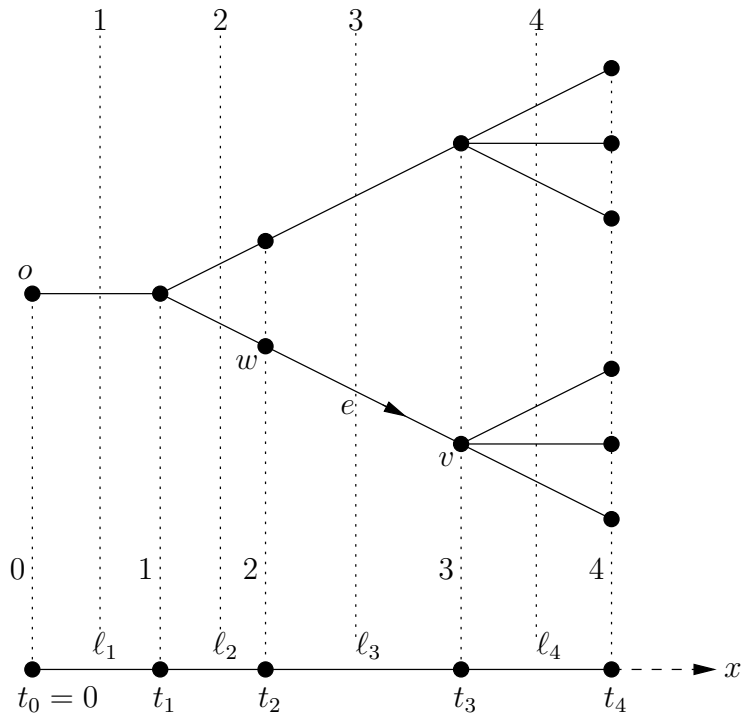}%
\end{picture}%
\setlength{\unitlength}{4144sp}%
\begingroup\makeatletter\ifx\SetFigFont\undefined%
\gdef\SetFigFont#1#2#3#4#5{%
  \reset@font\fontsize{#1}{#2pt}%
  \fontfamily{#3}\fontseries{#4}\fontshape{#5}%
  \selectfont}%
\fi\endgroup%
\begin{picture}(4551,4203)(271,-3481)
\put(3376,614){\makebox(0,0)[lb]{\smash{{\SetFigFont{12}{14.4}{\rmdefault}{\mddefault}{\updefault}{\color[rgb]{0,0,0}$4$}%
}}}}
\put(2836,-1996){\makebox(0,0)[lb]{\smash{{\SetFigFont{12}{14.4}{\rmdefault}{\mddefault}{\updefault}{\color[rgb]{0,0,0}$v$}%
}}}}
\put(2296,614){\makebox(0,0)[lb]{\smash{{\SetFigFont{12}{14.4}{\rmdefault}{\mddefault}{\updefault}{\color[rgb]{0,0,0}$3$}%
}}}}
\put(2251,-1681){\makebox(0,0)[lb]{\smash{{\SetFigFont{12}{14.4}{\rmdefault}{\mddefault}{\updefault}{\color[rgb]{0,0,0}$e$}%
}}}}
\put(766,614){\makebox(0,0)[lb]{\smash{{\SetFigFont{12}{14.4}{\rmdefault}{\mddefault}{\updefault}{\color[rgb]{0,0,0}$1$}%
}}}}
\put(1486,614){\makebox(0,0)[lb]{\smash{{\SetFigFont{12}{14.4}{\rmdefault}{\mddefault}{\updefault}{\color[rgb]{0,0,0}$2$}%
}}}}
\put(1621,-1411){\makebox(0,0)[lb]{\smash{{\SetFigFont{12}{14.4}{\rmdefault}{\mddefault}{\updefault}{\color[rgb]{0,0,0}$w$}%
}}}}
\put(316,-3436){\makebox(0,0)[lb]{\smash{{\SetFigFont{12}{14.4}{\rmdefault}{\mddefault}{\updefault}{\color[rgb]{0,0,0}$t_0=0$}%
}}}}
\put(1126,-3436){\makebox(0,0)[lb]{\smash{{\SetFigFont{12}{14.4}{\rmdefault}{\mddefault}{\updefault}{\color[rgb]{0,0,0}$t_1$}%
}}}}
\put(1756,-3436){\makebox(0,0)[lb]{\smash{{\SetFigFont{12}{14.4}{\rmdefault}{\mddefault}{\updefault}{\color[rgb]{0,0,0}$t_2$}%
}}}}
\put(2926,-3436){\makebox(0,0)[lb]{\smash{{\SetFigFont{12}{14.4}{\rmdefault}{\mddefault}{\updefault}{\color[rgb]{0,0,0}$t_3$}%
}}}}
\put(3826,-3436){\makebox(0,0)[lb]{\smash{{\SetFigFont{12}{14.4}{\rmdefault}{\mddefault}{\updefault}{\color[rgb]{0,0,0}$t_4$}%
}}}}
\put(271,-2671){\makebox(0,0)[lb]{\smash{{\SetFigFont{12}{14.4}{\rmdefault}{\mddefault}{\updefault}{\color[rgb]{0,0,0}$0$}%
}}}}
\put(1036,-2671){\makebox(0,0)[lb]{\smash{{\SetFigFont{12}{14.4}{\rmdefault}{\mddefault}{\updefault}{\color[rgb]{0,0,0}$1$}%
}}}}
\put(1666,-2671){\makebox(0,0)[lb]{\smash{{\SetFigFont{12}{14.4}{\rmdefault}{\mddefault}{\updefault}{\color[rgb]{0,0,0}$2$}%
}}}}
\put(2836,-2671){\makebox(0,0)[lb]{\smash{{\SetFigFont{12}{14.4}{\rmdefault}{\mddefault}{\updefault}{\color[rgb]{0,0,0}$3$}%
}}}}
\put(3736,-2671){\makebox(0,0)[lb]{\smash{{\SetFigFont{12}{14.4}{\rmdefault}{\mddefault}{\updefault}{\color[rgb]{0,0,0}$4$}%
}}}}
\put(766,-3121){\makebox(0,0)[lb]{\smash{{\SetFigFont{12}{14.4}{\rmdefault}{\mddefault}{\updefault}{\color[rgb]{0,0,0}$\ell_1$}%
}}}}
\put(1486,-3121){\makebox(0,0)[lb]{\smash{{\SetFigFont{12}{14.4}{\rmdefault}{\mddefault}{\updefault}{\color[rgb]{0,0,0}$\ell_2$}%
}}}}
\put(2341,-3121){\makebox(0,0)[lb]{\smash{{\SetFigFont{12}{14.4}{\rmdefault}{\mddefault}{\updefault}{\color[rgb]{0,0,0}$\ell_3$}%
}}}}
\put(3421,-3121){\makebox(0,0)[lb]{\smash{{\SetFigFont{12}{14.4}{\rmdefault}{\mddefault}{\updefault}{\color[rgb]{0,0,0}$\ell_4$}%
}}}}
\put(4546,-3256){\makebox(0,0)[lb]{\smash{{\SetFigFont{12}{14.4}{\rmdefault}{\mddefault}{\updefault}{\color[rgb]{0,0,0}$x$}%
}}}}
\put(316,-871){\makebox(0,0)[lb]{\smash{{\SetFigFont{12}{14.4}{\rmdefault}{\mddefault}{\updefault}{\color[rgb]{0,0,0}$o$}%
}}}}
\end{picture}%
%----------------------------------------------------------------------

  \caption{A radial tree graph with tree generations and branching
    numbers $b_0=1$, $b_1=2$, $b_2=1$, $b_3=3$; above the edge
    generation and below the vertex generation, e.g., the vertex $v$
    and the edge $e$ are in generation $3$. The bottom line is the
    corresponding half-line of the symmetry reduction.}
  \label{fig:tree}
\end{figure}

A rooted tree graph is \emph{radial} if the branching number $b(v)$ is
a function of the generation only, i.e., there exists a sequence
$(b_n)$ such that $b(v)=b_n$ for all $v \in V$ with $\gen v = n$
(cf.~\Fig{tree}).

A discrete tree graph $T \equiv (V,E,\bd)$ becomes a \emph{metric}
tree graph if there is a length function $\map \ell E {(0,\infty)}$
assigning a length $\ell_e$ to each edge $e \in E$. We identify each
edge $e$ with the interval $(0,\ell_e)$ turning $T$ into a
one-dimensional space with singularities at the vertices. In this way
we can define a \emph{continuous} distance function $d(x,y)$ for $x,y
\in T$ so that $T$ becomes a metric space.

A metric tree graph is \emph{radial} if it is a radial tree graph and
the length function depends only on the generation, i.e., if there is
a sequence $\{\ell_n\}_n$ such that $\ell_e=\ell_n$ for all edges $e$
in generation $n$. We assume that the lengths are bounded from below
and from above by finite, positive constants $\ell_\pm>0$, i.e.,
\begin{equation}
  \label{eq:len.bd}
  \ell_- \le \ell_n \le \ell_+
\end{equation}
for all $n \in \N$.  \emph{In the remaining parts of this and the next
  section (\Secs{rad.tree.gr}{ran.tg.loc}), we will only consider
  radial metric tree graphs.}  We will consider decorations of such
graphs in \Sec{gen.gr}.

%%%%%%%%%%%%%%%%%%%%%%%%%%%%%%%%%%%%%%%%%%%%%%%%%%%%%%%%%%%%%%%%%%%%%%%
%----------------------------------------------------------------------
\subsection{Radial Quantum Tree Graphs}
\label{sec:tree.qg}
%----------------------------------------------------------------------

We associate a Hilbert space $\Lsqr T$ with a general metric tree
graph by setting\footnote{Here and in the sequel, $\bigoplus_n \HS_n$
  always means the Hilbert space of all square-integrable sequences
  $\{f_n\}$, i.e., the \emph{closure} of the algebraic direct sum.}
$\Lsqr T := \bigoplus_{e \in E} \Lsqr e$, with norm given by
\begin{equation}
  \label{eq:norm}
  \normsqr f := \normsqr[T] f :=
  \sum_{e \in E} \int_e |f_e(x)|^2 \dd x.
\end{equation}
For \emph{radial} functions, i.e., functions depending only on
$d(o,x)$, this norm takes a simple form. Let $f_n$ denote the
restriction of the edge function $f_e$ to one of the edges at
generation $n$. We then have
\begin{equation}
\label{eq:norm2}
  \normsqr f =
  \sum_{n=1}^\infty ~\wt b_n \int_0^{\ell_n} |f_n (x)|^2 \dd x,
\end{equation}
where $f_n=f_e$ for an edge $e$ at generation $n$ and where $\wt b_n$
is the number of edges at generation $n$ and is a function of the
branching numbers $\{b_n\}_n$. For a radial tree graph with branching
number $b_n = b$ ($n \ge 1$) and $b_0=1$, often referred to as a Bethe
lattice, we have $\wt b_n = b^{n-1}$.

We next define our main operator on metric trees that make these trees
into \emph{quantum trees}.  The \emph{Dirichlet Hamiltonian}
$H=H(T,q)$, with strength $\map q V \R$, is defined by
\begin{equation}
  \label{eq:def.op}
  (Hf)_e = -f_e''
\end{equation}
on each edge for functions $f \in \dom H$ satisfying $f \in
\bigoplus_{e \in E} \Sob[2] e$ and satisfying two conditions. First,
the functions are continuous at each vertex,
\begin{equation}
\label{eq:bd.cond1}
    f_{e_1}(v) = f_{e_2}(v), \qquad \forall e_1, e_2 \in E_v.
\end{equation}
We will write $f(v)$ for the unique value.
Second, the functions
satisfy the Kirchhoff boundary conditions at each vertex,
\begin{equation}
\label{eq:bd.cond2}
    \sum_{j=1}^{b(v)} f_{e_j}' (\bd_- e_j) - f_{e_0}'(\bd_+ e_0)  =
    q(v) f(v),
\end{equation}
for all vertices $v \in V \setminus \{o\}$, where $e_0$ is the edge
preceding $v$ and $e_j$ label the $b(v)$ subsequent edges at the
vertex $v$.  \hiddenfootnote{We have chosen the sign of $q$ in order
  that the corresponding operator is nonnegative is $q \ge 0$: Note
  that on a single line
  \begin{multline*}
    0 \le
    \int (|f'|^2 + q|f(0)|^2) \dd x =
    \int (\conj f \, (-f'')) \dd x + [\conj f f']_{0+}^{0-} \\=
    \int (\conj f \, (-f'')) \dd x + \conj{f(0)}[f'(0-)-f'(0+)+q f(0)]
  \end{multline*}
  which forces $f'(0+)-f'(0-)=qf(0)$ for functions in the operator
  domain. One should not be confused about the different sign notion
  in $\bd_\pm e$: The starting vertex $\bd_-e$ corresponds to the
  limit from the right (i.e., to the point $0+$ on the associated
  interval), and similarly for the end point.}  For the root vertex we
impose a Dirichlet boundary condition, i.e.,
\begin{equation}
  \label{eq:bd.cond.0}
  f(o)=0.
\end{equation}
Without loss of generality, we suppose that there is only one edge
emanating from the root vertex, i.e., $b_0=1$, since otherwise, the
(radial) Dirichlet Hamiltonian $H$ decouples into $b_0$ many operators
on the edge subtrees of $o$.

We assume that $q$ is a radial function, i.e., there is a sequence
$\{q_n\}_n$ such that $q(v)=q_n$ for all vertices $v$ at generation
$n$.  In this case, we also say that the Hamiltonian $H$ is
\emph{radial}.  In addition, we assume that there are constants $q_\pm
\in \R$ such that
\begin{equation}
  \label{eq:pot.bd}
  q_- \le q_n \le q_+
\end{equation}
for all $n$.

The \emph{free} Hamiltonian or \emph{Kirchhoff Laplacian} $\laplacian
T$ on $T$ is the Hamiltonian without the potential $q$ at the
vertices, i.e., $\laplacian T := H(T,0)$.

In summary, a \emph{radial quantum tree graph} is a metric graph with
an operator $H(T,q)$ satisfying
\eqref{eq:def.op}--\eqref{eq:bd.cond.0}.  It is determined by the
branching numbers $\{b_n\}_n$, the edge lengths $\{\ell_n\}_n$, and
potentials $\{q_n\}_n$ that depend only on the generation.

%%%%%%%%%%%%%%%%%%%%%%%%%%%%%%%%%%%%%%%%%%%%%%%%%%%%%%%%%%%%%%%%%%%%%%%
%----------------------------------------------------------------------
\subsection{Reduction of Radial Quantum Tree Graphs}
\label{sec:red.qt}
%----------------------------------------------------------------------

For simplicity, we assume in this section, that $b_n=b$ for all $n \ge
1$.  We will show in a more abstract setting that under the
assumptions~\eqref{eq:len.bd} and \eqref{eq:pot.bd}, the operator $H$
is essentially self-adjoint on the space of functions $f \in \dom H$
with \emph{compact} support (cf.~\Lem{ess.sa}) and that $H$ is
relatively form-bounded with respect to $\laplacian T$ with relative
bound $0$ (cf.~\Lem{form.bdd}).

The distance from the root vertex $o$ to a vertex of generation $n$,
for $n \geq 1$, is denoted $t_n = \sum_{k=1}^n \ell_k$. We set $t_0 =
0$. The main reason why the analysis of \emph{radial} Dirichlet
Hamiltonians is much easier than the general case is the following
symmetry reduction (cf.~\cite{naimark-solomyak:00,
  sobolev-solomyak:02, solomyak:04}).  For completeness, we will give
a proof in \App{reduction}, also in a more general setting.  The
points $t_k$ play the role of vertices. We denote by $f(t_k \pm) :=
\lim_{s \rightarrow t_k \pm} f(s)$.
\begin{theorem}
  \label{thm:red.tree}
  The radial Hamiltonian $H$ on a radial quantum tree graph is
  unitarily equivalent to $H_1 \oplus \bigoplus_{n=2}^\infty
  ( \oplus b^{n-2}(b-1)) H_n$, where $(\oplus m) H_n$ means the
  $m$-fold copy of $H_n$. The operator
$H_n$ is the self-adjoint operator on
  $\Lsqr {[t_{n-1}, \infty)}$ given by $H_n f = -f''$ away from the
  points $t_k$ and with boundary conditions
  \begin{subequations}
  \label{eq:bd.cd.red}
  \begin{gather}
    \label{eq:bd.cd.red1}
    f(t_k-) = b^{-1/2} f(t_k+),\\
    \label{eq:bd.cd.red2}
    f'(t_k-) + q_k f(t_k-) = b^{1/2} f'(t_k+)
  \end{gather}
%\look{Check whether the $q_k$ etc. are on the right
%    side!}
%% Comment: Checked: seems to be OK! (2006-05-17)
%% 2006-07-12: I changed
                                %% the sign in front of $q$
  for all $k \ge n$ and
  \begin{equation}
    \label{eq:bd.cd.red0}
    f(t_{n-1}+)=0.
  \end{equation}
  \end{subequations}
\end{theorem}

We will refer to the reduced quantum graph, the half-line $[t_{n-1},
\infty )$, with boundary conditions at the vertices, as a
\emph{line-like quantum graph}.  This is particularly useful in the
discussion of decorated graphs, and we discuss this further in
\Sec{reduct.gen} and \Def{ll.graph.qg}.

\Thm{red.tree} is particularly useful in the ergodic case, cf.\ 
\Sec{ran.tg.loc}.  In this case, the operators $H_n$ are all simply
related. First, ergodicity implies that each $H_n ( \omega)$ has
almost sure spectrum. Secondly, we have the relation $H_n ( \tau_{n-1}
\omega) = H_1 ( \omega)$, for any configuration $\omega$. Since the
shift operator $\map{U_k} {\Lsqr{[t_{n-1}, \infty)}}
{\Lsqr{[t_{n+k-1}, \infty)}}$ is unitary, the operators are related as
$U_{n-1}^{-1} H_n (\omega)U_{n-1} = H_1 (\omega)$ and the operators
are unitarily equivalent.  Hence, the almost sure spectral components
are \emph{independent} of $n$, and it suffices to prove almost sure
pure point spectrum for $H_1$, for example.

\begin{remark}
  \label{rem:exp.growth}
  Note that the functions $f$ on $\Lsqr{[t_{n-1},\infty)}$ are
  obtained from functions on the tree graph satisfying certain
  invariance conditions together with a exponential \emph{weight}
  function reminiscent the fact that there are $b^{n-1}$ contributions
  from the edges at generation $n$. For example, the constant function
  $\1$ on the tree graph (not lying in either the domain of the
  Dirichlet Hamiltonian nor in $\Lsqr T$) is transformed in the step
  function $f(x)=b^{k/2}$ for $t_k< x < t_{k+1}$. In particular, $f$
  increases exponentially.

  On the other hand, suppose that $f_n$ is an eigenfunction of $H_n$,
  for $n \geq 2$, on $\Lsqr{[t_{n-1}, \infty)}$, with eigenvalue
  $\lambda$. We construct an eigenfunction $\tilde{f}_n$ of $H$ on
  $\Lsqr{T}$ with eigenvalue $\lambda$ as follows. The function
  $\tilde{f}_n$ will be supported on a subtree associated to any
  vertex of generation $(n-1)$ on the tree and equal zero outside of
  this subtree.  The eigenvalue $\lambda$ will have a multiplicity at
  least equal to the number of vertices at generation $(n-1)$.  Fixing
  $b = 2$ for simplicity, and a subtree of $T$ with vertex $o_{n-1}$,
  we construct $\tilde{f}_n$ at the first generation of the subtree by
  setting $\tilde{f}_n = ( 1 / \sqrt{2}) f_n \restr{[t_{n-1} , t_n]}$
  on one edge, and minus this value on the other. At the $m$th
  generation of the subtree, we use the weight $2^{-m/2}$ and $f_n$
  restricted to $[t_{m+n-1}, t_{m+n}]$ to construct the value of
  $\tilde{f}_n$ on the edges with coefficients assigned according to
  the $b$th roots of unit.  It is easy so see that
  \begin{equation}
  \label{norm-on-tree1}
     \normsqr[T] {\tilde f _n} =
     \sum_{m \geq 1}
      \frac 1 {2^m} \normsqr { f_n \restr{[t_{m+n-1},t_{m+n}]}}.
  \end{equation}
  In particular, if the eigenfunction $f_n$ of $H_n$ decays
  exponentially, that is, if $e^{\gamma d(0,x)} f_n \in
  \Lsqr{[t_{n-1}, \infty)}$, if follows from the fact that the
  distance function is a radial function and~\eqref{norm-on-tree1},
  that $\e^{\gamma d(o,x)} \tilde{f}_n \in \Lsqr{T}$.
\end{remark}

%%%%%%%%%%%%%%%%%%%%%%%%%%%%%%%%%%%%%%%%%%%%%%%%%%%%%%%%%%%%%%%%%%%%%%%
%----------------------------------------------------------------------
\subsection{Transfer Matrices and Generalized Eigenfunctions of the
  Reduced Operator $H_1$}
%----------------------------------------------------------------------
\label{sec:tm.tree}

We want to characterize the growth rate of the generalized
eigenfunctions $f$ of $H_n$. We will consider $H_1$ explicitly since
in the ergodic case the symmetry reduction in \Thm{red.tree} shows
that $H_1$ is unitarily equivalent to $H_n$.

We study functions $\map f {[0,\infty)} \C$ satisfying $-f''=\lambda
f$ away from the vertices $t_k$ and~\eqref{eq:bd.cd.red} at the
vertices $t_k$.  We assume here that $\lambda \ne 0$ (the case
$\lambda=0$ can be treated similarly, but it is unimportant for our
purposes).  If we know that $H_1 \ge 0$ (e.g., if $q\ge 0$), we may
assume here $\lambda>0$. In the definition of the Weyl-Titchmarsh
functions (see \Sec{weyl}) we also need generalized eigenfunctions for
\emph{complex} $z=\lambda+\im \eps$, $\eps>0$. In concrete examples,
it is often more convenient to use $\mu =\sqrt {|\lambda|}$ (or in the
complex case, $w = \sqrt z$, the branch with $\Im w>0$) as parameter.
We will switch between these two parameters without mentioning.

A basic fact that we use often is that the
existence of a generalized eigenfunction of $H_1$ solving $H_1 f =
\lambda f$ is equivalent with the existence of a nontrivial solution
of a discrete map $\map {{\vec F}_\lambda} \N {\C^2}$ since, on the open
interval $(t_{n-1}, t_n)$, the eigenfunction must have the form
\begin{equation}
\label{edgesoln1}
   f (x) =
   f(t_{n-1}^+) \cos ( \sqrt{\lambda} x) +
   \frac {1}{\sqrt{\lambda}} f'(t_{n-1}^+ ) \sin ( \sqrt{\lambda} x ),
  \quad \text{for $x \in (t_{n-1} , t_n)$,}
\end{equation}
for $\lambda>0$ (and similar for the other cases).  The infinite
family of coefficients
\begin{equation*}
  \{ f(t_{n-1}^+),  f'(t_{n-1}^+) \}_n
\end{equation*}
is determined iteratively by the map ${\vec F}_\lambda$ defined below
and the boundary conditions \eqref{eq:bd.cd.red}.

The discrete map ${\vec F}_\lambda$ is defined using the
transfer matrix as follows. The \emph{transfer matrix} $T_\lambda
(n)$ is given by
\begin{equation}
\label{eq:tm.tree}
  T_\lambda(n) =
  D(b) S(q_n)  R_{\sqrt \lambda}(\sqrt \lambda \ell_n)%\\=
\end{equation}
where the factors of the transfer matrix are the matrices
\begin{subequations}
  \label{eq:def.el.mat}
\begin{align}
  S(\kappa)&:= \begin{pmatrix} 1 & 0 \\ \kappa & 1 \end{pmatrix}, &
  D(b)     &:= \begin{pmatrix} b^{1/2} & 0 \\ 0  & b^{-1/2} \end{pmatrix}\\
  R_\mu(\phi)&:= \begin{pmatrix} \cos \phi & \dfrac {\sin \phi} \mu \\
    -\mu \sin \phi & \cos \phi \end{pmatrix}.
\end{align}
\end{subequations}
These are the standard matrices of shearing, dilation, and (elliptic)
rotation, respectively.  Note that $|\tr S(\kappa)|=2$, $|\tr D(b)|>2$
and $|\tr R_\mu(\phi)|<2$ (for real $\mu$ and $\phi$). A matrix $A \in
\SL_2(\R)$ is called \emph{parabolic, hyperbolic}, respectively,
\emph{elliptic}, if $|\tr A|=2$, $|\tr A|>2$, respectively, $|\tr
A|<2$.  For $\lambda<0$ we set $\mu := \sqrt{|\lambda|}$ and we obtain
the hyperbolic ``rotation'' matrix
\begin{equation*}
  R^{\mathrm h}_\mu (\phi) := R_{\im \mu}(\im \phi) =
  \begin{pmatrix}
    \cosh \phi & \frac 1 \mu \sinh \phi\\ \mu \sinh \phi &\cosh \phi
  \end{pmatrix}.
\end{equation*}

Given a vector ${\vec \alpha}_0 \in \C^2$, we obtain another vector
${\vec \alpha}_n$ by
\begin{equation}
  \label{abstract1}
  {\vec \alpha}_n =
  T_\lambda (n) T_\lambda (n-1) \ldots T_\lambda (1) {\vec \alpha}_0 .
\end{equation}
We define the map $\map {{\vec F}_\lambda} \N {\C^2}$ at site $n$ as
the product of transfer matrices acting on ${\vec \alpha}_0$,
\begin{equation}
\label{abstract2}
  {\vec F}_\lambda (n) =
  {\vec \alpha}_n =
  T_\lambda (n) T_\lambda (n-1) \ldots T_\lambda (1) {\vec \alpha}_0 .
\end{equation}
The map ${\vec F}_\lambda$ depends on the energy $\lambda \in \R$ and the
initial vector ${\vec \alpha}_0$.
We note that ${\vec F}_\lambda$ satisfies the condition
\begin{equation}
\label{eq:trans.mat}
  {\vec F}_\lambda (n) =
  T_\lambda(n) {\vec F}_\lambda (n-1), \quad\text{for $n\ge 1$.}
\end{equation}
Given an initial condition ${\vec \alpha}_0$ and the corresponding
sequence of coefficients ${\vec \alpha}_n$ obtained as
in~\eqref{abstract2}, we can construct a generalized eigenfunction $f$
for $H_1$ with eigenvalue $\lambda$, as in~\eqref{edgesoln1}, by using
the vector ${\vec \alpha}_n$ for the coefficients $\{ f( t_{n-1}^+),
f'(t_{n-1}^+) \}$.

Conversely, suppose we have a generalized eigenfunction $f$ of $H_1$
satisfying $H_1 f = \lambda f$, and Dirichlet boundary conditions $f
(0+) = 0$. Then, for each $n \geq 1$, it is easy to check that
\begin{equation}
  \label{eq:vec.f}
  {\vec F}_\lambda (n) :=
  \begin{pmatrix} f(t_n+)\\ f'(t_n+) \end{pmatrix},
\end{equation}
with the initial condition
\begin{equation}
  \label{eq:trans.mat.initial}
  {\vec F}_\lambda (0) =
  \begin{pmatrix} 0\\ f'(0+) \end{pmatrix}.
\end{equation}

We can interpret the \emph{transfer} or \emph{monodromy} matrix
$T_\lambda(n)$ as follows: Starting with the vector ${\vec F}_\lambda
(n-1)$ at the vertex $t_{n-1}$ we evolve the free eigenvalue equation
on the edge until the vertex $t_n$ (rotation matrix). The shearing
matrix corresponds to the delta-potential at $t_n$ and finally, the
dilation matrix encodes the jump condition at $t_n$ due to the
branching number. Note that $T_\lambda(n)$ is an unimodular matrix,
i.e., $\det T_\lambda(n)=1$.

We also need a control of the $\Lsymb_2$-norm of an (a priori)
generalized eigenfunction of $H_1$ in terms of the sequence ${\vec
  F}_\lambda (n)$. We write ${\vec F}_\lambda (n) = ( F_\lambda (n) ,
F_\lambda ' (n) )^{\tr}$, for the components of ${\vec F}_\lambda
(n)$, and define a norm $|{\vec F}_\lambda |_\lambda^2:=| F_\lambda
|^2 + \frac 1 {|\lambda|} |F_\lambda' |^2$.  Then, for $\lambda>0$, it
follows from~\eqref{edgesoln1} and~\eqref{eq:vec.f} that we have
\begin{subequations}
  \label{eq:l2.L2}
  \begin{multline}
    \label{eq:l2.L2.1}
    \normsqr[\Lsqr {t_{n-1},t_n}] f +
       \frac 1 \lambda \normsqr[\Lsqr {t_{n-1},t_n}] {f'} =
    \int_{t_{n-1}}^{t_n}
         | R_\mu(\mu x) {\vec F}_\lambda (n-1)|_\lambda^2 \dd x \\\leq
    \ell_+ | {\vec F}_\lambda (n-1)|_\lambda^2,
  \end{multline}
  due to~\eqref{eq:len.bd}.  Note that $R_\mu(\phi)$ (for real $\phi$)
  is orthogonal with respect to this norm. In addition, for
  $\lambda<0$ and $\mu := \sqrt{|\lambda|}$, we have
  \begin{multline}
    \label{eq:l2.L2.2}
    \normsqr[\Lsqr {t_{n-1},t_n}] f +
       \frac 1 {|\lambda|} \normsqr[\Lsqr {t_{n-1},t_n}] {f'} =
    \int_{t_{n-1}}^{t_n} | R_{\im \mu}(\im \mu x)
             {\vec F}_\lambda (n-1)|_\lambda^2 \dd x \\\le
    2\e^{2 \mu \ell_+} |{\vec F}_\lambda (n-1)|_\lambda^2 .
  \end{multline}
\end{subequations}

In particular, if $\{ {\vec F}_\lambda (n)\}_n \in \lsqr{\N,\C^2}$,
then the associated generalized eigenfunction $f$ and its derivative
$f'$ are indeed square-integrable, i.e., $f, f' \in \Lsqr {\R_+}$.
Since there is also a lower bound on $\ell_e$, we also have the
converse statement; in particular, a generalized eigenfunction $f$ is
in $\Lsqr {\R_+}$ if and only if ${\vec F}_\lambda$ is in
$\lsqr{\N,\C^2}$.

%%%%%%%%%%%%%%%%%%%%%%%%%%%%%%%%%%%%%%%%%%%%%%%%%%%%%%%%%%%%%%%%%%%%%%%
%----------------------------------------------------------------------
\subsection{The Spectrum of a Quantum Graph for the Free and Periodic
Problem}
\label{sec:free-per.op}
%----------------------------------------------------------------------

We first consider the simple periodic problem obtained when all the
parameters are constant, i.e., when the transfer matrices
$T_\lambda=T_\lambda(n)$ are independent of $n$.  In this case, it
follows from \Thm{red.tree} that all the reduced
Hamiltonians $H_n$ are unitarily equivalent.

\begin{theorem}
  \label{thm:sp.per.op}
  Suppose that the transfer matrices are independent of $n$ and that
  $\lambda \mapsto \tr T_\lambda$ is nonconstant.  Then, the spectrum
  of $H$ consists only of essential spectrum.  The spectrum is given
  by the set $\Sigma_{\mathrm{ac}}$ of $\lambda \in \R$ for which
  $T_\lambda$ is elliptic or parabolic (i.e., $|\tr T_\lambda| \le 2$)
  and the set $\Sigma_{\mathrm{pp}}$ of all energies $\lambda$ such
  that $(0,1)^{\mathrm tr}$ is an eigenvector of $T_\lambda$ with
  eigenvalue $\tau$ such that $|\tau|<1$.  The spectrum is purely
  absolutely continuous on $\Sigma_{\mathrm{ac}}$ and pure point on
  $\Sigma_{\mathrm{pp}}$.
\end{theorem}
\begin{proof}
  In the periodic case, $H$ is unitarily equivalent to infinitely many
  copies of $H_1$ by \Thm{red.tree}.  We let $\wt H_1$ be the periodic
  operator on $\Lsqr \R$ with $\wt H_1f=-f''$ on each edge and with
  boundary conditions~\eqref{eq:bd.cd.red1}--\eqref{eq:bd.cd.red2} on
  $t_k>0$ ($k>0$) and similarly for $t_{-k}=-t_k<0$ ($k \ge 0$) with
  $b^{1/2}$ replaced by $b^{-1/2}$. Let $H_{1,-}$ be the same operator
  as $H_1$, but on $\Lsqr {\R_-}$ (again, replacing $b^{1/2}$ by
  $b^{-1/2}$ in the boundary conditions, and with Dirichlet boundary
  condition at $0$). Then $H_{1,-} \oplus H_1$ is a rank one
  perturbation of $\wt H_1$, in particular, the absolutely continuous
  spectrum is the same. But the latter can be calculated by Floquet
  theory (cf.~\cite[Sec.~XIII.16]{reed-simon-4}) and consists of the
  set of $\lambda=\mu^2$ for which there exists $\theta \in [0,\pi)$
  such that $\tr T_\lambda=2 \cos \theta$. The latter equation
  determines the dispersion relation; since $\tr T_{\mu^2}$ is
  analytic (cf.~\eqref{eq:d2n.series} and~\eqref{eq:def.tm}) and
  nonconstant, the spectrum $\Sigma_{\mathrm{ac}}$ is purely
  absolutely continuous (cf.~\cite[Thm.~XIII.86]{reed-simon-4}). Note
  that $\Sigma_{\mathrm{ac}}$ and $\Sigma_{\mathrm{pp}}$ are always
  disjoint, since for parabolic or elliptic matrices, all eigenvalues
  $\tau$ satisfy $|\tau|=1$.

  The additional eigenvalues of $H_1$ are of multiplicity $1$ (and
  therefore of infinite multiplicity for $H$) and occur, if $T_\lambda
  (0,1)^{\tr}= \tau (0,1)^{\tr}$ with $|\tau|<1$.
\end{proof}

\begin{remark}
  In \Lemenum{exc.set}{tm.cont} we can express the eigenvalue $\tau$
  in terms of the Dirichlet eigenfunction $\phi_k$ provided
  $\lambda=\lambda_k$ is a simple eigenvalue of the Dirichlet problem
  and $\phi_k^\dag(o_i)\ne 0$ for both boundary points $i=0$ and $i=1$
  (for the notation we refer to \Sec{gen.gr}). Then $|\tau|<1$ if and
  only if $|\phi_k^\dag(o_1)| \le \sqrt b |\phi_k^\dag(o_0)|$.
\end{remark}
Our two primary models, the RKM and the RLM, were described in the
introduction and are presented in detail in \Sec{ran.tg.loc}.  We
apply \Thm{sp.per.op} to compute the spectrum of the periodic version
of the RKM when the vertex potential strength is a constant $q$,
independent of $n$, and of the periodic version of the RLM when the
edge length is a constant $\ell$.  We will use these results to
compute the deterministic spectra of these models in
\Thm{as.spec.mod}.

The spectrum of the periodic RLM is simply the spectrum of the free
Hamiltonian $\laplacian {T(\ell)}$ on a rooted, regular, radial tree
$T(\ell)$ with a fixed branching number $b \geq 1$ and constant edge
length $\ell$.  Let us define $\theta \equiv \arccos ( 2(b +
b^{-1})^{-1})$.  The identification of the spectrum is well-known
(e.g.~using \Thm{sp.per.op} and~\eqref{eq:tm.rlm} or
\cite{cattaneo:97}) and we refer to \cite{sobolev-solomyak:02} for a
nice discussion.  Carlson \cite{carlson:97} proved that the spectrum
is purely absolutely continuous away from the points $\set{\pi^2 k^2 /
  \ell^2 }{k \in \N}$.

\begin{theorem}
  \label{thm:sp.per.lm}
  The spectrum of the free Hamiltonian $\laplacian {T(\ell)}$ on a
  regular radial tree $T(\ell)$, with branching number $b \geq 1$ and
  constant edge length $\ell$ is a union of bands and points:
  \begin{equation}
  \label{eq:sp.per.lm}
     \spec{\laplacian  {T(\ell)} } =
     \bigcup_{k=1}^\infty \Bigl( \frac 1 {\ell^2} B_k \cup \Bigl\{
                  \frac{\pi^2 k^2}{\ell^2}\Bigr\} \Bigr),
          \quad \text{where} \quad
     B_k= \bigl[ (\pi (k-1) + \theta)^2, (\pi k - \theta )^2 \bigr],
  \end{equation}
  and is purely absolutely continuous on $\bigcup_k \frac 1 {\ell^2}
  B_k$. If $b>1$, all gaps are open.
\end{theorem}

Note that when $b=1$, $\theta = 0$, and the
spectrum~\eqref{eq:sp.per.lm} reduces to the known spectrum of the
free Laplacian on the half-line with Dirichlet boundary conditions at
zero. In this case, $\pi^2k^2 \in B_k=[\pi^2(k-1)^2,\pi^2 k^2]$ and
the spectrum is absolutely continuous on $\R_+$.

We next apply \Thm{sp.per.op} to compute the spectrum of the periodic
RKM when the vertex potential strength is a constant $q$, independent
of $n$. We fix the length edge to be one.

\begin{theorem}
  \label{thm:sp.per.km}
  For the Hamiltonian $H(q)$ with constant vertex potential $q \in \R$
  on a metric tree with constant length $\ell=1$ the spectrum is given
  by
  \begin{equation}
    \label{eq:sp.per.km1}
    \spec {H(q)} =
    \Bigset{\lambda \in \R}
      {\bigl|\xi_b(\sqrt \lambda,q)  \bigr| \le
        \frac {2 b^{1/2}}{b+1}} \cup \set {\pi^2 k^2}{k \in \N},
  \end{equation}
  where
  \begin{equation*}
    \xi_b(\mu,q) =
    \cos \mu +  \frac {q \sin \mu} {\mu  (b+1)},
        \qquad
    \xi_b(\im \mu,q) =
    \cosh \mu + \frac {q \sinh \mu} {\mu (b+1)}
  \end{equation*}
  for $\mu>0$ and $\xi_b(0,q) = 1 + q/(b+1)$. Furthermore,
  \begin{equation}
    \label{eq:sp.per.km2}
    \spec {H(q)} =
    \bigcup_{k=1}^\infty \bigl(B_k(q) \cup \{\pi^2 k^2\} \bigr)
  \end{equation}
  where $B_k(q)$ are closed intervals. In addition, the spectrum is
  purely absolutely continuous on $\bigcup_k B_k(q)$.

  The bands satisfy $B_k(q) \subset [(k-1)^2\pi)^2,k^2\pi^2]$ for $k
  \ge 2$.  In addition, $B_1(q) \subset [0,\pi^2]$ if and only if $q \ge
  -(b^{1/2}-1)^2$, and $B_1(q) \subset (-\infty,0)$ if and only if $q <
  -(b^{1/2}+1)^2$.
  If $b=1$ and $q \ne 0$, then the intervals $B_k(q)$ ($k \ge 2$)
  touch only one of the points $\pi^2k^2$ or $\pi^2(k-1)^2$. If $b>1$,
  the points $\pi^2k^2$ never lie in the union of the bands $\bigcup_k
  B_k(q)$. In particular, if $b>1$ or $b=1$ and $q \ne 0$, all gaps
  are open.
\end{theorem}
\begin{proof}
  The spectral characterization is an application of \Thm{sp.per.op}
  using~\eqref{eq:tm.rkm}. The case $b=1$ has been analyzed
  in~\cite[Thm.~2.3.3]{aghh:88}.
\end{proof}

%%%%%%%%%%%%%%%%%%%%%%%%%%%%%%%%%%%%%%%%%%%%%%%%%%%%%%%%%%%%%%%%%%%%%%%
%----------------------------------------------------------------------
%
\section{Random quantum tree graphs and localization}
\label{sec:ran.tg.loc}
%
%----------------------------------------------------------------------
%%%%%%%%%%%%%%%%%%%%%%%%%%%%%%%%%%%%%%%%%%%%%%%%%%%%%%%%%%%%%%%%%%%%%%%

%%%%%%%%%%%%%%%%%%%%%%%%%%%%%%%%%%%%%%%%%%%%%%%%%%%%%%%%%%%%%%%%%%%%%%%
%----------------------------------------------------------------------
\subsection{Random quantum tree graphs}
\label{sec:ran.tg}
%----------------------------------------------------------------------
We consider now random perturbations of the length sequence
$\{\ell_n\}$ or the vertex potential strength $\{q_n\}$. Let
$(\Omega_1, \Prob_1)$ be a probability space and $(\Omega, \Prob) :=
(\Omega_1, \Prob_1)^\N$ the product probability space. In our
applications, $\Omega_1$ will always be a compact interval. To exclude
unnecessary complications (see e.g.~\eqref{eq:spec.cd1}), we assume
that $\supp \Prob_1=\Omega_1$ where $\supp \Prob_1$ is the largest
closed subset such that the complement is of $\Prob_1$-measure $0$.

We can define the notion of \emph{ergodicity} on such spaces: There is
a canonical (right) shift function
$(\tau_{n_0}\omega)(n):=\omega_{n_0+n}$ preserving the probability
measure $\Prob$ on $\Omega$. Note that $\tau_n=\tau_1^{\circ n}$.
\begin{definition}
  \label{def:ran.mod}
  A measure preserving map $\map {\tau_1} \Omega \Omega$ is called
  \emph{ergodic} if any measurable set $A \in \mathcal F$ with
  $\tau_1(A)=A$ satisfies $\Prob(A) \in \{0,1\}$.
\end{definition}
{}From the Kolmogorov $0$-$1$ law it follows that the (right) shift is
an ergodic action on $\Omega$ (cf.~e.g.~\cite[p.~26]{simon:79b}).

\begin{definition}
  \label{def:ran.tree}
  The \textbf{Random Length Model} (RLM) is a \emph{random length}
  quantum tree graph defined by an iid~sequence $\{\ell_n\}$ of random
  variables $\map {\ell_n} {\Omega_1} {(0,\infty)}$ satisfying
  \eqref{eq:len.bd} $\Prob_1$-almost surely. We denote the
  corresponding family of quantum tree graphs and Laplacians by
  $\{T(\omega)\}$ and $\{\laplacian {T(\omega)} \}$.

  The \textbf{Random Kirchhoff Model} (RKM) is a \emph{random
    Hamiltonian} on a radial quantum tree graph $T$ given by an
  iid~sequence $\{q_n\}$ of random variables $\map {q_n} {\Omega_1}
  {(0,\infty)}$ satisfying \eqref{eq:pot.bd} $\Prob_1$-almost surely.
  We denote the corresponding family of Hamiltonians on the (fixed)
  quantum tree graph $T$ by $\{H(\omega)\}$. For simplicity, we assume
  that $\ell_n=1$ for all $n$.
\end{definition}
To unify the notation, we denote both operators by $H(\omega)$ acting
on $T(\omega)$. Since $H(\omega)$ is radial (for almost all $\omega$),
we can apply the symmetry reduction~\Thm{red.tree} and obtain a family
of random operators $H_n(\omega)$. As a consequence of the ergodicity,
we obtain:
\begin{theorem}
  \label{thm:as.spec}
  The spectral components of the spectrum of $H(\omega)$ are almost
  surely constant, i.e., there exist subsets $\Sigma_\bullet$ such
  that $\spec[\bullet] {H(\omega)} = \Sigma_\bullet$ for almost all
  $\omega \in \Omega$.  In addition, the spectral sets
  $\Sigma_\bullet$ are determined by the corresponding almost sure
  spectrum of the Hamiltonian $H_1(\omega)$ on $\Lsqr{\R_+}$. Here,
  $\bullet$ labels either the pure point ($\mathrm{pp}$), the
  absolutely continuous ($\mathrm{ac}$) or singularly continuous
  ($\mathrm{sc}$) spectrum.
\end{theorem}

\begin{proof}
  The first statement is standard for random operators (see
  e.g.~\cite{pastur-figotin:92}). The last statement follows easily
  from \Thm{red.tree} and the fact that $H_{n+1}(\omega)=H_1(\tau_n
  \omega)$ and $H_1(\omega)$ have the same almost sure spectral
  components for all $n$.
\end{proof}

\begin{theorem}
  \label{thm:as.spec.mod}
  The almost sure spectrum is given by
  \begin{equation*}
    \Sigma =
    \clo{\bigcup_{\omega_1 \in \Omega_1} \spec {H_1(\omega_1 \1)}},
  \end{equation*}
  where $\omega_1 \1 \in \Omega$ is the element with the same entry
  $\omega_1$ in each component and $H_1(\omega_1 \1)$ is periodic.
  Assuming that $\Omega_1$ is a compact interval, we have in the RLM
  \begin{multline}
    \label{eq:as.spec.rlm}
    \Sigma=
    \bigcup_{k=1}^\infty \bigcup_{\ell \in \Omega_1}
    \frac 1 {\ell^2} \bigl( B_k \cup \{\pi^2 k^2\} \bigr) \\=
    \bigcup_{k=1}^\infty
    \Bigr(
       \bigl[ \frac 1 {\ell_+^2} \min B_k,
              \frac 1 {\ell_-^2} \max B_k \bigr] \cup
       \bigl[ \frac {\pi^2 k^2} {\ell_+^2},
                    \frac {\pi^2 k^2} {\ell_-^2} \bigr] \Bigr),
  \end{multline}
  where $\Omega_1=[\ell_-,\ell_+]$ and the intervals $B_k$ are defined
  in~\eqref{eq:sp.per.lm}, and in the RKM, we have
  \begin{multline}
    \label{eq:as.spec.rkm}
    \Sigma=
    \bigcup_{k=1}^\infty \bigcup_{q \in \Omega_1}
       \bigr(B_k(q) \cup \{\pi^2 k^2\}\bigr) \\=
    \bigcup_{k=1}^\infty \Bigr(
       \bigl[ \min B_k(q_-), \max B_k(q_+) \bigr] \cup
       \{ \pi^2 k^2 \} \Bigl),
  \end{multline}
  where $\Omega_1=[q_-,q_+]$ and $B_k(q)$ is defined
  in~\eqref{eq:sp.per.km2}. If $b>1$ or $b=1$ and $0 \notin [q_-,q_+]$
  then $\Sigma$ has infinitely many gaps close to $\pi^2k^2$.
\end{theorem}
\begin{proof}
  The spectrum of the periodic operator was calculated in
  \Thms{sp.per.lm}{sp.per.km}. Note that in both models, the band
  edges depend continuously and monotonically on the random parameter
  and the union is locally finite, so the union of compact intervals
  is still a closed set.
\end{proof}

In order to prove that $H_1(\omega)$ has pure point spectrum almost
surely, we need to control the growth of generalized eigenfunctions.
We have already seen in the previous section, that it is enough to
control the growth of nontrivial solutions of the random discrete map
$\map {{\vec F}_\lambda = \vec F_\lambda(\omega,\cdot)} \N {\C^2}$
of~\eqref{eq:trans.mat}. The
random transfer matrix $T_\lambda(n)=T_\lambda(\omega_n)$ in the RLM
has the form, for $\lambda>0$,
\begin{subequations}
  \label{eq:tm.ran}
  \begin{equation}
    \label{eq:tm.rlm}
    T_\lambda(\omega_n) =
    D(b) R_\mu(\mu \ell(\omega_n))=
    \begin{pmatrix}
      b^{1/2} \cos(\mu \ell(\omega_n))   &
                   \dfrac{b^{1/2}} \mu \sin(\mu \ell(\omega_n))\\
      -b^{-1/2} \mu \sin (\mu \ell(\omega_n))&
                   b^{-1/2}\cos(\mu \ell(\omega_n))
    \end{pmatrix}
  \end{equation}
  where $\map \ell {\Omega_1} {(0,\infty)}$ is the single edge random
  length perturbation.  For the RKM, we have
  \begin{multline}
    \label{eq:tm.rkm}
    T_\lambda(\omega_n) = D(b) S\bigl(q(\omega_n)\bigr)
    R_\mu(\mu) \\=
    \begin{pmatrix}
      b^{1/2} \cos \mu & \dfrac{b^{1/2}} \mu \sin \mu\\
      b^{-1/2}\Bigl(-\mu \sin \mu + q(\omega_n) \cos \mu \Bigr)&
             b^{-1/2}
        \Bigl( \cos \mu + \dfrac {q(\omega_n)} \mu \sin \mu \Bigr)
    \end{pmatrix}
  \end{multline}
\end{subequations}
where the second equality holds for $\lambda>0$. Here, $\map q
{\Omega_1} {(0,\infty)}$ is the single site random potential
perturbation.  In the case $\lambda<0$, one has to replace $R_\mu(\mu)$
by $R^{\mathrm h}_\mu(\mu)$ with $\mu=\sqrt {|\lambda|}$. If
$\lambda=0$, then $R_0(0)= S(1)^\mathrm{tr}$.

%%%%%%%%%%%%%%%%%%%%%%%%%%%%%%%%%%%%%%%%%%%%%%%%%%%%%%%%%%%%%%%%%%%%%%%
%----------------------------------------------------------------------
\subsection{Lyapunov exponents}
\label{sec:lyapunov}
%----------------------------------------------------------------------
As we have seen we can control the growth of generalized
eigenfunctions via the growth of random matrices. We will provide
therefore some general results on Lyapunov exponents and exponentially
decaying solutions of recursion equations.

Assume that $\map T {\Omega_1 \times \Sigma_0} {\SL_2 (\R)}$,
$(\omega_1,\lambda) \mapsto T_\lambda(\omega_1)$ is measurable where
$\Sigma_0 \subset \R$ is a measurable set. We assume that
\begin{equation}
  \label{eq:int.cond}
  \Exp[_1]{\ln \norm{T_\lambda^{-1}}} < \infty.
\end{equation}
Note that $\norm A \ge 1$ for $A \in \SL_2(\R)$.  We set
\begin{equation}
  \label{eq:mat.u}
  U_\lambda(\omega,n) := T_\lambda(\omega_n) \cdot \ldots \cdot
                        T_\lambda(\omega_1), \qquad
  U_\lambda(\omega,0) := \1.
\end{equation}
Clearly,
\begin{equation}
  \label{eq:cycle}
  U_\lambda(\omega, n_1+n_0) = U_\lambda(\tau_{n_0}\omega, n_1)
                              U_\lambda(\omega, n_0),
\end{equation}
i.e., $U_\lambda$ is a \emph{multiplicative cocycle},
cf.~\cite[(11.23)]{pastur-figotin:92}.

We define the \emph{Lyapunov exponent}
\begin{equation}
  \label{eq:lyapunov.op}
  \gamma(\omega, \lambda) :=
  \lim_{n \to \infty} \frac 1 n \ln \norm{U_\lambda(\omega,n)}
\end{equation}
where $\norm \cdot$ is the operator norm of $2 \times 2$-matrices
defined by $\norm A := \sup_{v \in \R^2}{|Av|}/{|v|}$. The limit is
nonrandom:
\begin{lemma}
  \label{lem:lyapunov}
  Suppose that the single transfer matrix $T_\lambda(\cdot)$ satisfies
  the integrability condition~\eqref{eq:int.cond}. Then, there exists
  a measurable set $S_1 \subset \Omega \times \Sigma_0$ such that
  $S_1(\lambda):= \set {\omega} {(\omega,\lambda) \in S_1} \subset \Omega$
  has full measure, and the limit~\eqref{eq:lyapunov.op} exists and is
  finite for all $(\omega,\lambda) \in S_1$. In addition, the limit is
  nonrandom, i.e.,
  \begin{equation*}
    \gamma(\lambda) := \Exp {\gamma(\cdot,\lambda)} =
    \gamma(\omega,\lambda)
  \end{equation*}
  for all $\omega \in S_1(\lambda)$. Finally, $\gamma(\lambda) \ge 0$.
\end{lemma}
\begin{proof}
  We apply the subadditive ergodic
  theorem~\cite[Prop.~6.3]{pastur-figotin:92} and have to verify that
  \begin{equation*}
    \Exp {\ln \norm{U_\lambda(\cdot,n)}} \ge C_\lambda n
  \end{equation*} for $n \ge 0$ and some constant $C_\lambda \in
  \R$. A simple norm estimate using $\norm{AB} \ge (\norm{A^{-1}}
  \norm{B^{-1}})^{-1}$ shows that $C_\lambda=-\Exp[_1] {\ln
    \norm{T_\lambda^{-1}}}$ is enough. The measurability of $S_1$
  follows from the measurability of $(\omega_1,\lambda) \to
  T_\lambda(\omega_1)$.  \hiddenfootnote{In Kallenberg - Probability,
    (\texttt{kallenberg\_probability.97.pdf}, p. 160), the 0-set of
    the ergodic theorem is defined as $N_0(\lambda)=\bigcup_k
    N_k(\lambda)$ with $N_k(\lambda):=\set{\omega}{\limsup_n
      S_n(\omega,\lambda)/n>k}$. But clearly,
    $S_1:=\set{(\omega,\lambda)}{\omega \notin N_0(\lambda)}$ is
    measurable since $(\omega,\lambda) \to S_n(\omega,\lambda)$ is
    measurable (note that $(\omega_0,\lambda) \to T_\lambda(\omega)$
    is assumed to be measurable) and $S_n(\omega,\lambda)$ is
    constructed from $T_\lambda(\omega_n)$.}
\end{proof}
  We parameterize the set
of all directions in $\R^2$ (up to sign) by $\theta \in [0,\pi)$, or
more abstractly by points in the real projective line $\Proj {\R^1}$ and
sometimes write ${\vec F}_\lambda \sim \theta$ if the nonzero vector
${\vec F}_\lambda \in
\R^2$ is in the direction $\theta$, i.e., a multiple of $( \sin \theta,
\cos \theta)^{\tr}$, where $tr$ denotes transpose.

We denote
\begin{equation}
  \label{eq:sol.f}
  \vec F_\lambda(\omega, \theta, n) :=
  U_\lambda(\omega,n)
  \begin{pmatrix} \sin \theta \\ \cos \theta \end{pmatrix}
\end{equation}
the propagation of the initial vector $\vec F(0) \sim \theta$.  Clearly,
$\vec F_\lambda(\omega, \theta, \cdot)$ solves the recursion equation
\begin{equation}
  \label{eq:rec.eq}
  \vec F_\lambda(\omega, \theta, n+1) = T_\lambda(\omega_n)
  \vec F_\lambda(\omega, \theta, n), \quad
  \vec F(0) =
   \begin{pmatrix} \sin \theta \\ \cos \theta \end{pmatrix}.
\end{equation}
We want to turn the positivity of the Lyapunov exponent into
exponential bounds on the solution of the above recursion equation. To
do so, we need the following deterministic version of the Oseledec
theorem (cf.~\cite[Thm~IV.2.4]{carmona-lacroix:90}):
\begin{theorem}
\label{thm:oseledec}
  Suppose that $U(n) \in \SL_2 (\R)$ for all $n \ge 1$ such that
  \begin{enumerate}
  \item \label{lim.ex} $\lim_{n \to \infty} \frac 1 n \ln \norm{U(n)}
    = \gamma$ exists, $\gamma < \infty$ and
  \item
    \label{lim.tm.ex}
    $\lim_{n \to \infty} \frac 1 n \ln \norm{T(n)} = 0$
  \end{enumerate}
  where $T(n):=U(n)U(n-1)^{-1}$ is the single transition matrix. Then
  there exists a nonzero vector $\vec F(0) \in \R^2$ such that
  \begin{equation}
    \label{eq:vec.lyapunov}
    \lim_{n \to \infty} \frac 1 n \ln |U(n) \vec F(0)| = -\gamma
    \qquad \text{and}\qquad
    \lim_{n \to \infty} \frac 1 n \ln |U(n) \vec F| = \gamma
  \end{equation}
  where $\vec F$ is linearly independent of $\vec F(0)$ in the latter
  case. In particular, the solution $\vec F(n):= U(n) \vec F(0)$ of
  the recursion equation $\vec F(n+1)=T(n) \vec F(n)$ with initial
  vector $\vec F(0)$ has almost exponential decay rate
  $-\gamma$, i.e.,
  \begin{equation}
    \label{eq:exp.decay0}
    \forall \eps>0 \quad
    \exists C(\eps)>0: \quad
        |\vec F(n)| \le
        C(\eps) \e^{-(\gamma-\eps)n}.
  \end{equation}
\end{theorem}
\begin{remark}
  \label{rem:why.dir}
  The previous theorem already indicates that we cannot expect to show
  exponential decay directly for the initial condition $\theta=0$
  (corresponding to a Dirichlet boundary condition at $0$); moreover,
  we need the spectral averaging arguments of \App{sp.av}.  But the
  Dirichlet boundary condition is crucial in the symmetry reduction
  (see \Thm{red.tree} or \Thm{reduction}), not for the first reduction
  step, but for the subsequent ones.
\end{remark}
We will apply this theorem to $U(n)=U_\lambda(\omega,n)$ for fixed
$\omega$ and $\lambda$ in \Thm{kotani}. Clearly, in this case $\vec
F(0)$ and $C(\eps)$ also depend on $\lambda$ and $\omega$.

To ensure the positivity of the Lyapunov exponent we use the
Furstenberg theorem~\cite{furstenberg:63}:
\begin{theorem}
  \label{thm:furstenberg}
  Denote by $G_\lambda$ the smallest closed subgroup of $\SL_2 (\R)$
generated
  by all matrices $T_\lambda(\omega_1)$, $\omega_1 \in \Omega_1$. If $G$
  is noncompact and no subgroup of finite index is reducible then
  $\gamma(\lambda)>0$.
\end{theorem}
A sufficient condition for $\gamma(\lambda)>0$ is the following
(cf.~\cite[Thm.~4.1]{ishii:73}, \cite{ishii-matsuda:70}):
\begin{theorem}
  \label{thm:ishii}
  Suppose that $\set{T_\lambda(\omega_1)}{\omega_1 \in \Omega_1}
  \subset \SL_2 (\R)$ contains at least two elements with no common
  eigenvectors then $\gamma(\lambda)>0$.
\end{theorem}

The following lemma reduces the possibilities in our application,
since we are only interested in transfer matrices associated to
spectral parameters $\lambda$ in the almost sure spectrum:
\begin{lemma}
   \label{lem:crit.lyap.pos}
   Assume that the almost sure spectrum is the union of the periodic
   spectrum, i.e.,
   \begin{equation}
     \label{eq:spec.cd1}
     \Sigma =
     \bigcup_{\omega_1 \in \Omega_1} \spec{H(\omega_1 \1}.
   \end{equation}
   Suppose in addition, that the set
   \begin{equation}
     \label{eq:spec.cd2}
     N :=
     \set{(\omega_1,\lambda) \in \Omega_1 \times \Sigma}
        { |\tr T_\lambda(\omega_1)|=2}
   \end{equation}
   has $(\Prob_1 \otimes \leb)$-measure $0$, where $\leb$ denotes
   Lebesgue measure. Finally, suppose that there is a set $\Sigma_0
   \subset \Sigma$ so that for all $\lambda \in \Sigma_0$, there exist
   at least two different \emph{elliptic} matrices $T_1, T_2$ in
   $\set{T_\lambda(\omega_1)}{\omega_1 \in \Omega_1,
     \text{and}~\lambda \in \Sigma_0} \subset \SL_2 (\R)$ having no
   common eigenvectors.  Then $\gamma(\lambda)>0$ for all $\lambda \in
   \Sigma_0$.
\end{lemma}

\begin{proof}
  Due to the second assumption, for almost all $\lambda \in \Sigma$,
  the set
  \begin{equation*}
    N(\lambda) =\set{\omega_1}{ |\tr T_\lambda(\omega_1)|=2}
  \end{equation*}
  has probability $0$ so that the set of $\lambda$ such that
  $T_\lambda$ is elliptic or hyperbolic forms a support of $\Prob_1$.
  We have to show that there are at least two matrices in
  $\Omega_1=\supp \Prob_1$ with no common eigenvectors.  If both are
  elliptic, we are done due to our assumption. If one is elliptic and
  the other hyperbolic, they can never have a common eigenvector,
  since the eigenvectors of the first are nonreal, and the second are
  real. The case that both matrices are hyperbolic is not of interest,
  since $\lambda \in \Sigma$ implies that at least one of the matrices
  is not hyperbolic due to our first assumption.  The result now
  follows from \Thm{ishii}.
\end{proof}
In cases when the transfer matrix is complicated, the following
criteria is useful:
\begin{corollary}
   \label{cor:lyap.pos}
   Suppose that~\eqref{eq:spec.cd1} and~\eqref{eq:spec.cd2} are true.
   Assume in addition, that for all $\lambda \in \Sigma_0$ there exist
   two \emph{noncommuting} elliptic matrices in
   $\set{T_\lambda(\omega_1)}{\omega_1 \in \Omega_1} \subset
   \SL_2 (\R)$. Then $\gamma(\lambda)>0$ for all $\lambda \in
   \Sigma_0$.
\end{corollary}
\begin{proof}
  If the matrices $T_1$ and $T_2$ do not commute, they differ in at
  least one eigenspace. Since $T_1$ and $T_2$ are elliptic and real,
  all eigenvectors are nonreal, and the second eigenspace is obtained
  from the first one by conjugation. In particular, $T_1$ and $T_2$
  have no common eigenspace.
\end{proof}

%----------------------------------------------------------------------
\subsection{Lyapunov exponents for the RLM and RKM}
\label{sec:lyap.tree}
%----------------------------------------------------------------------
In this subsection we show that under suitable assumptions on the
single site random perturbation, the Lyapunov exponent of the transfer
matrices~\eqref{eq:tm.ran} are positive. In addition we show
that~\eqref{eq:int.cond} and Assumption~\eqref{lim.tm.ex} of
\Thm{oseledec} are fulfilled. We will need all these results in the
next subsection in order to prove exponential localization.

\begin{lemma}
  \label{lem:lyap.pos.rlm}
  Assume that $\lambda>0$ lies in the almost sure spectrum of
  $H_1(\omega)$ in the RLM. Suppose furthermore that the branching
  number $b>1$ and that there are at least two different values
  $\ell_1, \ell_2 \in \Omega_1$ such that $\mu(\ell_1 - \ell_2) \notin
  \pi \Z$.  Then $\gamma(\lambda)>0$. If $b=1$, then
  $\gamma(\lambda)=0$ for all $\lambda>0$.

  In particular, if $b>1$ and $\Omega_1$ contains at least two
  different length $\ell_1$ and $\ell_2$ then $\gamma(\lambda)>0$ for
  almost all $\lambda>0$.
\end{lemma}
\begin{proof}
  We want to apply \Lem{crit.lyap.pos}. The first two conditions are
  fulfilled and we only have to check that the eigenvectors of
  $T_i:=T_\lambda(\omega_i)$, i.e., $\{ \vec e_{1,+}; \vec e_{1,-}\}$
  and $\{ \vec e_{2,+}; \vec e_{2,-}\}$, never have an eigenspace in
  common in the elliptic case.  A simple
  calculation\hiddenfootnote{Suppose that $A \in \SL_2(\R)$ is
    elliptic. Then the eigenvalues of $A$ are
  \begin{equation*}
    \vec e_\pm =
    \begin{pmatrix}
      \frac12 \bigl( (a-d) \pm \im \sqrt {4-(\tr A)^2} \bigr)\\
      c
    \end{pmatrix}
    \quad \text{or} \quad
    \vec e_\pm =
    \begin{pmatrix}
      b \\
      \frac12 \bigl( (a-d) \mp \im \sqrt {4-(\tr A)^2} \bigr)
    \end{pmatrix}
    \end{equation*}
    where
    \begin{equation*}
    A =
    \begin{pmatrix}
      a & b\\ c & d
    \end{pmatrix}
    \end{equation*}
    (note: if $A \ne \pm \1$ is elliptic, then $c \ne 0$ and $b \ne
    0$! CHECK!!).  Furthermore, the eigenvectors are nonreal.} shows
  that the eigenvectors are linear dependent iff $\sin
  \mu(\ell_1-\ell_2) (b-1) = 0$, i.e., $\mu(\ell_1 - \ell_2) = k \pi$
  or $b=1$.  In the latter case we can calculate $\gamma(\lambda)=0$
  explicitly. The last statement follows since $\set{\mu^2}{
    \mu(\ell_1-\ell_2) \in \pi \Z}$ is a countable set iff $\ell_1 \ne
  \ell_2$.
\end{proof}

\begin{lemma}
  \label{lem:lyap.pos.rkm}
  Assume that there are $q_1, q_2 \in \Omega_1$ such that $q_1 \ne
  q_2$ and that $\lambda \in \Sigma$. If $\mu = \sqrt \lambda \notin
  \pi \N$ then $\gamma(\lambda)>0$. If $\mu \in \pi \N$ then
  $\gamma(\lambda)=\frac 12 \ln b$.  In particular,
  $\gamma(\lambda)>0$ for almost all $\lambda>0$.
\end{lemma}
\begin{proof}
  Again, we apply \Lem{crit.lyap.pos}. The first two assumptions are
  also satisfied in RKM. One can easily see that the eigenvectors of an
  \emph{elliptic} transfer matrix associated to $q_1$ are linearly
  dependent on the ones associated to $q_2$ iff $\sin \mu =0$ or
  $q_1=q_2$. The Lyapunov exponent for $\lambda=\mu^2$ with $\mu \in
  \pi \N$ can easily be calculated since $T_\lambda(q)=\pm D(b)$ and
  the largest eigenvalue of $U_\lambda(\omega,n)$ is always $b^n$.
\end{proof}

\begin{lemma}
  \label{lem:tm.cond.tree}
  In both models, the integrability condition~\eqref{eq:int.cond} and
  the condition~\eqref{lim.tm.ex} in~\Thm{oseledec} are fulfilled.
\end{lemma}
\begin{proof}
  The norm of the transfer matrix can be estimated by
  \begin{equation*}
    \norm{T_\lambda(\omega_n)} \le
    \norm{D(b)} \norm{R_+(\mu \ell_n)} \le b^{1/2}
  \end{equation*}
  in the random length model (here, we only need to consider
  $\lambda>0$ since $H=\laplacian {T(\omega)}\ge 0$). The same estimate
  holds for the inverse of $T_\lambda(\omega_n)$.  In the random
  potential model, we have
  \begin{equation*}
    \norm{T_\lambda(\omega_n)} \le
    \norm{D(b)} \norm{S(-q_n)} \norm{R_\pm(\mu)} \le
    b^{1/2} ( 1 + \max\{|q_-|,|q_+|\}) \e^\mu,
  \end{equation*}
  $\mu:=\sqrt{|\lambda|}$, and similarly for the inverse. Therefore,
  the norms are independent of $n$.  In
  particular,~\eqref{eq:int.cond} and Assumption~\eqref{lim.tm.ex} of
  \Thm{oseledec} are fulfilled for both models.
\end{proof}

%----------------------------------------------------------------------
\subsection{Exponential localization on the tree graph}
\label{sec:loc.tree}
%----------------------------------------------------------------------

Here, we show that in both random models of~\Def{ran.tree}
localization holds.  Denote by $H_1(\omega)$ the Hamiltonian on $\R_+$
with Dirichlet boundary condition $f(0)=0$.

\begin{theorem}[\cite{kotani:86}]
  \label{thm:kotani}
  Assume that $\gamma(\lambda)>0$ for Lebesgue-almost all $\lambda
  \in \Sigma_0$ and $\Sigma_0 \subset \R$. Assume in addition, that
  the spectral averaging formula~\eqref{eq:weyl.bdd} holds. Then
  $\spec {H_1(\omega)} \cap \Sigma_0$ is almost surely pure point, i.e.,
  if $\Sigma_\bullet$ denote the almost sure spectrum (respectively, almost
  sure spectral components) of $H_1(\omega)$, then
  \begin{equation*}
    \Sigma \cap \Sigma_0 =
    \Sigma_{\mathrm{pp}} \cap \Sigma_0
       \qquad \text{and} \qquad
    \Sigma_{\mathrm c} \cap \Sigma_0=\emptyset.
  \end{equation*}
  In addition, almost all eigenfunctions of $H_1(\omega)$ on the
  half-line $[0,\infty)$ decay with almost exponential decay rate
  $\gamma(\lambda)$ in the sense of~\eqref{eq:exp.decay}.
\end{theorem}
\begin{proof}
  Without loss of generality,
  we assume that $\gamma(\lambda)>0$ for \emph{all} $\lambda
  \in \Sigma_0$ (just exclude the exceptional set of measure $0$ from
  $\Sigma_0$). We decompose $\Omega$ into its first and remaining
  component, i.e., $\omega=(\omega_1,\hat \omega) \in \Omega_1 \times
  \hat \Omega = \Omega$ and set
  \begin{equation}
    \label{eq:set.lyap.pos}
    S :=
    \set{(\hat \omega,\lambda) \in \hat \Omega \times \Sigma_0}
      {\lim_n \frac 1 n \ln \norm{U_\lambda(\hat \omega,n)} > 0}.
  \end{equation}
  It follows from standard arguments that $S$ is measurable. In
  addition, $S(\hat\omega)=\set{\lambda \in \Sigma_0}{(\hat
    \omega,\lambda) \in S}$ is a tail event, i.e., $S(\hat \omega)$
  does not depend on a finite number of random variables.  From
  \Lem{lyapunov} and the assumption $\gamma(\lambda)>0$ we see that
  the set of energies $S_1$, defined in \Lem{lyapunov}, has full
  $(\hat \Prob \otimes \leb)$-measure. Since $S_1 \subset S$, the set
  $S$ has full $(\hat \Prob \otimes \leb)$-measure.  In particular,
  for $(\hat \omega,\lambda) \in S$, Assumption~\eqref{lim.ex} of
  \Thm{oseledec} is fulfilled. We have already seen that
  Assumption~\eqref{lim.tm.ex} is always fulfilled.  Therefore, there
  exists $\theta_0=\theta_0(\hat \omega,\lambda)$ such that
  \begin{equation}
    \label{eq:exp.decay2}
    \lim_{n \to \infty} \frac 1 n
       \ln |\vec F_\lambda(\hat \omega,n, \theta)| =
       \begin{cases}
         -\gamma(\lambda), & \theta=\theta_0\\
         \gamma(\lambda), & \theta \ne \theta_0
       \end{cases}
  \end{equation}
  where
  \begin{equation*}
    \vec F_\lambda(\hat \omega,n, \theta) =
    U_\lambda(\hat{\omega},n)
        \begin{pmatrix}
          \sin \theta \\ \cos \theta
        \end{pmatrix}.
  \end{equation*}
  Let $f$ be the generalized eigenfunction on $\R_+$ associated to
  $\vec F_\lambda(\hat \omega,\cdot, \theta_0)$. Since $\vec
  F_\lambda(\hat \omega,n, \theta_0)$ decays exponentially in $n$, we
  see from~\eqref{eq:l2.L2}, that then $f \in \Lsqr {\R_+}$. Now, the
  remaining point to show is, that $\theta_0=0$, i.e., that $f$
  satisfies a Dirichlet boundary condition at $0$.

  \sloppy Denote the measure associated to $H_1(\omega)$ in \Cor{desint}
  by $\rho_\omega$. Due to \Lem{m-fct}, the Weyl-Titchmarsh function
  $m$ associated to $H_1(\omega)$ is the Borel transform of the measure
  $\rho_\omega$ and we can apply the results on spectral averaging of
  \App{sp.av}.  In particular, using Fubini and the spectral averaging
  formula~\eqref{eq:sp.av}, we obtain
  \begin{multline}
    \label{eq:kotani.trick}
    \int_{\Omega_1} \int_{\hat \Omega}
      \rho_{(\omega_1,\hat \omega)} (\compl {S(\hat \omega)})
        \dd \hat \Prob(\hat \omega) \dd \Prob_1(\omega_1) =
    \int_{\hat \Omega}  \int_{\Omega_1}
      \rho_{(\omega_1,\hat \omega)} (\compl{S(\hat \omega)})
        \dd \Prob_1(\omega_1) \dd \hat \Prob(\hat \omega) \\ \le
    C_5 \int_{\hat \Omega} \leb(\compl{S(\hat \omega)})
        \dd \hat \Prob(\hat \omega) =
    C_5 (\hat \Prob \otimes \leb)(\compl S) = 0
  \end{multline}
  where $\leb$ denotes Lebesgue measure. This means that for
  $\Prob$-almost all $\omega=(\omega_1, \hat \omega)$, we have
  $\rho_\omega(\compl{S(\hat \omega)})=0$, i.e., $S(\hat \omega)$ is a
  support for the spectral measure $\rho_\omega$. Fix now such an $\omega$.

  We show in~\Thm{ptw.bd.ef} that the spectral measure is also
  supported on $\Sigma_\omega$, the set of eigenvalues having a
  polynomial bounded eigenfunction. The set of energies
  $S(\hat
  \omega) \cap \Sigma_\omega$ is a support for the spectral measure
$\rho_\omega$. For any $\lambda \in S(\hat
  \omega) \cap \Sigma_\omega$,  there is a generalized
  eigenfunction $\phi$ of $H_1(\omega)$ with eigenvalue $\lambda$ and
having polynomial growth. In
  addition, since $(\hat \omega,\lambda) \in S$, we have constructed an
  eigenfunction $f \in L^2$ from the coefficients $\vec F_\lambda(\hat
\omega,\cdot, \theta_0)$ as in~\eqref{eq:exp.decay2}. From \Lem{wronskian} we
  see that the Wronskian $W(f,\phi)(t_n+)$ of two generalized
  eigenfunctions is independent of $n$. Since $\phi(t_n+)$ and
  $\phi'(t_n+)$ are polynomially bounded in $n$ (cf.~\Thm{ptw.bd.ef})
  and since $f(t_n+)$ and $f'(t_n+)$ are almost exponentially decaying
  (cf.~\eqref{eq:exp.decay2}) we see that
  \begin{equation*}
    \lim_n W(f,\phi)(t_n+) =
    \lim_n \bigl( f'(t_n+) \phi(t_n+) -
                  f(t_n+)  \phi'(t_n+)\bigr) = 0.
  \end{equation*}
  In particular, $W(f,\phi)(0)=0$ and $f$, $\phi$ satisfy the same
  boundary condition at $0$, namely $\theta_0=0$, i.e., $f(0)=0$.

  Consequently, each $\lambda \in S(\hat \omega) \cap \Sigma_\omega$
  is an $\Lsymb_2$-eigenfunction of
  $H_1(\omega)$, i.e., that $\rho_\omega(\{\lambda\})>0$ for all
  $\lambda$ in a support of the spectral measure. Since a spectral
  measure is a Borel measure and the Hilbert space is separable, the
  support must be countable. This implies that
  the measure is pure point since a
  continuous measure cannot be supported on a countable
  set.
\end{proof}

\begin{remark}
  \label{rem:kotani.rkm}
  The spectral averaging used in~\eqref{eq:kotani.trick} is basically
  Kotani's trick. We may weaken the spectral averaging
  formula~\eqref{eq:weyl.bdd} in the following way: We assume
  that~\eqref{eq:weyl.bdd} is fulfilled for all $\lambda \in \Sigma_k
  \subset [\lambda_-,\lambda_+]=: \Sigma_0$ with an $k$-dependent
  constant $C_5=C_5(k)$ and where $\Sigma_k$ is an increasing
  sequence such that $\bigcup_k \Sigma_k =: \Sigma_\infty$ equals
  $\Sigma_0$ Lebesgue-almost everywhere. In the RKM, we will see that
  $\Sigma_k$ is just $\Sigma_0$ with a ``security'' distance from the
  points $k^2 \pi^2$ tending to $0$ as $k \to \infty$.

  We can still use Kotani's trick in this case: As
  in~\eqref{eq:kotani.trick} it follows that for each $k \in \N$ there
  is a set of full measure $\Omega(k)$ such that
  $\rho_\omega(\compl{S(\hat \omega)} \cap \Sigma_k) = 0$ for all
  $\omega \in \Omega(k)$. The intersection $\Omega(\infty)$ of all
  $\Omega(k)$ has still full measure, and for $\omega \in
  \Omega(\infty)$, we have
  \begin{equation*}
    \rho_\omega
          \bigl( \compl{S(\hat \omega)} \cap \Sigma_\infty \bigr) =
    \rho_\omega \Bigl (\bigcup_k
          \bigl(\compl{S(\hat \omega)} \cap \Sigma_k \bigr) \Bigr) \le
    \sum_k \rho_\omega
          \bigl(\compl{S(\hat \omega)} \cap \Sigma_k \bigr) =
    0.
  \end{equation*}
  The rest of the argument in the proof of \Thm{kotani} remains the
  same, replacing $\Sigma_0$ by $\Sigma_\infty$.
\end{remark}

On a tree graph, we need to precise the meaning of \emph{exponential
  decay}:
\begin{definition}
  \label{def:exp.decay}
  We say that a sufficiently smooth function $f$ on the tree graph $T$
  has \emph{almost exponential (pointwise) decay rate $\beta>0$} if for
  all $\eps>0$ there exists $C_\eps>0$ such that
  \begin{equation}
    \label{eq:exp.decay}
    |f(x)|+|f'(x)| \le C_\eps \e^{-(\beta-\eps)d(o,x)}
  \end{equation}
  for all $x \in T$ where $f'(x)$ is defined in~\eqref{eq:def.der} for
  $x \in V$.
\end{definition}
\begin{remark}
  \label{rem:exp.decay}
  \begin{enumerate}
  \item Due to the assumption~\eqref{eq:len.bd} and since a
    generalized eigenfunction has the form~\eqref{edgesoln1} on the
    edge it suffices to ensure
  \begin{equation*}
    |f(v)|+|f'(v)| \le C_\eps \e^{-(\beta-\eps)n}
  \end{equation*}
  for \emph{vertices} $v \in V$ at generation $n$ only.
\item \label{weight} Note that if $f$ has almost exponential pointwise
  decay rate $\gamma>0$ on the half-line, then the associated radial
  function $\wt f$ on the tree graph with constant branching number $b
  \ge 1$ has almost exponential pointwise decay rate $\gamma+(\ln
  b)/2$ due to the fact that in the symmetry reduction, we have the
  relation $f(d(o,x))=b^{(n-1)/2} \wt f(x)$ for $x$ in an edge at
  generation $n$.
\end{enumerate}
\end{remark}

Summarizing the results, we have shown:
\begin{theorem}
  \label{thm:main}
  Suppose that the random length quantum tree $T(\omega)$, respectively, the
  random Hamiltonian $H(\omega)$ on a radial quantum tree graph $T$
  with branching number $b$, have a single site random perturbation
  with absolutely continuous and bounded distribution $\eta$ on
  $\Omega_1=[\ell_-,\ell_+] \subset (0,\infty)$, respectively,
  $\Omega_1=[q_-,q_+] \subset \R$.  Suppose in addition, that $b>1$ in
  the random length model (RLM) and that $b \ge 1$ in the random
  Kirchhoff model (RKM).  Then the Kirchhoff Laplacian $\laplacian
  {T(\omega)}$, respectively, the Hamiltonian $H(\omega)$, has almost sure
  spectrum $\Sigma$ given in \Thm{as.spec.mod} and the spectrum is
  almost surely pure point.  In addition, the eigenfunctions have
  almost exponential decay rate $\gamma(\lambda)+(\ln b)/2$ where
  $\gamma(\lambda)$ denotes the Lyapunov exponent.
\end{theorem}
\begin{proof}
  Clearly, the assertion is local in energy. Let $\Sigma_0 \subset
  (0,\infty)$ be a bounded interval. Due to \Thm{as.spec} it suffices to
  consider $H_1(\omega)$ only. We have seen in
  \Lems{lyap.pos.rlm}{lyap.pos.rkm} that the Lyapunov exponent is
  positive almost everywhere on the almost sure spectrum $\Sigma$.
  Due to the assumptions on $\Omega_1$,~\eqref{eq:len.bd}
  and~\eqref{eq:pot.bd} are fulfilled, so that the results on bounds
  on generalized eigenfunctions of \App{ll.gr.gen.ef} apply.
  
  A proof of the spectral averaging assumption~\eqref{eq:weyl.bdd} is
  given in \Cors{sp.av.rlm}{sp.av.rkm} for the RLM and RKM,
  respectively. The exceptional set $\Sigma_k$ in the RKM consists of
  the zeros of $\sin (\sqrt \lambda)$, i.e., the Dirichlet spectrum of
  a single edge $e \cong (0,1)$ with a security distance of order
  $1/k$.  We finally apply \Thm{kotani} (taking \Rem{kotani.rkm} into
  account) and the result follows.
\end{proof}

\begin{remark}
  \begin{enumerate}
  \item \sloppy The case $b=1$ in the RKM has been considered by
    Ishii~\cite{ishii:73}.  In this case, the almost sure spectrum is
    $[\inf \Sigma,\infty)$ where $\inf \Sigma\ge0$ if $q \ge 0$ and
    $\inf \Sigma$ is given as the solution of $\tr T_{-\mu^2}(q_-)=2$,
    i.e., $2\cosh(\mu) + q_- \sinh(\mu)/ (\mu \sqrt 2) = 2$ where
    $q_-=\inf \Omega_1$ and localization holds everywhere in $\Sigma$.
    Localization has been shown by Delyon, Simon and Souillard
    (\cite[Thm.~1.3.~(i)]{dss:85}).
  \item The case $b=1$ in the RLM is of course uninteresting, since in
    this case, the tree Hamiltonian is the free Laplacian on
    $[0,\infty)$ with a Dirichlet boundary condition at $0$ and has
    therefore purely absolutely continuous spectrum (see also
    \Lem{lyap.pos.rlm}).
  \end{enumerate}
\end{remark}

%----------------------------------------------------------------------
%
\section{General tree-like graphs}
\label{sec:gen.gr}
%----------------------------------------------------------------------

In this section, we show that our methods also apply to a more general
class of metric graphs, namely to tree graphs, where an edge at
generation $n$ is replaced by a decoration graph $G_n$. In this case,
also the branching number $b=1$ is of interest, since it includes
line-like models like the necklace model considered in
\cite{kostrykin-schrader:04}. We only mention the necessary changes
and begin with a general definition of radial tree-like graphs.

%----------------------------------------------------------------------
\subsection{Tree-like graphs}
\label{sec:tl.graphs}
%----------------------------------------------------------------------

We will construct a radial tree-like graph from a radial tree-graph
$T=(V(T),E(T),\bd)$ by an edge decoration. We first need some notation
for the decoration graph:

Let $G_*=(V(G_*), E(G_*), \bd)$ be a compact quantum graph. We fix
two different vertices $o_0, o_1 \in V(G_*)$ sometimes called
\emph{boundary} or \emph{connecting vertices} of the decoration graph
$G_*$.  In addition, we denote by $V_0(G_*) := V(G_*)
\setminus \{o_0, o_1\}$ the the set of \emph{inner} vertices of $G_*$.

Here, and in the sequel we use the abbreviation
\begin{equation}
  \label{eq:def.der}
  f'(v) := f'_{G_*}(v):=\sum_{e \in E_v(G_*)} f_e'(v)
\end{equation}
for the sum over the \emph{inwards} derivative, i.e., $f'(v)$ is the
\emph{flux into the vertex} where $E_v(G_*)$ is defined
in~\eqref{eq:ed.vx} and
\begin{equation}
\label{eq:def.der2}
  f_e'(v) :=
  \begin{cases}
    -f_e'(0), & \text{if $v = \bd_-e$}\\
    f_e'(\ell_e), & \text {if $v = \bd_+e$}
  \end{cases}
\end{equation}
the \emph{inward} derivative of $f_e$ at $v$.  Note that $f'(v)$
depends on the graph; i.e., for a subgraph $S$ of $G_*$ or a graph $S$
containing $G_*$, we have in general $f'_S(v) \ne f'_{G_*}(v)$.

The Hilbert space $\Lsqr {G_*}$ associated to the decoration graph
$G_*$ is given by $\Lsqr {G_*} := \bigoplus_{e \in E(G_*)} \Lsqr e$
with norm given as in~\eqref{eq:norm}.  We define the Sobolev space of
order $1$ on $G_*$ as
\begin{equation}
  \label{eq:sob.1}
  \Sob{G_*} :=
  \Bigset{ f \in \bigoplus_{e \in E(G_*)} \Sob e }
  { f_{e_1}(v)=f_{e_2}(v), \quad
        \forall e_1,e_2 \in E(G_*), v \in V(G_*)}
\end{equation}
with norm given by
\begin{equation}
  \label{eq:norm.sob1}
  \normsqr[\Sob {G_*}] f :=
  \sum_{e \in E(G_*)} \bigl( \normsqr[e] f + \normsqr[e] {f'}\bigr).
\end{equation}
The Sobolev space of order $2$ on $G_*$ is
then
\begin{equation}
  \label{eq:sob.2}
    \Sob[2]{G_*} :=
    \Bigset{ f \in \bigoplus_{e \in E(G_*)}\Sob[2]e}
    { f \in \Sob {G_*}, \quad f'_{G_*} (v) = 0 \quad
            \forall v \in V_0(G_*)}.
\end{equation}
with norm defined via
\begin{equation}
  \label{eq:norm.sob2}
  \normsqr[{\Sob[2]{G_*}}] f :=
  \sum_{e \in E(G_*)}
      \bigl( \normsqr[e] f + \normsqr[e] {f'} + \normsqr[e] {f''} \bigr).
\end{equation}
In particular, we pose the boundary conditions only at the
\emph{inner} vertices, \emph{not} at the connecting vertices $o_0,
o_1$.  Hence, the differential operator $H_{G_*}$ acting on each edge
as in~\eqref{eq:def.op} with domain $\Sob[2]{G_*}$ is \emph{not}
self-adjoint.

\begin{figure}[h]
%----------------------------------------------------------------------
%  \centering \input{loc-graph-fig2.pstex_t}
\begin{picture}(0,0)%
\includegraphics{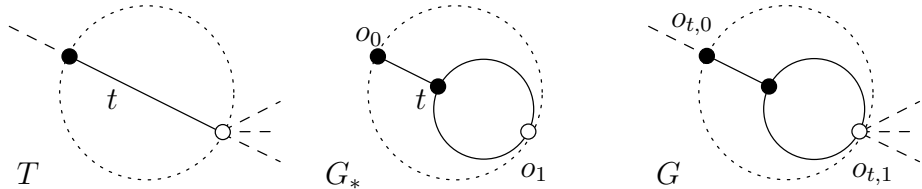}%
\end{picture}%
\setlength{\unitlength}{4144sp}%
\begingroup\makeatletter\ifx\SetFigFont\undefined%
\gdef\SetFigFont#1#2#3#4#5{%
  \reset@font\fontsize{#1}{#2pt}%
  \fontfamily{#3}\fontseries{#4}\fontshape{#5}%
  \selectfont}%
\fi\endgroup%
\begin{picture}(5790,1122)(349,-2041)
\put(946,-1546){\makebox(0,0)[lb]{\smash{{\SetFigFont{12}{14.4}{\rmdefault}{\mddefault}{\updefault}{\color[rgb]{0,0,0}$t$}%
}}}}
\put(406,-1996){\makebox(0,0)[lb]{\smash{{\SetFigFont{12}{14.4}{\rmdefault}{\mddefault}{\updefault}{\color[rgb]{0,0,0}$T$}%
}}}}
\put(5401,-1951){\makebox(0,0)[lb]{\smash{{\SetFigFont{12}{14.4}{\rmdefault}{\mddefault}{\updefault}{\color[rgb]{0,0,0}$o_{t,1}$}%
}}}}
\put(4231,-1996){\makebox(0,0)[lb]{\smash{{\SetFigFont{12}{14.4}{\rmdefault}{\mddefault}{\updefault}{\color[rgb]{0,0,0}$G$}%
}}}}
\put(4321,-1051){\makebox(0,0)[lb]{\smash{{\SetFigFont{12}{14.4}{\rmdefault}{\mddefault}{\updefault}{\color[rgb]{0,0,0}$o_{t,0}$}%
}}}}
\put(2431,-1141){\makebox(0,0)[lb]{\smash{{\SetFigFont{12}{14.4}{\rmdefault}{\mddefault}{\updefault}{\color[rgb]{0,0,0}$o_0$}%
}}}}
\put(3421,-1951){\makebox(0,0)[lb]{\smash{{\SetFigFont{12}{14.4}{\rmdefault}{\mddefault}{\updefault}{\color[rgb]{0,0,0}$o_1$}%
}}}}
\put(2791,-1546){\makebox(0,0)[lb]{\smash{{\SetFigFont{12}{14.4}{\rmdefault}{\mddefault}{\updefault}{\color[rgb]{0,0,0}$t$}%
}}}}
\put(2251,-1996){\makebox(0,0)[lb]{\smash{{\SetFigFont{12}{14.4}{\rmdefault}{\mddefault}{\updefault}{\color[rgb]{0,0,0}$G_*$}%
}}}}
\end{picture}%
%----------------------------------------------------------------------
  \caption{Decorating a graph $T$ with a graph $G_*$: The graph
    $T$ (solid and dashed) is decorated by replacing the edge $t \in
    E(T)$ with a graph $G_*$, and we call the new vertices $o_{t,j}$,
    $j=0,1$.}
  \label{fig:graph-deco}
\end{figure}

We now define the edge decoration:
\begin{definition}
\label{def:ed.deco}
  We say that a metric graph $G$ is obtained from a metric graph $T$
  by an
  \emph{edge decoration with a metric graph $G_*$ at the edge $t \in
    E(T)$} if we replace $t$ in $T$ by the graph $G_*$ where $\bd_\pm
  t \in V(T)$ is identified with two distinct vertices $o_0$, $o_1 \in
  V(G_*)$ ($o_0 \ne o_1$), i.e., $\bd_-t \cong o_0$ and $\bd_+t \cong
  o_1$ (see \Fig{graph-deco}).
\end{definition}
We embed $V(T) \hookrightarrow V(G)$ and $V(G_*) \hookrightarrow
V(G)$. If e.g.\ $G_*$ consists of a single edge $e$ only, the
edge decoration with $G_*$ does not change the original graph $T$.
\begin{definition}
  \label{def:tree-like.mg}
  A \emph{tree-like} metric graph associated to a tree graph $T$ is a
  graph $G$ obtained from a (generally infinite) tree graph $T$ by
  edge decoration with $G_*(t)$ at each tree edge $t \in E(T)$.  A
  \emph{radial tree-like} metric graph is a tree-like graph $G$ where
  the decoration graph $G_*(t)$ depends only on the generation of $t$,
  i.e., there exists a sequence of compact metric graphs $\{G_n\}_n$
  such that $G_n=G_*(t)$ for all $t \in E(T)$ with $\gen t=n$.
\end{definition}
We label $o_{t,0}=\bd_- t$ and $o_{t,1}=\bd_+ t$ the start/end vertex
of $t \in E(T)$ considered as vertices in the decoration graph
$G_*(t)$.  Obviously, a radial tree-like metric graph is determined by
the sequence of decoration graphs $\{G_n\}_n$, including the edge
lengths, and the sequence of branching numbers $\{b_n\}_n$.

The notion extends to \emph{quantum} graphs, i.e., metric graphs with
a Hamiltonian. Another notation for the right/left ``derivative'' at
the connecting vertices $o_0$ and $o_1$ of $G_*$ will be useful,
namely
\begin{subequations}
  \label{eq:gen.der}
  \begin{align}
     f^\dag(v)&:= -\sum_{e \in E_v(G_*)}  f'_e(v) = - f'(v)%- q(v) f(v)
    && \text{at $v=o_0$}\\
     f^\dag(v)&:= \sum_{e \in E_v(G_*)}  f'_e(v) +
    q(v) f(v) = f'(v)+q(v)f(v)&& \text{at $v=o_1$}
  \end{align}
\end{subequations}
with the notation $f'$ introduced
in~\eqref{eq:def.der}--\eqref{eq:def.der2}. Here $q(v)$ denotes the
vertex potential strength at the vertex $o_1$. For simplicity, we
assume that the vertex potential has support only at the vertex $o_1$,
i.e., $q$ is determined by the single number $q(o_1) \in \R$.  The
different signs for the vertex $o_0$ and $o_1$ are due to our
convention in~\eqref{eq:def.der2} considering always the \emph{inward}
derivative at a vertex. This notation allows us to express the
boundary condition for the Hamiltonian of a \emph{radial tree-like
  quantum graph} in a simple way (see also~\Rem{why.dag}):
\begin{definition}
  \label{def:tree-like.qg}
  \sloppy
  A \emph{radial tree-like quantum graph} is a radial tree-like metric
  graph $G=(V(G),E(G),\bd, \ell)$ together with a vertex potential
  strength $\map q {V(T)} \R$ such that there exists a sequence
  $\{q_n\}_n$ with $q(v)=q_n$ for all vertices $v \in V(T)$ in
  generation $n$ of the underlying tree.\footnote{\label{fn:vx.pot}
    For simplicity, we assume that there is only one vertex potential
    on each decoration graph $G_*(t)$, located at the ending point
    $o_{1,t}$.}  The corresponding Hamiltonian $H=H_G$ is given by
  \begin{equation}
    \label{eq:gen.def.op}
    (H f)_e = -f_e''
  \end{equation}
  on each edge, for functions $f \in \dom H_G$, where $\dom H_G$ is
  the set of those functions $f$ such that $f,f'' \in \Lsqr G =
  \bigoplus_{t \in E(T)} \Lsqr {G_*(t)}$ such that $f=\{f_t\}_t$ with
  $f_t:= f \restr {G_*(t)}$ satisfies
  \begin{equation}
    \label{eq:gen.bc1}
    f(o)=0 \qquad \text{and} \qquad
    f_t \in \Sob[2] {G_*(t)},\quad
t \in E(T)
  \end{equation}
  (in particular, $f_t$ satisfies the inner boundary conditions as
  in~\eqref{eq:sob.1} and~\eqref{eq:sob.2}), and
  \begin{equation}
    \label{eq:gen.bc2}
    f_{t_1}(v) = f_{t_2}(v) \qquad \text{and} \qquad
    f_{t_1}^\dag(v) = f_{t_2}^\dag(v)
  \end{equation}
  for all $t_1, t_2 \in E(T)$ meeting in a common \emph{tree} vertex
  $v \in E(T) \subset E(G)$.
\end{definition}
%% \begin{figure}[h]
%%   \centering \input{end-deco-tree-like.pstex_t}
%%   \caption{A tree-like graph $G$ with branching number $b=2$ and a
%%     onion decoration as in \Exenum{graph.deco}{end.point}. The random
%%     variable $\ell_n$ in each generation $n$ is the length of the
%%     initial edge of each decoration graph.  Below, the associated
%%     line-like graph $L$ is plotted (see~\Def{ll.mg}), where the edges
%%     $n_-$ and $n_+$ are drawn separately, although they are identified
%%     in $L$.}
%%   \label{fig:end-deco-tree-like}
%% \end{figure}
\begin{figure}[h]
%----------------------------------------------------------------------
%  \centering \input{loc-graph-fig3.pstex_t}
\begin{picture}(0,0)%
\includegraphics{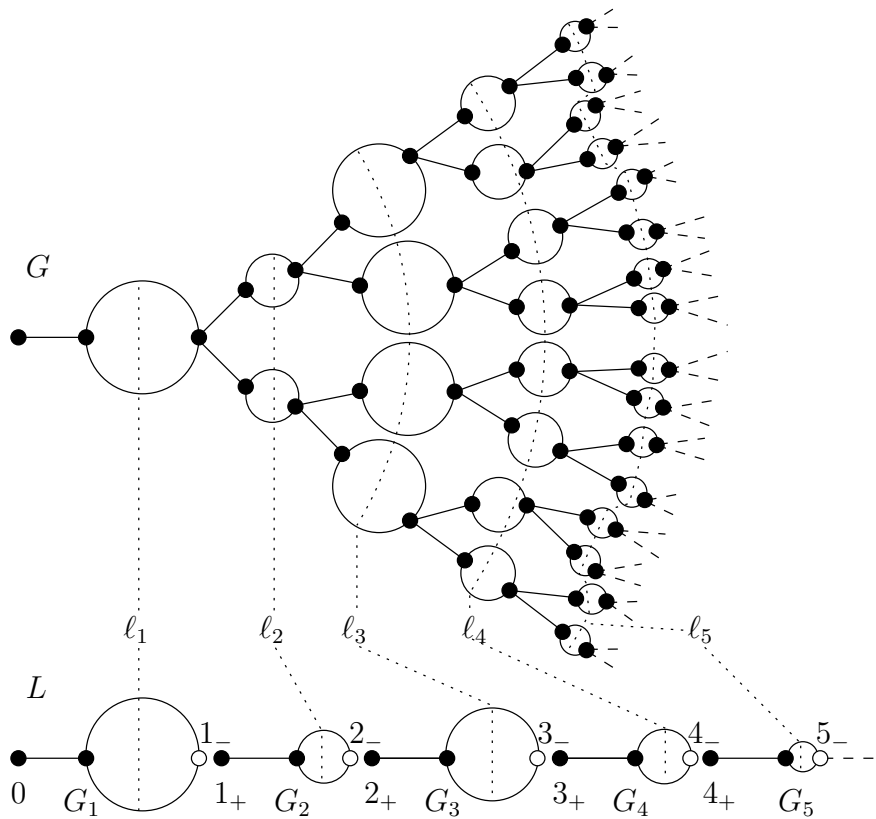}%
\end{picture}%
\setlength{\unitlength}{4144sp}%
\begingroup\makeatletter\ifx\SetFigFont\undefined%
\gdef\SetFigFont#1#2#3#4#5{%
  \reset@font\fontsize{#1}{#2pt}%
  \fontfamily{#3}\fontseries{#4}\fontshape{#5}%
  \selectfont}%
\fi\endgroup%
\begin{picture}(5257,4861)(214,-4291)
\put(316,-1051){\makebox(0,0)[lb]{\smash{{\SetFigFont{12}{14.4}{\rmdefault}{\mddefault}{\updefault}{\color[rgb]{0,0,0}$G$}%
}}}}
\put(901,-3211){\makebox(0,0)[lb]{\smash{{\SetFigFont{12}{14.4}{\rmdefault}{\mddefault}{\updefault}{\color[rgb]{0,0,0}$\ell_1$}%
}}}}
\put(1711,-3211){\makebox(0,0)[lb]{\smash{{\SetFigFont{12}{14.4}{\rmdefault}{\mddefault}{\updefault}{\color[rgb]{0,0,0}$\ell_2$}%
}}}}
\put(2206,-3211){\makebox(0,0)[lb]{\smash{{\SetFigFont{12}{14.4}{\rmdefault}{\mddefault}{\updefault}{\color[rgb]{0,0,0}$\ell_3$}%
}}}}
\put(2926,-3211){\makebox(0,0)[lb]{\smash{{\SetFigFont{12}{14.4}{\rmdefault}{\mddefault}{\updefault}{\color[rgb]{0,0,0}$\ell_4$}%
}}}}
\put(4276,-3211){\makebox(0,0)[lb]{\smash{{\SetFigFont{12}{14.4}{\rmdefault}{\mddefault}{\updefault}{\color[rgb]{0,0,0}$\ell_5$}%
}}}}
\put(316,-3571){\makebox(0,0)[lb]{\smash{{\SetFigFont{12}{14.4}{\rmdefault}{\mddefault}{\updefault}{\color[rgb]{0,0,0}$L$}%
}}}}
\put(226,-4201){\makebox(0,0)[lb]{\smash{{\SetFigFont{12}{14.4}{\rmdefault}{\mddefault}{\updefault}{\color[rgb]{0,0,0}$0$}%
}}}}
\put(1441,-4201){\makebox(0,0)[lb]{\smash{{\SetFigFont{12}{14.4}{\rmdefault}{\mddefault}{\updefault}{\color[rgb]{0,0,0}$1_+$}%
}}}}
\put(3466,-4201){\makebox(0,0)[lb]{\smash{{\SetFigFont{12}{14.4}{\rmdefault}{\mddefault}{\updefault}{\color[rgb]{0,0,0}$3_+$}%
}}}}
\put(4366,-4201){\makebox(0,0)[lb]{\smash{{\SetFigFont{12}{14.4}{\rmdefault}{\mddefault}{\updefault}{\color[rgb]{0,0,0}$4_+$}%
}}}}
\put(2251,-3841){\makebox(0,0)[lb]{\smash{{\SetFigFont{12}{14.4}{\rmdefault}{\mddefault}{\updefault}{\color[rgb]{0,0,0}$2_-$}%
}}}}
\put(3376,-3841){\makebox(0,0)[lb]{\smash{{\SetFigFont{12}{14.4}{\rmdefault}{\mddefault}{\updefault}{\color[rgb]{0,0,0}$3_-$}%
}}}}
\put(4276,-3841){\makebox(0,0)[lb]{\smash{{\SetFigFont{12}{14.4}{\rmdefault}{\mddefault}{\updefault}{\color[rgb]{0,0,0}$4_-$}%
}}}}
\put(5041,-3841){\makebox(0,0)[lb]{\smash{{\SetFigFont{12}{14.4}{\rmdefault}{\mddefault}{\updefault}{\color[rgb]{0,0,0}$5_-$}%
}}}}
\put(1351,-3841){\makebox(0,0)[lb]{\smash{{\SetFigFont{12}{14.4}{\rmdefault}{\mddefault}{\updefault}{\color[rgb]{0,0,0}$1_-$}%
}}}}
\put(2341,-4201){\makebox(0,0)[lb]{\smash{{\SetFigFont{12}{14.4}{\rmdefault}{\mddefault}{\updefault}{\color[rgb]{0,0,0}$2_+$}%
}}}}
\put(541,-4246){\makebox(0,0)[lb]{\smash{{\SetFigFont{12}{14.4}{\rmdefault}{\mddefault}{\updefault}{\color[rgb]{0,0,0}$G_1$}%
}}}}
\put(1801,-4246){\makebox(0,0)[lb]{\smash{{\SetFigFont{12}{14.4}{\rmdefault}{\mddefault}{\updefault}{\color[rgb]{0,0,0}$G_2$}%
}}}}
\put(2701,-4246){\makebox(0,0)[lb]{\smash{{\SetFigFont{12}{14.4}{\rmdefault}{\mddefault}{\updefault}{\color[rgb]{0,0,0}$G_3$}%
}}}}
\put(3826,-4246){\makebox(0,0)[lb]{\smash{{\SetFigFont{12}{14.4}{\rmdefault}{\mddefault}{\updefault}{\color[rgb]{0,0,0}$G_4$}%
}}}}
\put(4816,-4246){\makebox(0,0)[lb]{\smash{{\SetFigFont{12}{14.4}{\rmdefault}{\mddefault}{\updefault}{\color[rgb]{0,0,0}$G_5$}%
}}}}
\end{picture}%
%----------------------------------------------------------------------
  \caption{A tree-like graph $G$ with branching number $b=2$ and a
    necklace decoration with $p=2$ as in
    \Exenum{graph.deco}{necklace}. The random variable $\ell_n$ in
    each generation $n$ is the length of the edges of the necklace
    decoration.}
  \label{fig:necklace-tree-like}
\end{figure}
\begin{remark}
  \label{rem:why.dag}
  The previous characterization of the boundary condition explains why
  we introduced the notion~\eqref{eq:gen.der}. Note that the vertex
  potential strength is hidden in the notation. In the case when each
  decoration graph $G_*$ is a single edge $(0,1)$ without vertex
  potentials, $ f^\dag(v)$ for $v=o_0$ and $v=o_1$ is just the usual
  right and left derivative of $f$, respectively.
\end{remark}
Clearly, a radial tree-like quantum graph is determined by the
sequence of quantum graphs $\{G_n\}$, the sequence of vertex potential
strength $\{q_n\}_n$ and the sequence of branching numbers $\{b_n\}$.
We mention some examples falling into the class of radial tree-like
metric graphs:
\begin{example}
  \label{ex:graph.deco}
  \begin{enumerate}
  \item \label{simple.tree} \emph{Simple tree graphs:} The simplest
    example of a tree-like graph is of course a tree graph itself. A
    radial tree graph is completely determined by the sequences of
    edge lengths $\{\ell_n\}$, vertex potential strengths $\{q_n\}$
    and branching numbers $\{b_n\}_n$ where $b_n \ge 1$.
  \item \label{end.point} \emph{Graph decoration at the ending point:}
    (a)~Let $\hat G_*$ be a finite graph. If we attach an edge $e$ of
    length $\ell_e \ge 0$ to $\hat G_*$ we obtain a decoration graph
    $G_*=\hat G_* \cup \{e\}$ with starting point $o_0$ being the free
    end of the attached edge and with ending point being any vertex of
    the decoration graph (even the other vertex of the attached edge).

    (b)~For example, if $\hat G_*$ consists of a loop of length $1$,
    we obtain a decoration of the radial tree graph with base edge of
    length $\ell_n$ and a decoration loop of length $1$ at each
    generation $n$.

    We refer to this model as the \emph{loop decoration model}.
  \item \label{necklace} \emph{Necklace or onion decoration:} If $G_*$
    consists of an edge $e_0$ of length $1$ starting at $o_0$ together
    with $p$ edges of length $\ell$ joining the ending point of $e_0$
    with the ending vertex $o_1$, we obtain a \emph{(branched,
      half-line) onion} or \emph{necklace decoration model} ($p=2$).
    Clearly, the decorated tree graph is determined by the sequence of
    lengths $\{\ell_n\}_n$, the (constant) edge number $p$ and the
    (constant) branching number $b$ (see \Fig{necklace-tree-like}). We
    will allow that the length of the loop is $0$, i.e., $\ell_n \in
    [0,\ell_+]$, in the sense that the loop degenerates to a single
    vertex.
  \item \label{line.graphs} \emph{Line graphs:} If the branching
    numbers $b_n$ all equal to $1$, we obtain a line-like graph. For
    example, the previous (half-line) necklace decoration model is
    similar to the model already considered
    in~\cite{kostrykin-schrader:04} (see also \Sec{necklace}).
  \item \label{kirchhoff}
    \emph{Kirchhoff models:} We can add a vertex potential at the
    ending vertex $o_1$ of a fixed decoration graph $G_*$. The
    corresponding decorated tree graph has the same decoration graph
    at all steps, but a sequence of vertex potential strength
    $\{q_n\}_n$ at a vertex of generation $n$.
  \end{enumerate}
\end{example}
\begin{figure}[h]
%----------------------------------------------------------------------
%  \centering \input{loc-graph-fig4.pstex_t}
\begin{picture}(0,0)%
\includegraphics{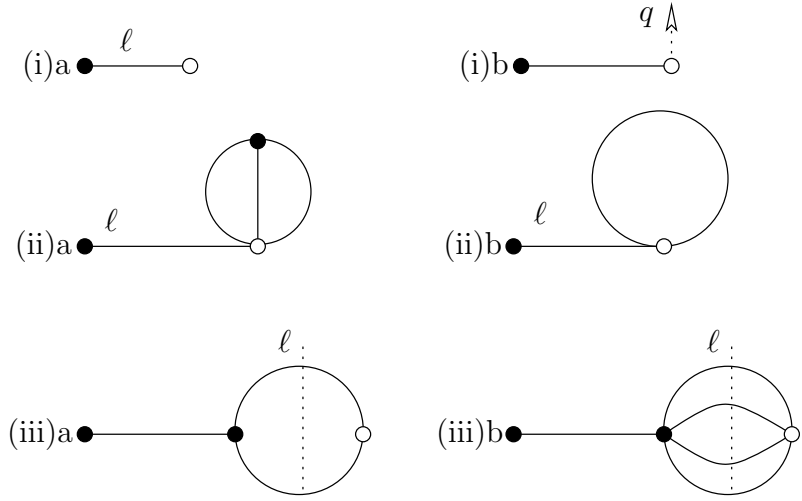}%
\end{picture}%
\setlength{\unitlength}{4144sp}%
\begingroup\makeatletter\ifx\SetFigFont\undefined%
\gdef\SetFigFont#1#2#3#4#5{%
  \reset@font\fontsize{#1}{#2pt}%
  \fontfamily{#3}\fontseries{#4}\fontshape{#5}%
  \selectfont}%
\fi\endgroup%
\begin{picture}(4733,2994)(1,-2188)
\put(676,479){\makebox(0,0)[lb]{\smash{{\SetFigFont{12}{14.4}{\rmdefault}{\mddefault}{\updefault}{\color[rgb]{0,0,0}$\ell$}%
}}}}
\put(3151,-556){\makebox(0,0)[lb]{\smash{{\SetFigFont{12}{14.4}{\rmdefault}{\mddefault}{\updefault}{\color[rgb]{0,0,0}$\ell$}%
}}}}
\put(586,-601){\makebox(0,0)[lb]{\smash{{\SetFigFont{12}{14.4}{\rmdefault}{\mddefault}{\updefault}{\color[rgb]{0,0,0}$\ell$}%
}}}}
\put(1621,-1321){\makebox(0,0)[lb]{\smash{{\SetFigFont{12}{14.4}{\rmdefault}{\mddefault}{\updefault}{\color[rgb]{0,0,0}$\ell$}%
}}}}
\put(4186,-1321){\makebox(0,0)[lb]{\smash{{\SetFigFont{12}{14.4}{\rmdefault}{\mddefault}{\updefault}{\color[rgb]{0,0,0}$\ell$}%
}}}}
\put(2701,344){\makebox(0,0)[lb]{\smash{{\SetFigFont{12}{14.4}{\rmdefault}{\mddefault}{\updefault}{\color[rgb]{0,0,0}(i)b}%
}}}}
\put(3781,659){\makebox(0,0)[lb]{\smash{{\SetFigFont{12}{14.4}{\rmdefault}{\mddefault}{\updefault}{\color[rgb]{0,0,0}$q$}%
}}}}
\put( 91,344){\makebox(0,0)[lb]{\smash{{\SetFigFont{12}{14.4}{\rmdefault}{\mddefault}{\updefault}{\color[rgb]{0,0,0}(i)a}%
}}}}
\put(2611,-736){\makebox(0,0)[lb]{\smash{{\SetFigFont{12}{14.4}{\rmdefault}{\mddefault}{\updefault}{\color[rgb]{0,0,0}(ii)b}%
}}}}
\put( 46,-736){\makebox(0,0)[lb]{\smash{{\SetFigFont{12}{14.4}{\rmdefault}{\mddefault}{\updefault}{\color[rgb]{0,0,0}(ii)a}%
}}}}
\put(  1,-1861){\makebox(0,0)[lb]{\smash{{\SetFigFont{12}{14.4}{\rmdefault}{\mddefault}{\updefault}{\color[rgb]{0,0,0}(iii)a}%
}}}}
\put(2566,-1861){\makebox(0,0)[lb]{\smash{{\SetFigFont{12}{14.4}{\rmdefault}{\mddefault}{\updefault}{\color[rgb]{0,0,0}(iii)b}%
}}}}
\end{picture}%
%----------------------------------------------------------------------
  \caption{The decoration graphs of \Ex{graph.deco} together with the
    (random) parameter: (i)~the simple RLM resp.\ RKM model;
    (ii)a~decoration with a onion; (ii)b~decoration with a loop at the
    ending point; (iii)a necklace/onion decoration.}
  \label{fig:deco-graph}
\end{figure}

We will give some natural conditions on the parameters in order that
our examples satisfy the needed assumptions on the decoration graphs
$\{G_n\}$ given in \Ass{deco.gr}.

%----------------------------------------------------------------------
\subsection{Line-like graphs and symmetry reduction}
\label{sec:reduct.gen}
%----------------------------------------------------------------------

As on a simple quantum tree graph, we can profit from the symmetry
reduction (see~\Sec{reduct.gen}). To do so, we need the notion of a
\emph{line-like} graph associated to the sequence $\{G_n\}_n$ of
decoration graphs and the sequence of branching numbers $\{b_n\}_n$.

On each decoration graph $G_n$, we specify two different vertices
$(n-1)_+=o_{n,0}$, $n_-=o_{n,1} \in V(G_n)$. We sometimes simply write
$n-1=(n-1)_+$ or $n=n_-$ if it is clear that they belong to $V(G_n)$
(e.g., $0=0_+$).

\begin{definition}
  \label{def:ll.mg}
  A \emph{line-like metric graph} $L_n=L_{n,\infty}$ starting at $n$
  is obtained from the union of $G_k$, $n<k$ by identifying $k_+ \in
  V(G_k)$ with $k_-\in V(G_{k+1})$ for $n<k$. Similarly, we denote by
  $L_{n_0,n_1}$ the line-like graph obtained as concatenation of
  $G_k$, $n_0<k \le n_1$, and set $L:=L_0$ for the entire line-like
  graph.
\end{definition}
The norm on $\Lsqr{L_n}$ is defined by
\begin{equation*}
  \normsqr[L_n] f := \sum_{k > n} \normsqr[G_k] f.
\end{equation*}
Clearly, $L_n$ is determined by the sequence of graphs $\{G_k\}_{k>n}$.
Similarly, the notion of a line-like quantum graph can be defined:

\begin{definition}
  \label{def:ll.graph.qg}
  A \emph{line-like quantum graph} is given by a line-like metric
  graph $L_n=(V(L_n),E(L_n),\bd,\ell)$ together with a sequence
  $\{b_k\}_{k >n}$ of positive numbers and a sequence of vertex
  potential strength $\{q_k\}_{k > n} \subset \R$. The corresponding
  Hamiltonian $H_{L_n}$ acts on each edge as in~\eqref{eq:gen.def.op}
  for functions $f \in \dom H_{L_n}$ where $\dom H_{L_n}$ is the set
  of those functions $f$ such that $f, f'' \in \Lsqr {L_n} =
  \bigoplus_{k > n} \Lsqr {G_k}$ such that $f=\{f_k\}_k$ with $f_k:= f
  \restr {G_k}$ satisfies
  \begin{equation}
    \label{eq:ll.bc1}
    f_1(n)=0 \qquad \text{and} \qquad
    f_k \in \Sob[2] {G_k}, \quad k>n
  \end{equation}
  (in particular, $f_k$ satisfies the inner boundary conditions as
  in~\eqref{eq:sob.1} and~\eqref{eq:sob.2}), and
  \begin{equation}
    \label{eq:ll.bc2}
    f_k(k_-) = b_k^{-1/2} f_{k+1}(k_+) \qquad \text{and} \qquad
    f_k^\dag(k_-) = b_k^{1/2} f_{k+1}^\dag(k_+)
  \end{equation}
  for all $k > n$. Again, the vertex potential strength $q_k$ is
  hidden in~\eqref{eq:ll.bc2} in the symbol $f_k^\dag(k_-)$
  (cf.~\eqref{eq:gen.der}).
\end{definition}

We can now associate a line-like metric graph $L=L_0$ to a radial
tree-like metric graph: Let $\{t_k\} \subset E(T)$ be an infinite path
in the tree graph $T$ such that $\gen t_n=n$. In particular, the path
starts at $\bd_-t_1=o$. We denote $L_0$ the quantum subgraph of $G$
corresponding to the path $\{t_k\}$ in $T$. Similarly, let $L_n$ be
the quantum subgraph of $L_0$ starting at generation $n$. For example
if $G$ is a simple tree graph, $L_n$ is isometric to the half line
$[0,\infty)$. In general, $L_n$ is isometric to the concatenation of
the decoration graphs $G_k:=G_*(t_k)$ where $k_-=o_{t_k,1} \cong
o_{t_{k+1},0}=k_+$ are identified ($k > n$). Clearly, $L_n$ is a
line-like graph. Similarly, a tree-like quantum graph, i.e., a
tree-like metric graph $G$ with Hamiltonian $H_G$ determines uniquely
a sequence of line-like quantum graphs $\{L_n\}_n$ with Hamiltonians
$\{H_{L_n}\}_n$.

We will see that the converse is also true: The family $\{H_{L_n}\}_n$
of Hamiltonians on the line-like graphs $L_n$ determines uniquely the
(spectral) behavior of the quantum graph $G$ with Laplacian $H_G$.
Namely, due to the symmetry reduction in \Thm{reduction}, we can
reduce the spectral analysis of $H_G$ on $G$ to the analysis of the
family $\{H_{L_n}\}_n$ on the line-like graphs $L_n$. Note that as in
the simple tree graph case, the functions on the tree-like graph and
on the line-like graph differ by a weight factor (although we denoted
both by $f$), see \Rem{exp.growth}.

We will need some assumptions on the decoration graphs $\{G_n\}$ and
the vertex potential strength --- like the assumptions ~\eqref{eq:len.bd}
and~\eqref{eq:pot.bd} for a simple tree graph --- for example to
ensure the self-adjointness of $H_G$ and $H_L$, and the bounds on
generalized eigenfunctions.

\begin{assumption}
  \label{ass:deco.gr}
  We say that the sequence of quantum decoration graphs $\{G_n\}_n$ is
  \emph{uniform} if there exist finite constants $\ell_\pm>0$,
  $0<\kappa\le 1$ and $q_\pm>0$ such that each member
  $G_*=(V,E,\bd,\ell,q(o_1)) \in \{G_n\}_n$ satisfies the following
  conditions:
  \begin{subequations}
    \label{eq:ass.graphs}
    \begin{align}
      \label{eq:lower.len}
      d_{G_*}\bigr(o_0, o_1\bigl) &\ge \ell_-,\\
      \label{eq:start.vx}
      \ell_e &\ge \kappa \ell_-,
      \qquad e \in E_{o_0}\\
      \label{eq:start.vx2}
      \deg o_0 &= 1\\
     \label{eq:vol.bd}
     \ell(G_*) &:=\sum_{e \in E(G_*)} \ell_e \le \ell_+ \\
      \label{eq:q.bdd}
      q_- \le q(o_1) &\le q_+.
    \end{align}
  \end{subequations}
\end{assumption}
The first three assumptions assure that each decoration graph $G_*$ is
``long'' enough and does not branch at the starting vertex $o_0$ (this
will be needed in order to calculate the Green's function,
cf.~\Lem{green}).  The fourth condition is a global upper bound on the
decoration graph (cf.~\eqref{eq:len.bd}). Assumption~\eqref{eq:q.bdd}
is a global bound on the strength of the vertex potential
(cf.~\eqref{eq:pot.bd}).

Note that all our assumptions are fulfilled on a simple tree graph,
i.e., when $G_n$ consists of a single edge with vertex potential at
the ending vertex (\Exenum{graph.deco}{simple.tree}). The same is true
for \Exenum{graph.deco}{necklace}. In addition, in
\Exenum{graph.deco}{end.point}, the assumptions are fulfilled once
there is a lower bound on the base edge length $\ell_n \ge \ell_- >0$
or the end vertex $n_-$ of $G_n$ does not lie on the base edge.  In
\Exenum{graph.deco}{kirchhoff} we only need to assure that in the
(constant) decoration graph $G_*$ the vertex $o_0$ has degree $1$ and
a bounded vertex potential (cf.~\eqref{eq:q.bdd}).

We summarize the various results needed later which are proven in the
appendix (cf.~\Thm{reduction}, \Lem{ess.sa},
\Thms{sp.meas.ef}{ptw.bd.ef}).
\begin{theorem}
  \label{thm:op.result}
  Assume that the sequence of decoration graphs $\{G_n\}$ is uniform
  and of polynomial length growth (i.e, it satisfies
  Assumptions~\eqref{eq:ass.graphs}). Assume in addition that $G$ is
  the radial tree-like quantum graph with decoration graphs $\{G_n\}$
  and branching number sequence $\{b_n\}$.  Denote by $L_n$ the
  associated line-like graph $L_n$ starting at vertex $n$ and by
  $H_{L_n}$ its Hamiltonian.  Then $H_G$ defined in~\Def{tree-like.qg}
  is self-adjoint on $\dom H_G$.  Furthermore,
  \begin{equation*}
    H_G \cong  H_1 \oplus
    \bigoplus_{n=2}^\infty
        (\oplus b_0 \cdot \ldots \cdot b_{n-2}(b_{n-1}-1)) H_n
  \end{equation*}
  where $(\oplus m) H_n$ means the $m$-fold copy of $H_n$.  Each
  operator $H_n = H_{L_{n-1}}$ is self-adjoint on $\dom H_{L_n}$ as
  defined in \Def{ll.graph.qg}.
  
  In addition, the spectrum of $H_n$ is supported by polynomially
  bounded generalized eigenfunctions $\phi$. More precisely,
  $\phi(k_+)$ and $\phi^\dag(k_+)$ are bounded by $k$ times a constant
  depending only on the constants of~\eqref{eq:ass.graphs} and the
  eigenvalue.
\end{theorem}

%----------------------------------------------------------------------
\subsection{Transfer matrices}
\label{sec:tm.gen}
%----------------------------------------------------------------------

As in the tree graph case, we need control over the growth of
generalized eigenfunctions of the Hamiltonian $H=H_L$ of a line-like
graph.  A generalized eigenfunction here is a function satisfying
\begin{equation}
  \label{eq:ew1}
  H f = -f'' = \lambda f
\end{equation}
on each edge such that $f$ satisfies all inner boundary conditions
(i.e., $f \restr {G_n} \in \Sob[2] {G_n}$) and all connecting boundary
conditions~\eqref{eq:ll.bc2} except at $0$ (and there is no
integrability condition at $\infty$).

We can calculate the solutions explicitly, since on each edge, the
solution still has the form~\eqref{edgesoln1} with coefficients
determined by the boundary conditions. Namely, we can define the
\emph{transfer} or \emph{monodromy matrix} $T_\lambda(n)$ for the
decoration graph $G_n$ as follows: For a given $\vec F(n-1) =
(F_{n-1},F'_{n-1})^{\tr} \in \C^2$ let $f$ be a solution of~\eqref{eq:ew1}
such that $f \in \Sob[2] {G_n}$, i.e., $f$ satisfies all inner
boundary conditions (cf.~\eqref{eq:sob.2}) and $f((n-1)_+)=F_{n-1}$
and $f^\dag((n-1)_+)=F_{n-1}'$. The transfer matrix is then defined as
in~\eqref{eq:trans.mat} via
\begin{equation}
  \label{eq:trans.mat2}
   \vec F(n) =
   T_\lambda(n) \vec F(n-1),
    \qquad \text{and} \qquad
   \vec F(0) \in \C \begin{pmatrix} 0\\ f^\dag(0+) \end{pmatrix}.
\end{equation}
where
\begin{equation}
  \label{eq:vec.f2}
  \vec F(n) := \vec F(n_+) :=
  \begin{pmatrix}
    f(n_+) \\ f^\dag (n_+)
  \end{pmatrix}.
\end{equation}
We sometimes write
\begin{equation}
  \label{eq:tm.sol}
  T_\lambda(x,G_n) \vec F(n-1)=f(x)
\end{equation}
for the solution $f$ of the eigenvalue equation on $G_n$ with initial
data $\vec F(n-1)$.

Note that in contrast to the simple tree graph case where $G_n$ is a
single edge, the transfer matrix might not be defined for all energies
$\lambda \in \C$. We specify an exceptional set $E(G_n)$ in
~\eqref{eq:exc.g} for which the transfer matrix might not be defined.
The set $E(G_n)$ roughly consists of the spectrum of the Dirichlet
operator on $G_n$, i.e., the self-adjoint operator $H_{G_n}^\Dir$ with
boundary condition $f((n-1)_+)=0$ and $f(n_-)=0$. In addition, there
might be more exceptional energies expressed via the
Dirichlet-to-Neumann map on $G_n$. We call the values in $E(G_n)$ the
\emph{exceptional energies} of $G_n$.  The exceptional set $E(L)$ of
the line-like graph $L$ consisting of the concatenation of all $G_n$'s
is the union of all exceptional sets. In particular, if $\lambda
\notin E(L)$, then the transfer matrix $T_\lambda(n)$ is uniquely
defined as below and has determinant $1$ (\Lem{exc.set}).

In some concrete examples, it is easier to directly determine the set
of values $\lambda$ for which the transfer matrix is not defined. The
direct calculation has the advantage, that the set of values for which
$T_\lambda(n)$ is not defined may be smaller than the set $E(L)$
defined abstractly in \Def{non.sep.l}.  This phenomena occurs for the
simple tree graph: The abstract setting would yield the Dirichlet
spectrum of a single edge, namely $E(L)=\set{\pi^2k^2/\ell_n^2}{k \in
  \N, n \in \N}$, but the direct calculation of \Sec{tm.tree} shows,
that the transfer is defined for all values of $\lambda$ (cf.\ also
\Lem{exc.set}).

We will give the transfer matrices and Dirichlet-to-Neumann maps for
the examples cited below:
\begin{example}
  \label{ex:graph.tm}
  \begin{enumerate}
    \setcounter{enumi}{1}
  \item \emph{Graph decoration at the ending point:}
    The transfer matrix of the decoration graph associated to the
    energy $\lambda$ is given by
    \begin{equation}
      \label{eq:gr.deco.2}
      T_\lambda(n)=
      D(b) T_\lambda(\hat G_*) R_{\mu}(\mu \ell_n)
    \end{equation}
    where $T_\lambda(\hat G_*)$ is the transfer matrix with respect to
    the decoration graph $G_*$ and where $\mu = \sqrt \lambda$.

    If $G_*$ is a graph attached to the end point of the edge (i.e.,
    the connecting points $(n-1)_+$ and $n_-$ lie on the base edge),
    then $T_\lambda(\hat G_*) = S(r_\lambda)$ where
    $r_\lambda=r_\lambda(\hat G_*)$ is the Dirichlet-to-Neumann map
    associated to the graph $\hat G_*$ with $o_1$ as single boundary
    point, i.e., $r_\lambda=\phi^\dag(o_1)$ where $\phi$ is the unique
    solution of $H_{G_*} \phi = \lambda \phi$ with $\phi(o_1)=1$. The
    Dirichlet-to-Neumann map is defined for all $\lambda \notin
    E(G_*)$ where $E(G_*)$ is the spectrum of the Dirichlet
    Hamiltonian $H_{G_*}^\Dir$ (with Dirichlet boundary condition at
    $o_1 \in V(\hat G_*)$). The transfer matrix is similar to the one
    of the RKM, i.e.,
    \begin{multline}
      \label{eq:gr.deco.2b1}
      \qquad \qquad T_\lambda(n)=
      D(b) S(r_\lambda) R_\mu(\mu \ell_n)\\=
       \begin{pmatrix}
          b^{1/2} \cos (\mu \ell_n) &
             \dfrac{b^{1/2}} \mu \sin (\mu \ell_n)\\
          \dfrac{-\mu \sin (\mu \ell_n) +
             r_\lambda \cos (\mu\ell_n)}{b^{1/2}}&
          \dfrac{\cos (\mu \ell_n)
            + \dfrac {r_\lambda} \mu \sin (\mu \ell_n)}{b^{1/2}}
       \end{pmatrix},
    \end{multline}
    but now with an energy depending vertex potential strength and the
    random parameter being a length perturbation.  The
    Dirichlet-to-Neumann map is
    \begin{equation}
      \label{eq:d2n.2}
      \Lambda(\ell,\lambda) = \mu
      \begin{pmatrix}
        -\cot \mu \ell & \dfrac 1 {\sin \mu \ell}\\
        -\dfrac 1 {\sin \mu \ell} &
             \cot \mu \ell  + \dfrac {r_\lambda} \mu
      \end{pmatrix}.
    \end{equation}
    Concretely, in the loop decoration model (with a loop of length
    $1$), we have
    \begin{equation}
      \label{eq:gr.deco.2b2}
      T_\lambda(n)=D(b) S(r_\lambda) R_\mu(\mu\ell_n)
      \quad \text{with} \quad
      r_\lambda = -2\mu \tan(\mu/2)
    \end{equation}
    with exceptional set $E(\ell)=\set{\pi^2 k^2}{k \in \N}$
    independent of $\ell$.
  \item \emph{Necklace or onion decoration:} Here, the transfer matrix
    is
    \begin{multline}
      \label{eq:gr.deco.3}
      \qquad \qquad T_\lambda(n)=
       D(b) R_{p \mu}(\ell_n \mu)
            R_{\mu} (\mu) \\=
      \begin{pmatrix}
        b^{1/2} \cos_{\frac 1p}(\mu,\ell_n) &
                \dfrac{b^{1/2} \sin_p(\mu,\ell_n)} {p \mu}\\
       -b^{-1/2} p \mu \sin_{\frac 1p}(\mu,\ell_n) &
        b^{-1/2} \cos_p(\mu,\ell)
      \end{pmatrix}
    \end{multline}
    defined for all $\lambda>0$ where
    \begin{subequations}
      \label{eq:sink.cosk}
    \begin{align}
      \sin_p(\mu,\ell) & :=
      \sin(\mu \ell) \cos \mu + p \cos(\mu \ell) \sin \mu\\
      \cos_p(\mu,\ell) & := \cos(\mu \ell) \cos \mu - p \sin(\mu \ell)
      \sin \mu.
    \end{align}
  \end{subequations}
The Dirichlet-to-Neumann map is
    \begin{equation}
      \label{eq:d2n.3}
      \Lambda(\ell,\lambda):= \frac{p\mu}{\sin_p (\mu,\ell)}
      \begin{pmatrix}
        - \cos_{\frac 1p}(\mu,\ell) & 1\\
        -1  & \cos_p (\mu,\ell)
      \end{pmatrix}
    \end{equation}
    with exceptional set $E(\ell)$ consisting of the Dirichlet
    spectrum of the decoration graph (cf.~\Fig{necklace-dir}), i.e.,
    of those $\lambda=\mu^2$ such that $\sin_p(\mu,\ell) = 0$ or
    $\sin(\mu \ell)=0$.\footnote{\label{fn:red.dir} Note that the
      Dirichlet spectrum of the necklace decoration graph consists of
      the squares $\lambda=\mu^2$ of the zeros of $\sin_p(\mu,\ell)=0$
      \emph{and} of $\sin (\mu \ell)=0$ (cf.~\Fig{necklace-dir}).  The
      latter zeros correspond to eigenfunctions living only on the
      loop.  These zeros do not appear as poles in the
      Dirichlet-to-Neumann map, since in its definition, the end
      vertices of the loop edges are identified as one vertex $o_1$
      (cf.\ also \Lem{d2n}).}
    \begin{figure}[h]
%----------------------------------------------------------------------
%        \centering \input{loc-graph-fig5.pstex_t}
\begin{picture}(0,0)%
\includegraphics{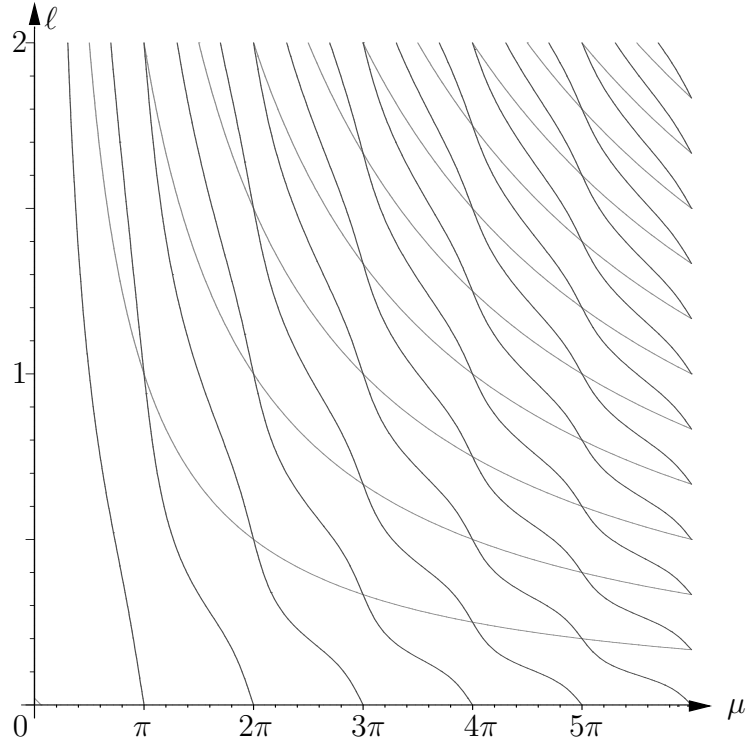}%
\end{picture}%
\setlength{\unitlength}{4144sp}%
\begingroup\makeatletter\ifx\SetFigFont\undefined%
\gdef\SetFigFont#1#2#3#4#5{%
  \reset@font\fontsize{#1}{#2pt}%
  \fontfamily{#3}\fontseries{#4}\fontshape{#5}%
  \selectfont}%
\fi\endgroup%
\begin{picture}(4635,4635)(1,-3616)
\put(4636,-3256){\makebox(0,0)[lb]{\smash{{\SetFigFont{12}{14.4}{\rmdefault}{\mddefault}{\updefault}{\color[rgb]{0,0,0}$\mu$}%
}}}}
\put(551,844){\makebox(0,0)[lb]{\smash{{\SetFigFont{12}{14.4}{\rmdefault}{\mddefault}{\updefault}{\color[rgb]{0,0,0}$\ell$}%
}}}}
\put(355,-1278){\makebox(0,0)[lb]{\smash{{\SetFigFont{12}{14.4}{\rmdefault}{\mddefault}{\updefault}{\color[rgb]{0,0,0}$1$}%
}}}}
\put(355,704){\makebox(0,0)[lb]{\smash{{\SetFigFont{12}{14.4}{\rmdefault}{\mddefault}{\updefault}{\color[rgb]{0,0,0}$2$}%
}}}}
\put(361,-3391){\makebox(0,0)[lb]{\smash{{\SetFigFont{12}{14.4}{\rmdefault}{\mddefault}{\updefault}{\color[rgb]{0,0,0}$0$}%
}}}}
\put(1081,-3391){\makebox(0,0)[lb]{\smash{{\SetFigFont{12}{14.4}{\rmdefault}{\mddefault}{\updefault}{\color[rgb]{0,0,0}$\pi$}%
}}}}
\put(1711,-3391){\makebox(0,0)[lb]{\smash{{\SetFigFont{12}{14.4}{\rmdefault}{\mddefault}{\updefault}{\color[rgb]{0,0,0}$2\pi$}%
}}}}
\put(2386,-3391){\makebox(0,0)[lb]{\smash{{\SetFigFont{12}{14.4}{\rmdefault}{\mddefault}{\updefault}{\color[rgb]{0,0,0}$3\pi$}%
}}}}
\put(3061,-3391){\makebox(0,0)[lb]{\smash{{\SetFigFont{12}{14.4}{\rmdefault}{\mddefault}{\updefault}{\color[rgb]{0,0,0}$4\pi$}%
}}}}
\put(3691,-3391){\makebox(0,0)[lb]{\smash{{\SetFigFont{12}{14.4}{\rmdefault}{\mddefault}{\updefault}{\color[rgb]{0,0,0}$5\pi$}%
}}}}
\end{picture}%
%----------------------------------------------------------------------
       \caption{The Dirichlet spectrum of the necklace decoration
         ($p=2$). The zeros of $\sin_p(\mu,\ell)$ are plotted in dark
         grey, the zeros of $\sin(\mu \ell)$ are plotted in light grey.}
      \label{fig:necklace-dir}
    \end{figure}
  \item \emph{Line-like graphs:} Here, we have
    \begin{equation}
      \label{eq:gr.deco.4}
      T_\lambda(n) =
      T_\lambda(G_n)
    \end{equation}
    where $T_\lambda(G_n)$ is the transfer matrix of the (random)
    decoration graph $G_n$.
  \item \emph{Kirchhoff models:} The transfer matrix is just
    \begin{equation}
      \label{eq:gr.deco.5}
      T_\lambda(n)=
       D(b) S(q_n) \hat T_\lambda(G_*)
    \end{equation}
    where $q_n$ denotes the strength of the vertex potential at $n_-$
    and where $\hat T_\lambda(G_*)$ is the transfer matrix of the fixed
    decoration graph $G_*$.
  \end{enumerate}
\end{example}

We end this section with a typical example for the \emph{periodic}
spectrum.  Note that the onion decoration was also considered
in~\cite{ael:94} as a line-like graph ($b=1$) considering the band-gap
ratio of the periodic operator.
\begin{theorem}
  \label{thm:per.necklace}
  Suppose that all length are the same $\ell_n=\ell$ in the
  necklace/onion decoration model \Exenum{graph.deco}{necklace} with
  branching number $b\ge 1$ and edge decoration number $p \ge 2$. Then
  the spectrum of the corresponding Laplacian on the decoration graph
  $G=G(\ell)$ is given by
  \begin{multline}
    \label{eq:per.sp.necklace}
    \spec{\laplacian{G(\ell)}} =
    \bigset{\mu^2} {|(b^{1/2}+b^{-1/2}) \cos_\kappa(\mu,\ell)| \le 2}
       \cup
    \bigset{\mu^2} { \sin_p(\mu,\ell)=0} \\=
    \bigcup_{k=1}^\infty \bigl(B_k(\ell) \cup \{\lambda_k(\ell)\} \bigr),
    \qquad
     \kappa := \frac{b+p^2}{p(b+1)},
  \end{multline}
  where $B_k(\ell)$ are compact intervals and $\lambda_k(\ell)$ is the
  $k$th Dirichlet eigenvalue of a single decoration graph with length
  $\ell$ (cf.~\eqref{eq:sink.cosk} for the notation $\cos_p$ etc.).
\end{theorem}
\begin{proof}
  The spectral characterization is a simple consequence of
  \Thm{sp.per.op} and~\eqref{eq:gr.deco.3}. Note that $\tr
  T_\lambda$ is nonconstant.
\end{proof}
\begin{remark}
  \label{rem:necklace}
  \begin{enumerate}
    \label{monotone}
  \item The square roots of the band edges as functions of $\ell$
    (i.e., the solutions $\mu=\mu_k^\pm(\ell)$ of the equation
    $(b^{1/2}+b^{-1/2})\cos_\kappa(\mu,\ell)=\pm 2$) satisfy
  \begin{equation*}
    \mu_k'(\ell)= - \mu
       \Bigl(\ell + \kappa \dfrac{\sin_{\frac 1 \kappa}(\mu ,\ell)}
          {\sin_\kappa(\mu,\ell)}\Bigr)^{-1}.
  \end{equation*}
  Numerical examples show that $\mu_k'(\ell)<0$, i.e, that the band
  edges are monotonically decreasing in $\ell$. Furthermore, if $b$ or
  $p$ are very large, the bands are very narrow. In addition, for
  small $b$ and large $p$ (i.e, if $\kappa \gg 1$), the bands are
  almost constant if $\ell$ is not an integer.
\item
  \label{equal}
  \sloppy In the case $b=p$, i.e., if the branching number equals the
  number of decoration edges in the loop, we have an interesting
  phenomena: First, $\kappa=1$ and $\mu_k'(\ell)=-\mu(\ell +
  1)^{-1}<0$.  Furthermore,
  $\cos_{\kappa}(\mu,\ell)=\cos(\mu(\ell+1))$ and
  $\sin_\kappa(\mu,\ell)=\sin(\mu(\ell+1))$, i.e., the absolutely
  continuous spectrum is exactly the same as for the RLM with length
  $\ell+1$ (cf.~\Thm{sp.per.lm}). In this sense, the transport
  properties of the branched necklace model and the simple RLM are the
  same, i.e., for the transport properties, it is irrelevant, whether
  there are loops or the loops are opened at the end point (in order
  to obtain a RLM with length $\ell+1$).
\end{enumerate}
\end{remark}
%----------------------------------------------------------------------
%
\section{Localization for random tree-like quantum graphs}
\label{sec:ran.graphs}
%----------------------------------------------------------------------

%----------------------------------------------------------------------

%----------------------------------------------------------------------
\subsection{General random models}
\label{sec:ran.graphs.gen}
%----------------------------------------------------------------------

Here, we assume that the symmetric, radial tree-like quantum graph
$G=(V,E,\bd,\ell,q)$ which is completely determined by the sequence of
decoration graphs $\{G_n\}_n$ together with the sequence of branching
numbers $\{b_n\}$ is random in the following sense:
\begin{definition}
   \label{def:rand.graph}
   Let $\mathcal G$ be a family of % uniform
   compact quantum decoration graphs.  We say that the radial
   tree-like quantum graph $G$ is constructed \emph{randomly} from the
   set $\mathcal G$, if there is an iid\ sequence of random variables
   $\{G_n,b_n\}_n$ with values in $\mathcal G$ and $\{b_-, \dots,
   b_+\}$, respectively, such that $G(\omega)$ has the decoration
   graph $G_n(\omega)$ and the branching number $b_n(\omega)$ at
   generation $n$. Similarly, the sequence of iid random variables
   $\{G_n,b_n\}$ determines a \emph{random line-like quantum graph}.
\end{definition}

We fix a probability measure $\Prob_1$ on $\Omega_1:= \mathcal G
\times \{b_-, \dots, b_+\}$. Clearly, we can consider a random radial
tree-like or line-like quantum graph $G(\omega)$ or $L(\omega)$ as a
random variable on the product measure space $(\Omega,\Prob):=
(\Omega_1,\Prob_1)^\N$.

We are mostly interested in \emph{minimal} random models, since one
expects localization at least for high disorder. In all our
application, the class of decoration quantum graphs $\mathcal G$ will
depend only on one real parameter. For example in the RLM, $\Omega_1$
consists of quantum graphs $G=G(\ell)$ of a single edge and fixed
branching number $b$. The random parameter is the length, so we can
set $\ell \in [\ell_-,\ell_+]=:\Omega_1$.  In the RKM, we have a
similar model, now $\Omega_1 := [q_-,q_+]$.

In order to copy the proof of localization of Kotani as in
\Thm{kotani}, we need some further adaptations, mainly due to the
fact, that several constants tend to $\infty$ if we approach the
exceptional set. Here, and in the sequel,
$\Lambda_{ij}(\omega_1,\lambda)$ are the components of the
Dirichlet-to-Neumann map of the decoration graph $G(\omega_1)$
defined in \Def{d2n}.

We need more assumptions for the random model. Let $\Sigma_0 \subset
\R$ be a bounded interval.
\begin{assumption}
   \label{ass:ran.graph}
   We say that the random radial tree-like graph $G=G(\omega)$ with
   decoration graphs in $\mathcal G$ is \emph{good in the compact
     spectral interval $\Sigma_0$} if the following conditions are
   fulfilled:
   \begin{enumerate}
     \item
       \label{deco.gr}
       The family of decoration graphs $\mathcal G$ is \emph{uniform},
       i.e., each decoration graph $G(\omega_1) \in \mathcal G$
       satisfies~\eqref{eq:ass.graphs} $\Prob_1$-almost surely with
       \emph{uniform} constants.
     \item
       \label{comp}
       The single site probability space $\Omega_1$ is the union of
       finitely many compact intervals with its Borel
       $\sigma$-algebra, and the eigenvalues of the Dirichlet
       Laplacian $\laplacianD {G(\omega_1)}$ depend piecewise
       analytically on $\omega_1$. In addition, the
       Dirichlet-to-Neumann map $\Lambda(\omega_1,\lambda)$ depend
       analytically on $\omega_1$ whenever $\lambda \notin
       \spec{\laplacianD {G(\omega_1)}}$. Both maps are assumed to be
       continuous up to the border of $\Omega_1$.
     \item
       \label{exc.set}
       We assume that the exceptional set $E(G(\omega_1)) \subset \R$
       of each decoration graph (cf.~\eqref{eq:exc.g}) is discrete.
     \item
       \label{int.cond}
       There is a constant $C=C(\lambda)$ such that\footnote{Note that
         $\norm A=\norm{A^{-1}}$ for $A \in \SL_2 (\R)$ (see
         also~\eqref{eq:int.cond}).}
       \begin{equation*}
         \Exp[_1] {\ln \norm{T_\lambda(\cdot)}} \le C_\lambda
       \end{equation*}
    \item
      \label{sp.av} There exists an increasing sequence of real
      numbers $\{\lambda_k\}$ and for each $k$ a sequence
      $\{\delta_{k,n}\}_n$, $\delta_{k,n} \to 0$ as $n \to \infty$
      such that the spectral averaging formula~\eqref{eq:weyl.bdd}
      holds in the compact energy interval
      $[\lambda_k+\delta_{k,n},\lambda_{k+1}-\delta_{k+1,n}] \cap
      \Sigma_0$.
  \end{enumerate}
\end{assumption}

\begin{remark}
  \label{rem:ran.graph}
  \begin{enumerate}
  \item \label{exc.set.real2} We believe that
    \Assenum{ran.graph}{exc.set} is generally true (under some mild
    conditions), although we are not aware of a proof. Since this
    condition is always satisfied in our examples, we state it as an
    assumption (see also \Remenum{exc.set}{exc.set.real}).
  \item \Assenum{ran.graph}{sp.av} is usually fulfilled only for
    models if the single site random distribution is absolutely
    continuous, i.e., if there is a nonnegative function $\eta \in
    \Lp[\infty]{\Omega_1}$ such that $\dd \Prob_1(\omega_1)=
    \eta(\omega_1) \dd \omega_1$.

    Typically, the sequence $\{\lambda_k\}_k$ consists of the
    Dirichlet spectrum of the decoration graph (with length $\ell \in
    \bd \Omega_1$ in random lengths models). In some random lengths
    models, the exceptional set is not needed, e.g.\ in
    \Exenum{graph.deco}{end.point} or the RLM of \Sec{ran.tg.loc}.
  \end{enumerate}
\end{remark}

We set
\begin{equation}
  \label{eq:exc.set}
  E_0 := \set {(\omega_1,\lambda)} {\lambda \in E(G_*(\omega_1))}.
\end{equation}
The next lemma assures that $E_0$ is still a ``small'' set:
\begin{lemma}
  \label{lem:borel.cantelli}
  There exists $a>0$ such that
   \begin{equation*}
      E_k := \bigset{(\omega_1,\lambda) \in \Omega_1 \times \Sigma_0}
         {\dist(\lambda, \spec{H_{G(\omega_1)}^\Dir}) < \eta_k \text{ or }
          |\Lambda_{01}(\omega_1,\lambda)|< \eta_k}
   \end{equation*}
   with $\eta_k=k^{-2a}$ fulfills $\sum_k (\Prob_1 \otimes \leb)(E_k) <
   \infty$. Furthermore, $E_0 := \bigcap_k E_k$ and $(\Prob_1 \otimes
   \leb) (E_0)=0$.
\end{lemma}
Note that $E(G_*(\omega_1))$ is a \emph{closed} set (and therefore
measurable) and that it consists of $\spec{H_{G(\omega_1)}^\Dir}$ and
those $\lambda \in\Sigma_0$ such that $\Lambda_{01}(\omega_1,\lambda)=0$.
\begin{proof}
  By assumption, the Dirichlet eigenvalues $\lambda_k(\omega_1)$
  depend piecewise analytically on $\omega_1$ and that
  $\Lambda_{01}(\omega_1,\lambda)$ is analytic (by assumption it is
  analytic in $\omega_1$ and by the series
  representation~\eqref{eq:d2n.series} it is also analytic in
  $\lambda$). Therefore, the thickened exceptional set $E_k$ lies in a
  strip of order $k^{-2}$ around $E_0$ if we choose $\eta_k=k^{-2a}$
  for some $a>0$. Since $\Omega_1 \times \Sigma_0$ is compact, the sum
  over the measures is finite. The second assertion is an easy
  consequence of the first Borel-Cantelli lemma, see for
  example~\cite[Thm.~3.1]{simon:79b}.
\end{proof}

Next, we need several lemmas ensuring that we have a \emph{global}
$\Lsymb_2$-estimate as in~\eqref{eq:l2.L2} on the eigenfunction
$T_\lambda(\cdot, G(\omega_1)) \vec F_0$ defined in~\eqref{eq:tm.sol}
on a sufficiently large subset of $\Omega \times \Sigma_0$:
\begin{lemma}
  \label{lem:exc.set1}
  There exists a sequence $\{C_k'\}_k$ growing at most polynomially
  such that
  \begin{equation}
    \label{eq:exc.set1}
         E_k':= \Bigset{(\omega_1,\lambda) \in \Omega_1 \times \Sigma_0
           \setminus E_0}
        {\exists \vec F_0 \ne \vec 0 \colon\;
             \norm{T_\lambda\bigl(\cdot,G(\omega_1)\bigr) \vec F_0} >
             C_k' |\vec F_0|}
  \end{equation}
  satisfies $E_k' \subset E_k$. In particular, $E_0' :=
  \bigcap_k E_k' \subset E_0$ has $(\Prob_1 \otimes \leb)$-measure
  $0$.
\end{lemma}
\begin{proof}
  Let
  \begin{equation}
    \label{eq:exc.set3}
         \wt E_k :=
          \bigcup_{i,j=0,1}
            \bigset{(\omega_1,\lambda) \in \Omega_1 \times \Sigma_0
                        \setminus E_0}
        {|\Lambda_{ij}(\omega_1,\lambda)| > \wt C_k}
  \end{equation}
  where
  \begin{equation}
    \label{eq:d2n.bdd}
    \wt C_k := \sup \bigset{|\Lambda_{ij}(\omega_1,\lambda))|}
      {(\omega_1,\lambda) \in \Omega_1 \times \Sigma_0 \setminus E_k,
         \quad i,j=0,1}.
  \end{equation}
  Note that the supremum $\wt C_k$ exists since $E_k$ is an open set
  and $\Omega_1 \times \Sigma_0$ is compact by
  \Assenum{ran.graph}{comp}.  In addition, $\wt C_k$ is bounded by the
  supremum of the entries of the Dirichlet-to-Neumann map on the set
  $K_k$ of $(\omega_1,\lambda)$ with
  $\dist(\lambda,\spec{H_{G(\omega_1)}^\Dir} \ge \eta_k$ only. But
  since $K_k$ is compact, and since the Dirichlet-to-Neumann map is
  meromorphic with simple poles (see~\eqref{eq:d2n.series}), we have
  $|\Lambda_{ij}(\omega,\lambda)| \le \wt C/\eta_k$ for a constant
  $\wt C>0$ independent of $k$. In particular, $\wt C_k \le \wt
  C/\eta_k=O(k^{2a})$ as in \Lem{borel.cantelli}.

  By definition, we have $\compl{(E_k)} \subset \compl{(\wt E_k)}$.
  Furthermore, we can bound the norm of $T_\lambda(\cdot,
  G(\omega_1))$ estimated in~\eqref{eq:norm.sol.map} by
  \begin{equation*}
    C_k':=
    \Bigl( 1+ \frac {1+ \sup \Sigma_0} {\eta_k}\Bigr)
    \norm E
    \Bigl( 1 + \frac {\wt C_k+1}{\eta_k}\Bigr)
  \end{equation*}
  for $(\omega_1,\lambda) \in \compl{(E_k)}$.  Therefore,
  $\compl{(E_k)} \subset \compl{(E_k')}$ and $C_k'=O(k^{6a})$ follows.
\end{proof}

Similarly, we can show that the set where the norm of the transfer
matrix is not bounded, is small:
\begin{lemma}
  \label{lem:exc.set2}
  There exists a sequence $\{C_k''\}_k$ growing at most polynomially
  such that
  \begin{equation}
    \label{eq:exc.set2}
    E_k'':=
     \bigset{(\omega_1,\lambda) \in
                      \Omega_1 \times \Sigma_0 \setminus E_0}
        {\norm{T_\lambda(\omega_1)} > C_k''}
  \end{equation}
  satisfies $E_k'' \subset E_k$. In particular, $E_0'' := \bigcap_k
  E_k'' \subset E_0$ has $(\Prob_1 \otimes \leb)$-measure $0$.
\end{lemma}
\begin{proof}
  The transfer matrix has been expressed in terms of the
  Dirichlet-to-Neumann map in~\eqref{eq:def.tm}. Using an appropriate
  matrix norm, we see that $\norm{T_\lambda(\omega_1)}$ can be
  estimated by $C_k'' = p(\wt C_k)/\eta_k$ for $(\omega_1,\lambda) \in
  \compl{(E_k)}$ where $p(C)$ is a universal polynomial of degree $2$,
  monotone in $C$. As in the previous lemma, $C_k''=O(k^{6a})$ and again,
  $\compl{(E_k)} \subset \compl{(E_k'')}$.
\end{proof}

Let $\map {\pi_n}{\Omega \times \Sigma_0}{\Omega_1 \times \Sigma_0}$,
$(\omega,\lambda) \mapsto (\omega_n,\lambda)$ be the projection onto
the $n$th component.  Furthermore, we set
\begin{align*}
  S_0 &:= \bigcap_n \pi_n^-(\compl{(E_0)}) =
  \set{(\omega,\lambda) \in \Omega \times \Sigma_0}
  {(\omega_n,\lambda) \notin E_0 \text{ for all $n$}},\\
  S_k &:= \pi_k^-(\compl{(E_k)}) =
  \set{(\omega,\lambda) \in S_0}
  {(\omega_k,\lambda) \notin E_k}.
\end{align*}
\begin{lemma}
  \label{lem:borel.cantelli2}
  The sets $S_0$ and $S_k$ are measurable. Furthermore, $S_0$ and
  $\underline S_\infty:=\liminf S_k := \bigcup_{n \in \N} \bigcap_{k
    \ge n} S_k$ have full $(\Prob \otimes \leb)$-measure. In
  particular, for $(\omega,\lambda)=(\omega_1,\omega_2,\dots,\lambda)
  \in S_0$, the transfer matrix $T_\lambda(\omega_n)$ is defined for
  all $n$ and for $(\omega, \lambda) \in \underline S_\infty$ there
  exists $n \in \N$ such that
  \begin{equation}
    \label{eq:norm.est}
    \norm{T_\lambda(\cdot, G(\omega_k))} \le C_k'
        \qquad \text{and} \qquad
    \norm{T_\lambda(\omega_k)} \le C_k''
  \end{equation}
  for all $k \ge n$, i.e., the norm of the solution operator and the
  transfer matrix is bounded by constants depending on $k$ and which
  are of polynomial growth.
\end{lemma}
\begin{proof}
  Clearly, $S_0$ and $S_k$ are measurable since $E_0$ and $E_k$ are
  (by \Assenum{ran.graph}{comp}). Furthermore, $(\omega,\lambda) \in
  S_0$ iff $\lambda \notin E(G(\omega_n))$ for all $n$, i.e., for
  these $\omega$ and $\lambda$, the transfer matrix
  $T_\lambda(\omega_n)$ is defined for all $n$. In addition,
  \begin{equation*}
    (\Prob \otimes \leb)(\compl{(S_0)}) =
    \lim_n (\Prob \otimes \leb)
       \bigl( \bigcup_{k \le n} \pi_k^-(E_0) \bigr) \le
    \sum_n \bigl( (\Prob_1 \otimes \leb)(E_0) \bigr) = 0
  \end{equation*}
  due to the continuity of the measure and \Lem{borel.cantelli}.

  Next, we have
  \begin{equation*}
    \sum_k (\Prob \otimes \leb)(\compl{(S_k)}) =
    \sum_k (\Prob_1 \otimes \leb) (E_k) < \infty
  \end{equation*}
  due to the independence of the family $\{S_k\}_k$ and
  \Lem{borel.cantelli}. It follows from the Borel-Cantelli
  lemma for the complement $\compl{(\underline S_\infty)}$ that
  $\underline S_\infty$ has full measure in $\Omega \times \Sigma_0$.
  The norm estimates are simple consequences of the definitions of
  $E_k'$, respectively, $E_k''$, and the fact that $E_k', E_k'' \subset E_k$
  (see \Lems{exc.set1}{exc.set2}).
\end{proof}
Now, the results of \Sec{lyapunov} extends to the case when the
transfer matrices are only defined on $\compl{(E_0)}$ instead of
$\Omega_0 \times \Sigma_0$. Similarly, the product transfer matrix
$U_\lambda(\omega,n)$ is defined for $(\omega,\lambda) \in S_0$
instead of the full product $\Omega \times \Sigma_0$.

\begin{theorem}
  \label{thm:kotani2}
  Let $H(\omega)$ be the random Hamiltonian on a random tree-like
  graph $G(\omega)$ with constant branching number $b \ge 1$ such that
  \Ass{ran.graph} is fulfilled. Assume in addition, that the Lyapunov
  exponent satisfies $\gamma(\lambda)>0$ for almost all $\lambda \in
  \Sigma_0$.  Then localization holds for all energies in the almost
  sure spectrum, i.e., $\spec {H(\omega)} \cap \Sigma_0$ is almost
  surely pure point.

  In addition, there exists a set $S_0 \subset \Omega \times \Sigma_0$
  of full $(\Prob \otimes \leb)$-measure such that all eigenfunctions
  associated to $\lambda$ and $H(\omega)$ on the tree-like graph with
  $(\omega, \lambda) \in S_0$ decay with almost exponential decay rate
  $\beta:=\gamma(\lambda)+(\ln b)/2$ of an eigenfunction $f$ on the
  tree-like graph in the sense that for each $\eps>0$ there exists
  $C_\eps>0$ such that
  \begin{equation}
    \label{eq:exp.decay3}
    |f(x)| \le C_\eps \e^{-(\beta-\eps)d(o,x)}
  \end{equation}
  for all $x \in T$.
\end{theorem}
\begin{remark}
  We expect that also for the exceptional values $\compl{(S_0)}$ we
  have exponential decaying or even compactly supported
  eigenfunctions; this can be seen in most examples directly. A
  general proof would need more analysis on the behavior in the
  exceptional set (see also \Sec{trans.mat}).
\end{remark}
\begin{proof}
  We argue as in the proof of \Thm{kotani} and stress only the needed
  changes here. We define the set $S$ as in~\eqref{eq:set.lyap.pos},
  but intersected with $\underline S_\infty$ (in particular, all
  transfer matrices are defined).  {}From \Lem{lyapunov} (note
  \Assenum{ran.graph}{int.cond}), we see that there exists a set of
  full measure $S_1 \subset S_0$ such that $\gamma(\lambda)$ exists
  for $\omega \in S_1(\lambda)$ a.s.  Now, together with the
  assumption $\gamma(\lambda)>0$ it follows that $S_1 \subset S$ and
  in particular, $S$ has full measure in $\Omega \times \Sigma_0$ (or
  in $\hat \Omega \times \Sigma_0$, what is the same). For $(\hat
  \omega,\lambda) \in S$, Assumption~\eqref{lim.ex} of \Thm{oseledec}
  is fulfilled. Next, Assumption~\eqref{lim.tm.ex} follows from
  \Lem{borel.cantelli2}: since $C_k''$ has polynomial growth, we have
  \begin{equation*}
    \lim_k \frac 1k \ln \norm{T_\lambda(\omega_k)} \le
    \lim_k \frac 1k \ln C_k'' = 0
  \end{equation*}
  provided $k$ is large enough.
  We therefore get a nontrivial solution $\vec F(\hat \omega, \cdot,
  \theta_0) \in \lsqr {\N,\C^2}$ of the discretized eigenvalue
  equation. To see that the associated eigenfunction $f$ on the
  line-like graph is in $\Lsqr L$, we note that
   \begin{equation*}
    \normsqr[L] f =
    \sum_{k=1}^\infty
      \normsqr[G(\omega_k)]
      {T_\lambda(\cdot,G(\omega_k)) \vec F(\hat \omega, k-1, \theta_0)}
  \end{equation*}
  and estimate the norms by $C_k'|\vec F(\hat \omega, k-1, \theta_0)|$
  if $k \ge n$ for some $n \in \N$ large enough due to
  \Lem{borel.cantelli2}. Since the convergence only depends on the
  behavior of the tail of the sum and since $|\vec F(\hat
  \omega,k-1,\theta_0)|$ decays exponentially in $k$
  (see~\eqref{eq:exp.decay2}), we have $f \in \Lsqr L$.

  If $\rho_\omega$ denotes the spectral measure of $H(\omega)$, note
  that due to \Lem{m-fct}, the Weyl-Titchmarsh function $m$ associated
  to $H(\omega)$ is the Borel transform of a measure $\hat
  \rho_\omega$ and the spectral measure has the decomposition
  $\rho_\omega=\hat \rho_\omega + \rho_{\omega,\mathrm{pp}}$ into
  disjoint measures where $\rho_{\omega,\mathrm{pp}}$ is already pure
  point and has support in $E(L(\omega))=\compl{(S_0(\omega))}$.  The
  rest of the localization proof is similar to the proof of
  \Thm{kotani} replacing the measure $\rho_\omega$ by $\hat
  \rho_\omega$.  Note \Rem{kotani.rkm} for the weaker
  \Assenum{ran.graph}{sp.av} on the spectral averaging condition.

  From the almost exponential decay of $\vec F(\hat \omega,n,0)$
  and~\eqref{eq:norm.est} it follows that $\Phi_\eps f \in \Lsqr L$
  for $\eps>0$ where $\Phi_\eps$ is the exponential weight
  $\Phi_\eps(x)=\e^{(\gamma(\lambda)-\eps)n}$ for $x \in G_n$ and $f$
  is the associated eigenfunction on the line-like graph.
  \Thm{ptw.bd.ef} implies the almost exponential pointwise decay of
  $f$ on $L$ in the sense that for $\eps>0$ there exists $C_\eps>0$
  such that $|f(x)| \le C_\eps \Phi_\eps(x)^{-1}$ for $x \in G_n$, $n
  \in \N$.  Finally, from~\eqref{eq:lower.len} and~\eqref{eq:vol.bd}
  it follows that we can replace the discontinuous weight function
  $\Phi_\eps$ by $\e^{-(\gamma(\lambda)-\eps)d(0,x)}$ on the line-like
  graph. The additional exponential decay $(\ln b)/2$ for an
  eigenfunction on the \emph{tree}-like graph comes from the symmetry
  reduction (see \Remenum{exp.decay}{weight}).
\end{proof}

%----------------------------------------------------------------------
\subsection{Examples}
\label{sec:examples}
%----------------------------------------------------------------------

We are now able to check the assumptions in our concrete examples:
\begin{theorem}
  \label{thm:loc.end.point}
  Suppose that the decoration graph consists of a single edge of
  length $\ell$ with a fixed graph $\hat G_*$ attached at the ending
  point (\Exenum{graph.deco}{end.point}(b),
  \Figenum{deco-graph}{end.point}(b)). Assume that the single site
  perturbation of the decoration at the ending point model with
  branching number $b \ge 1$ has an absolutely continuous distribution
  with bounded density $\dd \Prob_1(\ell)= \eta(\ell) \dd \ell$ and
  support in $\Omega_1:=[\ell_-,\ell_+]$ for $0<\ell_-<\ell_+$. Then
  localization holds for all energies in the almost sure spectrum and
  the eigenfunctions decay almost exponentially with rate
  $\gamma(\lambda)+(\ln b)/2$ in the sense of~\eqref{eq:exp.decay3}.
\end{theorem}
\begin{proof}
  Fix a compact spectral interval $\Sigma_0$. We have to check that
  the decoration graphs are ``good'' in $\Sigma_0$ in the sense of
  \Ass{ran.graph}. Clearly, the decoration graphs satisfy the
  uniformity assumptions~\eqref{eq:ass.graphs}.  Furthermore, the
  dependence of the Dirichlet eigenvalues and the Dirichlet-to-Neumann
  map (cf.~\eqref{eq:d2n.2}) on the random parameter $\ell$ is
  (piecewise) analytic.

  The exceptional set $E(\ell)$ consists only of the Dirichlet
  spectrum of a single decoration graph $\hat G_*$ and of the values
  $\lambda=\mu^2$ such that $\sin(\mu \ell)=0$. In particular,
  $E(\ell)$ is discrete.  The integrability condition is fulfilled
  since the norm of the transfer matrix $T_\lambda(n)$
  (cf.~\eqref{eq:gr.deco.2b1}) can easily be estimated by a constant
  depending only on $\lambda \notin E(\ell)$.  The spectral averaging
  is established in \Cor{sp.av.rlm}.

  The Lyapunov exponent is positive: It is easy to see (due to the
  analytic dependence on $\ell$) that the two
  assumptions~\eqref{eq:spec.cd1} and~\eqref{eq:spec.cd2} of
  \Cor{lyap.pos} are fulfilled. The third condition is also satisfied
  since one can always find two noncommuting matrices
  $T_\lambda(\ell_i)$, $\ell_i \in [\ell_-,\ell_+]$.
\end{proof}

\begin{theorem}
  \label{thm:loc.necklace}
  Assume that the single site perturbation of the necklace or onion
  model (with $p \ge 2$ loop edges) of \Exenum{graph.deco}{necklace}
  (see also \Figenum{deco-graph}{necklace}) with branching number $b
  \ge 1$ has an absolutely continuous distribution with bounded
  density $\dd \Prob_1(\ell)= \eta(\ell) \dd \ell$ and support in
  $\Omega_1:=[0,\ell_+]$. Then localization holds for all energies in
  the almost sure spectrum and the eigenfunctions decay almost
  exponentially with rate $\gamma(\lambda)+(\ln b)/2$ in the sense
  of~\eqref{eq:exp.decay3}.
\end{theorem}
\begin{proof}
  Fix a compact spectral interval $\Sigma_0$. Again, we have to check
  that the decoration graphs are ``good'' in $\Sigma_0$ in the sense
  of \Ass{ran.graph}. Clearly, the necklace, respectively, onion, decoration
  graphs satisfy the uniformity assumptions~\eqref{eq:ass.graphs}.
  Furthermore, the dependence of the Dirichlet eigenvalues and the
  Dirichlet-to-Neumann map (cf.~\eqref{eq:d2n.3}) on the random
  parameter $\ell$ is (piecewise) analytic; in addition,
  $\Lambda(0,\lambda)$ corresponds to the Dirichlet-to-Neumann map of
  a single edge (i.e., the case when the loop of length $\ell$
  degenerates to a point), hence the dependence is continuous up to
  the border of $\Omega_1=[0,\ell_+]$.

  The exceptional set consists only of the Dirichlet spectrum of a
  single decoration graph $G_*(\ell)$ and is therefore a discrete
  subset of $\R$.

  The integrability condition is fulfilled since the norm of the
  transfer matrix $T_\lambda(n)$ (cf.~\eqref{eq:gr.deco.3}) can easily
  be estimated by a constant depending only on $\Sigma_0$ and $p$.

  For the spectral averaging, we use \Lem{sp.av.int}: Note that the
  representation of the M\"obius transformation of the inverse
  transfer matrix $\hat T_z(\ell)^{-1}=R_w(-w) R_{pw}(-\ell w)$
  (cf.~\eqref{eq:integrand}) holds with
  \begin{equation*}
    A_w := \frac p {p^2 \sin^2 w + \cos^2 w}
          \qquad \text{and} \qquad
    B_w := -\frac {(p^2-1) w \sin w \cos w} {p^2 \sin^2 w + \cos^2 w}
  \end{equation*}
  where $w^2=z$. Now,
  \begin{equation*}
    A_w = \frac p {p^2 \sin^2 \mu + \cos^2 \mu} -
      \im \eps \, \frac {2p(p^2-1)\sin \mu \cos \mu }
                     {(p^2 \sin^2 \mu + \cos^2 \mu)^2} + O(\eps^2)
  \end{equation*}
  for $w=\mu+\im \eps$ ($0<\eps\le \eps_0$). Therefore, $\Re A_w =
  O(1)$, $\Im A_w = O(\eps)$ and similarly, $B_w=O(1)$ with constants
  depending only on $\Sigma_0$ and $\eps_0$. The winding number of the
  denominator of the M\"obius transformation is bounded since $\ell
  \in [0,\ell_+]$ and the values of $\mu$ also lie in a compact
  interval. Here, the exceptional values $\{\lambda_k\}$ consists of
  the union of the Dirichlet spectrum $\laplacianD {G_*(\ell)}$ for
  the end points, i.e., $\ell=0$ and $\ell=\ell_+$.

  The Lyapunov exponent is positive: It is easy to see (due to the
  analytic dependence on $\ell$) that the two
  assumptions~\eqref{eq:spec.cd1} and~\eqref{eq:spec.cd2} of
  \Cor{lyap.pos} are fulfilled. The third condition is also satisfied
  since one can always find two noncommuting matrices
  $T_\lambda(\ell_i)$, $\ell_i \in [0,\ell_+]$.
\end{proof}

\begin{theorem}
  \label{thm:loc.kirchhoff}
  Assume that we have a fixed decoration graph $G_*$ in each
  generation satisfying~\eqref{eq:start.vx2} and that the set of zeros
  of the Dirichlet-to-Neumann matrix element $\Lambda_{01}(z)$ is a
  discrete subset of $\R$ (e.g.~if $G_*$ is a necklace decoration).
  Assume in addition, that the single site perturbation is a vertex
  potential at the end point of each decoration graph with range $q
  \in \Omega_1:=[q_-,q_+] \subset \R$.  Then localization holds for
  all energies in the almost sure spectrum with eigenfunctions having
  almost exponential pointwise decay rate $\gamma(\lambda)+(\ln b)/2$
  on the tree-like graph in the sense of~\eqref{eq:exp.decay3}.
\end{theorem}
\begin{proof}
  We argue as in the previous proof.
  Assumptions~\eqref{eq:ass.graphs} are clear. The
  Dirichlet-to-Neumann map in this case does not depend on the random
  parameter $q$. The condition on the exceptional set (here also
  independent of $q$) is fulfilled by assumption, as well as the
  integrability condition (since $q$ has its range in a compact
  interval). The spectral averaging holds due to \Cor{sp.av.rkm}. To
  show that the Lyapunov exponent is positive we apply again
  \Cor{lyap.pos}. One can always find two noncommuting transfer
  matrices $T_\lambda(q_i)=D(b) S(q_i) T_\lambda(G_*)$ provided
  $T_\lambda(G_*)$ is not of the form $D(b)S(\kappa)$. Note that the
  latter can only happen for a countable set of $\lambda$'s since
  generally, a transfer matrix contains the rotation matrices
  $R_{p\mu}(\mu\ell_0)$ ($\lambda>0$).
\end{proof}

\subsubsection*{Mixed examples}

We can also mix the examples, for example an edge decoration with
$b=1$, random length $\ell_1 \in [0,\ell_+]$ and a simple edge of
random length $\ell_2 \in [\ell_-,\ell_+]$ and branching number $b=2$.
The probability space now consists of two components $\Omega_1=
\Omega_{1,1} \cup \Omega_{1,2}$.

%----------------------------------------------------------------------
\subsection{Full-line models}
\label{sec:necklace}
%----------------------------------------------------------------------
So far, we only considered \emph{rooted} radial quantum trees which
lead to \emph{half} line-like graphs. Our results on localization
extend to \emph{unrooted} trees leading to \emph{full} line-like
graphs. We do not present the details, but we illustrate the result in
the case of a line-like graph, i.e., the branching number is $b=1$, and
the necklace decoration of \Exenum{graph.deco}{necklace} (see also
\Figenum{deco-graph}{necklace}). The random necklace model was
originally treated by Kostrykin and Schrader in
\cite{kostrykin-schrader:04} where the authors showed discontinuity of
the integrated density of states.  We complete this study by proving
Anderson localization for the random necklace model.

For $n \in \Z$, let $G_n(\omega)=G_*(\omega_n)$, be the necklace
decoration of \Exenum{graph.deco}{necklace} (with $p=2$ arcs of length
$\ell_n=\omega_n$ forming the loop). Let $L(\omega)$, $\omega \in
\Omega:=\Omega_1^\Z$, be the line-like graph obtained by joining the
decoration graph in a line unbounded in both directions. All the
results of the random half-line models extend to full-line models. The
spectrum of the \emph{periodic} full model on $L=L(\ell)$ (with
constant length $\ell=\ell_n$) is purely absolutely continuous and is
given by
\begin{equation}
  \label{eq:per.sp.necklace2}
    \spec{\laplacian{L(\ell)}} =
    \bigset{\mu^2}
       { |\cos(\mu \ell) \cos \mu -
          {\textstyle \frac 54} \sin(\mu \ell) \sin \mu| \le 1} =
    \bigcup_{k=1}^\infty B_k(\ell)
\end{equation}
(cf.~\eqref{eq:per.sp.necklace}).

Our result on localization in this situation reads as follows:
\begin{theorem}
  \label{thm:loc.necklace2}
  Assume that the single site perturbation of the full-line necklace
  model (with $p=2$ loop edges) of \Exenum{graph.deco}{necklace} has
  an absolutely continuous distribution with bounded density $\dd
  \Prob_1(\ell)= \eta(\ell) \dd \ell$ and support in $\Omega_1 :=
  [0,\ell_+]$.  Then localization holds for all energies in the almost
  sure spectrum $\Sigma=\bigcup_{\ell \in \Omega_1}
  \spec{\laplacian{L(\ell)}}$ with eigenfunctions having almost
  exponential decay rate $\gamma_\pm(\lambda)$ for $x \to \pm \infty$
  in the sense of~\eqref{eq:exp.decay3}.
\end{theorem}
\begin{proof}
  The proof in the full-line model (cf.~\cite{kotani-simon:87}) is
  similar to the proof of the half-line model, so we give only a
  sketch of the proof: We have already seen that the Lyapunov exponent
  for $n \to +\infty$ is positive; for $n \ge 0$ the transfer matrix
  from generation $0$ to $-n$ is $U_\lambda(\omega,-n) =
  T_\lambda(\omega_{n-1})^{-1}\cdot \ldots \cdot
  T_\lambda(\omega_0)^{-1}$, and the same argument as for $n \to
  \infty$ shows that the Lyapunov exponent is positive also for $n \to
  -\infty$.  As before, from the Oseledec theorem it follows that
  there exist generalized eigenfunctions $f_\pm$ on the positive,
  respective, negative, half-line model decaying exponentially where
  $(f_\pm,f_\pm^\dag)(0) \sim
  \theta_0^\pm=\theta_0^\pm(\omega,\lambda)$ for a set $S \subset
  \Omega \times \Sigma$ of full measure. Here, we have to show that
  $\theta_0^+=\theta_0^-$ in order to assure that $f_+$ and $cf_-$ are
  the restrictions of an eigenfunction $f$ in the domain of
  $H(\omega)$ for a suitable constant $c \in \C$.

  From the spectral averaging argument, we see that $S(\omega) \subset
  \Sigma$ is a support of the spectral measure (component) $\hat
  \rho_\omega$. Note that we need the estimate~\eqref{eq:weyl.bdd} for
  $T_z(\omega_1)$ (see the proof of \Thm{loc.necklace}) \emph{and}
  $T_z(\omega_1)^{-1}$ (cf.~\Cor{sp.av.rlm}). The spectral measure is
  also supported on the set of eigenvalues $\lambda$ with polynomially
  bounded (generalized) eigenfunctions $\phi$. The Wronskian of $\phi$
  and $f_+$, respectively, $f_-$, is constant, and $0$ in the limit $n \to 
\pm
  \infty$, so that $f_\pm$ and $\phi$ satisfy the same condition at
  $0$, namely $\theta_0^+=\theta_0^-$.
\end{proof}

%----------------------------------------------------------------------
%
% Appendix
%
%----------------------------------------------------------------------

% We want the Appendix numbers A.1 etc.
\setcounter{section}{0}
\renewcommand{\thesection}{\Alph{section}}

%----------------------------------------------------------------------
%
\section{Symmetry reduction}
\label{app:reduction}
%----------------------------------------------------------------------

\sloppy
For radial tree-like graphs $G=(V,E,\bd, \ell, q)$ associated to a
tree graph $T=(V(T),E(T),\bd)$, we can profit from the symmetric
structure of $G$ (for a definition of a radial tree-like graph we
refer to \Def{tree-like.mg}). The argument used here follows closely
the symmetry reduction for the simple tree graph $T$
(cf.~\cite{naimark-solomyak:00,sobolev-solomyak:02,
  solomyak:04}).

On a rooted tree, we can define a partial order $\succeq$ on the set
of vertices and edges as follows: If a vertex $v \in V(T)$ lies on the
shortest path from $o$ to $v' \in V(T)$ we say that $v'$
\emph{succeeds} $v$ ($v' \succeq v$).  Similarly, an edge $t \in E(T)$
succeeds $v$ iff its start vertex succeeds $v$, i.e., $\bd_-t \succeq
v$. The \emph{vertex subtree} $T_{\succeq v}$ succeeding $v \in V$ is
the graph of all edges and vertices succeeding $v$.  The \emph{edge
  subtree} $T_{\succeq t}$ is the subgraph of all edges and vertices
succeeding $\bd_+t$ together with the root $\bd_-t$ of the subtree. In
particular, a vertex subtree $T_{\succeq v}$ is the union of all edge
subtrees $T_{\succeq t}$ with $v=\bd_-t$.

Similarly, let $G_{\succeq v}$, respectively, $G_{\succeq t}$, be the
\emph{vertex}, respectively, \emph{edge}, subgraph of the tree-like graph 
$G$
corresponding to the underlying tree subgraph $T_{\succeq v}$,
respectively, $T_{\succeq t}$, i.e., $G_{\succeq v}$,
respectively, $G_{\succeq t}$, consists
of all decoration graphs $G_*(t')$ with $t' \in E(T_{\succeq v})$,
respectively, $t' \in E(T_{\succeq t})$.  We can associate a line-like graph
$L_n$ to the radial tree-like graph $G_{\succeq v}$, $\gen v=n$, as in
\Sec{reduct.gen}, consisting of the sequence $\{G_k\}_{k>n}$ of
decoration graphs with branching number sequence $\{b_k\}_{k \ge n}$

Let $b=b_n$ be the branching number of $v \in V(T)$. The cyclic group
$\Z_b$ acts on the vertex subgraph $G_{\succeq v}$ by shifting the $b$
succeeding edge subgraphs $G_{\succeq t}$, $\bd_-t=v$, in a cyclic
way.  The group action on $G_{\succeq v}$ lifts naturally to an
unitary action on $\Lsqr {G_{\succeq v}}$. We denote the action of $1
\in \Z_b$ by $Q_v$. Since $1$ generates $\Z_b$, the operator $Q_b$
also generates the action on $\Lsqr {G_{\succeq v}}$. Furthermore,
$Q_v^b=\1$ and the eigenvalues of $Q_v$ are the $b$th unit roots
$e^s_b$ of $1$, $s=0, \dots, b-1$. The corresponding eigenspaces are
denoted by
\begin{equation*}
  \Lsqr[s] {G_{\succeq v}}:= \ker (Q_v^b - e^s_b \1).
\end{equation*}
\begin{definition}
  A function $f \in \Lsqr {G_{\succeq v}}$ is called \emph{$s$-radial
    at the tree vertex $v \in V(T)$} iff $f \in \Lsqr[s] {G_{\succeq
      v}}$ and if $f \in \Lsqr[0]{G_{\succeq v'}}$ for all succeeding
  tree vertices $v' \succ v$. The set of all $s$-radial functions is
  denoted by $\Lsqr[s,\rad] {G_{\succeq v}}$. A $0$-radial function is
  simply called \emph{radial}.
\end{definition}
In other words, a function $f \in \Lsqr{G_{\succeq v}}$ is $s$-radial
iff $Q_v f= e_b^s f$ for $b=b_n$, $n=\gen v$, and if $f$ is
invariant under the group action $Q_{v'}$ on the subsequent subgraph
$G_{\succeq v'}$ for all $v' \succ v$.  Clearly, such a function is
completely determined by its value on the line-like graph $L_n$.
We therefore define
\begin{equation}
  \label{eq:def.jt}
  \map {J_v^s} {\Lsqr[s,\rad]{G_{\succeq v}}} {\Lsqr{L_n}} \qquad
  J_v^s f :=
     \bigoplus_{k > n} (b_n \cdot \ldots \cdot b_{k-1})^{1/2}
     f \restr {G_k}.
\end{equation}

\begin{lemma}
  The operator $J_v^s$ is unitary.
\end{lemma}
\begin{proof}
  We have
  \begin{equation*}
    \normsqr[L_n] {J_v^s f} =
    \sum_{k > n} (b_n \cdot \ldots \cdot b_{k-1}) \normsqr[G_k] f
    = \normsqr[G_{\succeq v}] f
  \end{equation*}
  since $b_n \cdot \ldots \cdot b_{k-1}$ is the total number of copies
  of $G_k$ contained in $G_{\succeq v}$ at generation $k$ and since
  for $f \in \Lsqr[s,\rad]{G_{\succeq v}}$ the value $\norm[G_k] f$ is
  independent of the choice of the decoration graphs $G_*$ at
  generation $k$. Finally, it can easily be seen that $J_v^s$ is
  surjective.
\end{proof}

\begin{lemma}
  \label{lem:decomp}
  We have the decomposition
  \begin{equation}
    \label{eq:decomp}
    \Lsqr G = \Lsqr[0,\rad] {G_{\succeq o}} \oplus
     \bigoplus_{n=1}^\infty \bigoplus_{\substack{v \in V(T)\\ \gen v=n}}
     \bigoplus_{s=1}^{b_n-1}   \Lsqr[s,\rad]{G_{\succeq v}}.
  \end{equation}
\end{lemma}
\begin{proof}
  Since $b_0=1$ we can split off the first decoration graph $G_1$ and
  consider only $\Lsqr{G_{\succeq o_1}}$ where $o_1$ denotes the
  vertex of generation $1$. Note that $\Lsqr[0,\rad]{G_{\succeq
      o}}=\Lsqr {G_1} \oplus \Lsqr[0,\rad]{G_{\succeq o_1}}$.  {}From
  the eigenspace decomposition of $Q_{o_1}$ we obtain
  \begin{equation*}
    \Lsqr {G_{\succeq o_1}} =
    \bigoplus_{s=0}^{b_1-1}   \Lsqr[s]{G_{\succeq o_1}}
  \end{equation*}
  since $G=G_1 \cup G_{\succeq o_1}$ where $G_1$ is the decoration
  graph at generation $1$. Next, we have
  \begin{equation}
    \label{eq:decomp1}
    \Lsqr[s] {G_{\succeq o_1}} =
    \Lsqr[s,\rad]{G_{\succeq o_1}} \oplus
       \bigoplus_{\substack{v \in V(T)\\ \gen t=2}}
          \Bigl( \Lsqr {G_{\succeq v}} \ominus
              \Lsqr[0,\rad]{G_{\succeq v}} \Bigr)
  \end{equation}
  since functions in $\Lsqr[s] {G_{\succeq o_1}} \ominus \Lsqr[s,\rad]
  {G_{\succeq o_1}}$ vanish on the decoration graphs $G_*(t)$ of
  generation $\gen t = 2$. In addition, the radial component of
  functions on the subtrees $G_{\succeq v}$ is already contained in
  $\Lsqr[s,\rad]{G_{\succeq o_1}}$, therefore, $\Lsqr {G_{\succeq v}}
  \ominus \Lsqr[0,\rad]{G_{\succeq v}} = \bigoplus_{s=1}^{b_2-1}
  \Lsqr[s]{G_{\succeq v}}$ so that
  \begin{equation*}
    \Lsqr G =
    \Lsqr[0,\rad]{G_{\succeq o}} \oplus
    \bigoplus_{s=1}^{b_1-1} \Lsqr[s,\rad]{G_{\succeq o_1}} \oplus
       \bigoplus_{\substack{v \in V(T)\\ \gen v=2}}
     \bigoplus_{s=1}^{b_2-1} \Lsqr[s]{G_{\succeq v}}.
  \end{equation*}
  Now we can decompose $\Lsqr[s] {G_{\succeq v}}$ as
  in~\eqref{eq:decomp1} and obtain the desired
  formula~\eqref{eq:decomp} recursively. It remains to show that a
  function $f$ orthogonal to the direct sum in the right hand side
  of~\eqref{eq:decomp} vanishes. Clearly, such a function vanishes on
  the first decoration graph $G_1$. In addition, such a function must
  also vanish on the decoration graphs $G_*(2):= \bigcup_{t \in E(T),
    \gen t=2} G_*(t)$ of generation $2$ since $\Lsqr[s,\rad]
  {G_{\succeq o_1}} \cap \Lsqr{G_*(2)} = \Lsqr[s] {G_{\succeq o_1}}
  \cap \Lsqr{G_*(2)}$.  The same arguments holds for any subgraph
  $G_{\succeq v}$ so that $f$ vanishes on all decoration graphs
  $G_*(t)$, i.e., $f=0$.
\end{proof}

We assume that the decoration graphs $G_*(t)$ satisfy the uniformity
assumptions~\eqref{eq:ass.graphs}. We then define a quadratic form on
the subgraph $G_{\succeq v}$ with domain
\begin{equation}
    \label{eq:dom.qf.ll.gr}
    \dom \qf h_{G_{\succeq v}} =
    \bigset{f \in \bigoplus_{t \in E(T_{\succeq v})} \Sob{G_*(t)}}
    {f \in \Contn {G_{\succeq v}}}
\end{equation}
where $\Contn {G_{\succeq v}}$ denotes the space of continuous on
$G_{\succeq v}$ vanishing at the root vertex $v$ of $G_{\succeq v}$.
The quadratic form is defined as
\begin{equation}
    \label{eq:qf.ll.gr}
    \qf h_{G_{\succeq v}} (f) =
    \sum_{t \in E(T_{\succeq v})}
       \bigl( \normsqr[G_*(t)]{f'} + q(\bd_+t) |f(t)|^2 \bigr).
\end{equation}
As in \Lem{form.bdd} it follows that $\qf h_{G_{\succeq v}}$ is a
closed quadratic form and relatively form-bounded w.r.t.\ the free
form $\qf d_{G_{\succeq v}}$ (where $q(v)=0$) with relative bound $0$.
We denote the self-adjoint operator associated to $\qf h_{G_{\succeq
    v}}$ by $H_{G_{\succeq v}}$. We now show that the orthogonal
composition of the previous lemma also decomposes $H=H_G$:
\begin{lemma}
  \label{lem:decomp.quad}
  The components of the decomposition~\eqref{eq:decomp} are invariant
  subspaces of the Hamiltonian $H$ on $G$.
\end{lemma}
\begin{proof}
  We want to show that the domain of $\qf h=\qf h_G$ decomposes into
  \begin{equation*}
    \dom \qf h =
    \dom \qf h^{0,\rad}_o \oplus
    \bigoplus_{n=1}^\infty \bigoplus_{\substack{v \in V(T)\\ \gen v=n}}
    \bigoplus_{s=1}^{b_n-1}   \dom \qf h^{s,\rad}_v
  \end{equation*}
  where $\dom \qf h^{s,\rad}_v$ is the space of all functions $f_v^s \in
  \Lsqr[s,\rad] {G_{\succeq v}} \cap \Contn {G_{\succeq v}}$ such that
  \begin{equation}
    \label{eq:def.qf.fin}
    \qf h_{G_{\succeq v}}(f_v^s) < \infty.
  \end{equation}
  In order to do so, we have to verify that if $f=\{f_v^s\}$ is the
  orthogonal decomposition of $f \in \dom \qf h$
  w.r.t.~\eqref{eq:decomp} then $f_v^s \in \dom \qf h^{s,\rad}_v$: By
  definition, $f_v^s \in \Lsqr[s,\rad]{G_{\succeq v}}$ and $f_v^s \in
  \Contn {G_{\succeq v}}$ follows from the continuity of $f \in \dom
  \qf h$ and the fact that $f_v^s$ vanishes on $\compl{(G_{\succeq
      v})}$. Furthermore, if $f_v^s \in \dom \qf h_{G_{\succeq v}}$
  then $Q_v f_v^s \in \dom \qf h_{G_{\succeq v}}$ and $\qf
  h_{G_{\succeq v}}(Q_v f_v^s)=h_{G_{\succeq v}}(f_v^s)$, i.e., $Q_v$
  leaves $\qf h_{G_{\succeq v}}$ invariant; in
  particular,~\eqref{eq:def.qf.fin} is fulfilled.
\end{proof}
Note that functions vanishing outside a subgraph $G_{\succeq v}$ must
satisfy a Dirichlet condition at the root vertex $v$ since functions
in $\dom \qf h$ are continuous.

We now want to compare the radial quadratic form $\qf h^{s,\rad}_v$
with a form on the line-like graph $L_n$, $n=\gen v$.
\begin{lemma}
  \label{lem:qf.rad}
  The quadratic form $\qf h^{s,\rad}_v$ is unitary equivalent to $\qf
  h_{L_n}$ where
  \begin{equation}
    \label{eq:qf.ll2}
    \qf h_{L_n}(g) :=
    \sum_{k>n} \bigl( \normsqr[G_k]{g'} + q_k |f(k_-)|^2\bigr)
  \end{equation}
  and
  \begin{equation}
    \label{eq:qf.ll2.dom}
    \dom \qf h_{L_n} =
    \set{ f \in \bigoplus_{k>n} \Sob{G_k}}
           {\forall k > n: \, f(k_-)=b_k^{-1/2}f(k_+), \quad f(0)=0},
  \end{equation}
  i.e.,
  \begin{equation*}
   J^s_v(\dom \qf h^{s,\rad}_v)=\dom \qf h_{L_n} \qquad \text{and} \qquad
   \qf h^{s,\rad}_v(f)=\qf h_{L_n}(J^s_v f).
\end{equation*}
\end{lemma}
\begin{proof}
  The proof follows from a straightforward calculation.
\end{proof}

Summarizing the results, we obtain (denoting $H_n=H_{L_{n-1}}$ the
operator on $L_{n-1}$ associated to the quadratic form $\qf
h_{L_{n-1}}$):
\begin{theorem}
  \label{thm:reduction}
  The Hamiltonian $H$ on a symmetric, radial tree-like quantum graph
  $G$ is unitary equivalent to
  \begin{equation*}
    H \cong  H_1 \oplus
    \bigoplus_{n=2}^\infty
        (\oplus b_0 \cdot \ldots \cdot b_{n-2}(b_{n-1}-1)) H_n
  \end{equation*}
  where $(\oplus m) H_n$ means the $m$-fold copy of $H_n$.
\end{theorem}
The domain of $H_n=H_{L_{n-1}}$ is given in~\Lem{ess.sa}.

%----------------------------------------------------------------------
%
\section{Bounds on generalized eigenfunctions}
\label{app:bd.ef}
%----------------------------------------------------------------------

In this section we provide $\Lsymb_2$-bounds on generalized
eigenfunctions on quantum graphs.  We start with a slightly more
general setting. Let $X$ be a metric, $\sigma$-finite measure space
with measure $\mu$. We usually denote the measure by $\dd x = \dd
\mu(x)$. A slightly different setting using only Hilbert space
arguments can be found in~\cite{psw:89,poerschke-stolz:93}.
\begin{assumption}
  \label{ass:op}
  Let $H$ be a semibounded, self-adjoint operator $H \ge \lambda_0$ in
  $\Lsqr X=\Lsqr{X,\dd\mu}$.  We assume that $H$ is \emph{local},
  i.e., $\supp H f \subset \supp f$ for any $f \in \dom H$. In
  addition, we assume that the space of functions $f \in \dom H$ with
  compact support are dense in $\Lsqr X$.  Denote
  $K:=(H-\lambda_0+1)^{-m/2}$ the $\frac m2$-th power of the resolvent at
  $\lambda_0 \in \R$ ($m>0$).  Our main assumption in this section
  assures that $K$ is a Carleman operator, i.e.,
  \begin{equation*}
    \map K {\Lsqr X} {\Linfty X}
  \end{equation*}
  is bounded, or, equivalently,
  \begin{equation}
    \label{eq:carleman}
    \norm[\Linfty X] f \le
    C_1 \norm[\Lsqr X] {(H-\lambda_0+1)^{m/2}f}.
  \end{equation}
\end{assumption}
We will prove in the next section that on a quantum graph or a line
graph, the above conditions are met with $m=1$ for the graph
Hamiltonian under suitable conditions on the model.

Carleman operators
have a measurable kernel $\map k {X \times X} \C$
(cf.~\cite[Cor.~A.1.2]{simon:82}), i.e.,
\begin{equation*}
  (K f)(x) = \int_X k(x,y) f(y) \dd y
\end{equation*}
satisfying
\begin{equation*}
  \norm[\Lsymb_2 \to \Lsymb_\infty] K =
  \sup_{x \in X} \norm[\Lsqr X]{k(x,\cdot)} \le C_1 < \infty.
\end{equation*}

We will show that $K$ and certain other functions of $H$ have an
integral kernel also in a \emph{weighted} $\Lsymb_2$-space.
\begin{assumption}
\label{ass:weight}
Let $\Phi \in \Lsqr X \cap \Linfty X$ be a bounded, square-integrable
and \emph{positive} function such that $\Phi$ is bounded away from $0$
on any compact set.
\end{assumption}
To $\Phi$ we associate the Hilbert scaling
(cf.~\cite{berezin-shubin:91})
\begin{equation*}
  \HS_+ := \Lsqr {X,\dd\mu_+} \hookrightarrow
  \HS   := \Lsqr {X,\dd\mu} \hookrightarrow
  \HS_- := \Lsqr {X,\dd\mu_-}
\end{equation*}
where $\dd\mu_\pm := \Phi^{\mp 2} \dd \mu$ are weighted measures and
$\HS_\pm$ are normed by $\norm[\pm] f := \norm{\Phi^{\mp 1} f}$. Then
the inner product $\map{\iprod \cdot \cdot} {\HS \times \HS} \C$
extends to a dual (sesquilinear) pairing $\map {(\cdot,\cdot)} {\HS_-
  \times \HS_+} \C$. In particular, $\HS_-$ can be interpreted as the
dual $(\HS_+)^*$ with respect to this pairing. In addition, the
multiplication with $\Phi$, respectively, $\Phi^{-1}$, becomes an isometry,
i.e.,
\begin{equation*}
  \HS_+
  \begin{array}{c}
   \scriptstyle {\Phi^{-1}}\\[-1ex]
   \rightleftarrows\\[-1ex]
   \scriptstyle \Phi
  \end{array}
  \HS
  \begin{array}{c}
   \scriptstyle {\Phi^{-1}}\\[-1ex]
   \rightleftarrows\\[-1ex]
   \scriptstyle \Phi
  \end{array}
  \HS_-
\end{equation*}
and $(f,g)=\iprod{\Phi f}{\Phi^{-1} g}$ for $f \in \HS_-$, $g\in
\HS_+$.  Since $\Phi$ is bounded away from $0$ on any compact set, the
norms in $\HS$ and $\HS_\pm$ are equivalent for functions with support
in fixed compact subset of $X$.

Our aim is to show that $T:=\Phi K^2 \Phi$ or more generally,
$T_\phi:=\Phi \phi(H)^2 \Phi$ for fast enough decaying functions
$\phi$ (in particular, $\phi=\1_I$, $I$ bounded) are of trace class as
operators from $\HS$ to $\HS$ and have an integral kernel $t_\phi$.
Our Hilbert scaling allows us to consider $\wt T_\phi := \phi(H)^2$ as
map $\HS_+ \to \HS_-$. It is still of trace class as product of the
Hilbert Schmidt operators $\map{\phi(H)}{\HS_+} \HS$ and
$\map{\phi(H)}\HS {\HS_-}$ (cf.~the lemma below) and has integral
kernel $\wt t_\phi(x,y)=\Phi(x)^{-1} t_\phi(x,y) \Phi(y)^{-1}$ with
respect to the pairing $(\cdot,\cdot)$, i.e.,
\begin{equation*}
  (\wt T_\phi f, g) =
  \int_X \int_X \wt t_\phi(x,y) \conj{f(x)} g(x) \dd x \dd y
\end{equation*}
for $f,g \in \HS_+$. In a second step, apply the above considerations
to $\phi=\1_I$ and disintegrate the spectral resolution $\1_I(H)$ with
respect to a spectral measure of $H$.

We start with showing that $\Phi K$ is a Hilbert-Schmidt operator
(cf.~\cite[Prop.~II.3.11]{carmona-lacroix:90}):
\begin{lemma}
  \label{lem:hs}
  Suppose that $\Phi \in \Lsqr X \cap \Linfty X$ and
  that~\eqref{eq:carleman} is fulfilled. Then for any measurable
  function $\map \phi \R \R$ such that
  \begin{equation}
    \label{eq:hs}
    |\phi(\lambda)| \le C_2 (\lambda - \lambda_0 + 1)^{-m/2}
  \end{equation}
  for all $\lambda \ge \lambda_0$ the operator $\Phi \phi(H)$ is
  Hilbert-Schmidt with Hilbert-Schmidt norm bounded by
  $C_3:=\norm[\Lsqr X] \Phi C_1 C_2$. In addition,
  \begin{equation*}
    \map{\Phi \phi(H)^2 \Phi} \HS \HS
  \end{equation*}
  is trace class with trace bounded by $C_3^2$.
\end{lemma}
\begin{proof}
  The kernel of $\Phi K$ is $(x,y) \mapsto \Phi(x) k(x,y)$.
  By~\eqref{eq:carleman}, its $\Lsqr{X \times X}$-norm is bounded by
  $\norm[\Lsqr X] \Phi C_1$ so that $\Phi K$ is Hilbert-Schmidt. In
  particular, $T = \Phi K (\Phi K)^*$ is trace class and
  \begin{equation*}
    \norm[\TRsymb]{\Phi K^2 \Phi} =
    \tr (\Phi K^2 \Phi) \le
    \normsqr[\HSsymb]{\Phi K} \le
    \normsqr[\Lsqr X] \Phi C_1^2.
  \end{equation*}
  To pass to a general function $\phi$, note that $\phi(H)^2 \le C_2^2
  K^2$ by the spectral theorem. The result follows from the
  monotonicity of the trace and $0 \le \Phi \phi(H)^2 \Phi \le C_2^2
  \Phi K^2 \Phi$.
\end{proof}

\sloppy In particular, the above lemma applies for the characteristic
function $\1_I$ of a bounded, measurable set $I \subset \R$ with $C_2
:= (\sup I - \lambda_0+1)^{m/2}$. Therefore, one can show that $E(I):=
\Phi \1_I(H) \Phi$ defines a nonnegative, trace-class-operator-valued,
strongly $\sigma$-additive measure, i.e, (i)~$E(I) \ge 0$, (ii)~$E(I)$
is trace class for all bounded and measurable $I \subset \R$ and
(iii)~$E(\bigdcup I_n) = \slim \sum_n E(I_n)$.

We use the following lemma to disintegrate $E(I)$
(cf.~\cite[Thm.~C.5.1]{simon:82}:%[Prob.~I.7.14]{carmona-lacroix:90}):
\begin{lemma}
  \label{lem:desint}
  Suppose $E(\cdot)$ is a nonnegative, trace-class-operator-valued,
  strongly $\sigma$-additive measure. Then there exists a Borel
  measure $\rho$ (i.e., a measure on the Borel sets of $\R$, finite on
  all compact sets) and a measurable function $\map E \R
  {\TRsymb(\HS)}$ such that $E(\lambda) \ge 0$,
  \begin{equation*}
    E(I) = \wint_I E(\lambda) \dd \lambda \qquad\text{and}\qquad
    \tr E(\lambda) = 1 \quad \text{$\rho$-a.e.}
  \end{equation*}
\end{lemma}
\begin{proof}
  Set $\rho(I):= \tr E(I)$ and $\rho_{ij}(I):=\iprod
  {\phi_i}{E(I)\phi_j}$ for an orthonormal basis $\{\phi_i\}_i$ of
  $\HS$. Clearly, $\rho$ is a Borel measure, as well as $\rho_{ij}$
  are $\C$-valued Borel measures. Furthermore, $(\rho_{ij}(I))_{i,j
    \in J}$ is a nonnegative matrix for any finite subset $J \subset
  \N$. In addition, $\rho_{ij}$ is absolutely continuous w.r.t.\
  $\rho$, so by the Radon-Nikodym theorem there exists a measurable
  function $e_{ij}$ such that $\dd \rho_{ij}(\lambda)=e_{ij}(\lambda)
  \dd \rho(\lambda)$ for all $i,j$. Using the fact that
  $\rho(I)=\sum_i\rho_{ii}(I)$ one sees that $\sum_i e_{ii}(\lambda) =
  1$ a.e.

  Define $E(\lambda)$ as the operator with associated matrix
  $(e_{ij}(\lambda))_{ij}$ in the basis $\{\phi_i\}_i$. Clearly,
  $E(\lambda)$ has trace $1$ and a limit argument\hiddenfootnote{Why are we
    allowed to interchange the sum and the integral? Monotone
    convergence? Do we have strong convergence in the integral?} shows
  that
  \begin{equation*}
    \iprod {E(I) f} g =
    \sum_{ij} \conj {f_i} \, g_j \, \rho_{ij}(I) =
    \int \sum_{ij} \conj {f_i} \, g_j \, e_{ij}(\lambda) \dd
        \rho(\lambda) =
    \int \sum_{ij} \iprod {E(\lambda) f} g \dd \rho(\lambda)
  \end{equation*}
  where $f_i=\iprod {\phi_i} f$ and $g_j=\iprod {\phi_j} g$.
\end{proof}

As a corollary, we obtain
\begin{corollary}
  \label{cor:desint}
  The measure $\rho(I):= \tr E(I)$ associated to the
  trace-class-operator-valued, strongly $\sigma$-additive measure
  $E(I):=\Phi \1_I(H) \Phi$ is a spectral measure for $H$, i.e.,
  $\rho(I)=0$ iff $1_I(H)=0$.  If $\phi$ satisfies~\eqref{eq:hs} then
  $\int_\R |\phi(\lambda)|^2 \dd \rho(\lambda)<\infty$.

  Furthermore, the disintegrated operator $\map{E(\lambda)} \HS \HS$
  is also Hilbert-Schmidt and has a kernel $e_\lambda \in \Lsqr{X
    \times X}$ and $\norm[\Lsqr{X \times X}] {e_\lambda} \le 1$.  In
  addition,
  \begin{equation*}
    \map{\wt E(\lambda):= \Phi^{-1} E(\lambda) \Phi^{-1}}
       {\HS_+} {\HS_-}
  \end{equation*}
  has the kernel
  \begin{equation*}
    \wt e_\lambda(x,y) := \Phi(x)^{-1} e_\lambda(x,y) \Phi(y)^{-1}
  \end{equation*}
  and allows the disintegration formula
  \begin{equation*}
    ( \1_I(H) f,g ) =
    \int_I ( \wt E(\lambda) f, g) \dd \rho (\lambda) =
    \int_I \int_X \int_X \wt e_\lambda(x,y) \conj{f(x)}\, g(y)
          \dd x \dd y \dd \rho(\lambda)
  \end{equation*}
  for all $f,g \in \HS_+$ with $\wt e_\lambda \in \HS_- \otimes \HS_-$
  and $\norm[\HS_- \otimes \HS_-]{\wt e_\lambda} \le 1$.
\end{corollary}
\begin{proof}
  Since $\ker \Phi = \{0\}$ and $\Phi^*=\Phi$ as multiplication
  operator, $\rho(I)=0$ implies $\Phi \1_I(H) \Phi=0$ and therefore
  $\1_I(H)=0$. In particular, $\rho$ is a spectral measure. The fact
  that $\phi \in \Lsqr{\R,\dd \rho}$ follows from \Lem{hs}, the
  definition of $\rho$ and the spectral calculus.

  The remaining assertions are almost obvious: Trace class operators
  are also Hilbert-Schmidt and
  \begin{equation*}
    \norm[\Lsqr {X \times X}] {e_\lambda } =
    \norm[\HSsymb] {E(\lambda)} \le
    \norm[\TRsymb] {E(\lambda)} =
    \tr E(\lambda) = 1
  \end{equation*}
  a.e.
\end{proof}

Under the same assumptions as before, we can now pass to more general
functions of $H$ using a standard approximation argument.
\begin{lemma}
  \label{lem:desint.fct}
  Let $\map \phi \R \C$ be bounded and measurable. Then
  \begin{multline*}
    ( \phi(H) f,g ) =
    \int_\R \phi(\lambda) ( \wt E(\lambda) f, g) \dd \rho (\lambda) \\=
    \int_\R \int_X \int_X \phi(\lambda)\wt e_\lambda(x,y) \conj{f(x)}\, g(y)
          \dd x \dd y \dd \rho(\lambda)
  \end{multline*}
  for all $f,g \in \HS_+$.
\end{lemma}

We now want to show that $\wt e_\lambda(\cdot , y)$ solves the
eigenvalue equation $(H-\lambda) u = 0$ in a generalized sense. To do
so, set
\begin{equation*}
  \dom H_+ := \set{ f \in \HS_+} {f \in \dom H, \quad Hf \in \HS_+},
  \qquad
  H_+ f:= Hf.
\end{equation*}
The operator $H_+$ is a closed operator in $\HS_+$. Since the space of
functions $f \in \dom H$ with compact support is dense in $\HS$ and
since the norms on $\HS$ and $\HS_+$ are equivalent on a fixed compact
set, $\dom H_+$ is dense in $\HS_+$. Therefore, we can define the
adjoint $H_-:=(H_+)^*$ w.r.t.\ $(\cdot, \cdot)$, i.e., $f \in \dom
\HS_-$ iff $f \in \HS_-$ and if there exist $H_-f \in \HS_-$ such that
\begin{equation*}
  (H_- f, g) = (f, H_+ g)
\end{equation*}
for all $g \in \dom H_+$.
\begin{definition}
  \label{def:gen.ef.weight}
  A function $u \in \HS_-$ is a \emph{generalized eigenfunction} of
  $H$ with $\Lsymb_2$-growth rate $\Phi^{-2}$ for the eigenvalue
  $\lambda$ , if $H_- u = \lambda u$, i.e.,
  \begin{equation*}
    (u, (H_+-\lambda) g) = 0
  \end{equation*}
  for all $g \in \dom H_+$.
\end{definition}

The next lemma assures that the integral kernel is a generalized
eigenfunction (cf.~\cite[Thm.~C.5.2]{simon:82}):
\begin{lemma}
  \label{lem:gen.ef}
  We have $\wt e_\lambda(\cdot, y) \in \dom H_-$ and
  \begin{equation*}
    h(y):=(\wt e_\lambda (\cdot, y), (H_+-\lambda) g) = 0
  \end{equation*}
  for all $g \in \dom H_+$, $\mu$-almost all $y \in X$ and
  $\rho$-almost all $\lambda \in \R$.
\end{lemma}
\begin{proof}
  Since $\wt e_\lambda \in \HS_- \otimes \HS_-$, we have $h \in \HS_-$
  and the above equation is equivalent to
  \begin{equation}
    \label{eq:gen.ef}
    \int_X \int_X \wt e_\lambda(x,y) \conj{f(x)}\, (H_+ - \lambda)g(y)
        \dd x \dd y =
   ( \wt E(\lambda) f, (H_+-\lambda)g) = 0
  \end{equation}
  for all $f \in \HS_+$ and $g \in \dom H_+$ using the kernel
  representation of $\wt E(\lambda)$ in \Cor{desint} and the fact that
  $H_+ g$, $g \in \HS_+$. Now, we define a signed measure $\wt \rho(I)
  = \int_I(\wt E(\lambda') f, (H_+-\lambda) g) \dd \rho(\lambda')$ and
  obtain\hiddenfootnote{Do we have the following expression for the
    density? Let $\nu$ be a.c.  w.r.t. $\mu$ ($\mu$ a $\C$-valued,
    $\nu$ a nonnegative measure on $\R$). Then $\dd
    \nu(\lambda)=f(\lambda)\dd \mu(\lambda)$ and
    \begin{equation}
      \label{eq:rad.nik}
      f(\lambda)= \lim_{I \searrow \{\lambda\}} \frac {\nu(I)}{\mu(I)}
  \end{equation}
  with the convention $0/0 := 0$. Yes, this is the definition of the
  Radon-Nikodym derivative!}
  \begin{equation*}
   ( \wt E(\lambda) f, (H_+-\lambda)g) =
   \lim_{I \searrow \{\lambda\}} \frac {\wt \rho(I)} {\rho(I)} =
   \lim_{I \searrow \{\lambda\}} \frac 1 {\rho(I)}
   \bigl(\1_I(H) f, (H_+ - \lambda) g\bigr)
  \end{equation*}
  using the Radon-Nikodym derivative and \Cor{desint}. The left hand side
  equals
\begin{equation*}
   \lim_{I \searrow \{\lambda\}} \frac 1 {\rho(I)}
   \bigiprod {\1_I(H) f} {(H - \lambda) g} =
   \lim_{I \searrow \{\lambda\}} \frac 1 {\rho(I)}
   \bigiprod {\phi_\lambda(H) f} g
  \end{equation*}
  since $(\cdot,\cdot)=\iprod \cdot \cdot$ on $\HS$ and by the
  functional calculus with
  $\phi_\lambda(\lambda')=(\lambda'-\lambda)\1_I(\lambda')$. Finally,
  the latter term equals
\begin{equation*}
   \lim_{I \searrow \{\lambda\}} \frac 1 {\rho(I)}
   \int_\R \phi_\lambda(\lambda')
       \bigl(\wt E(\lambda') f, g \bigr) \dd \rho(\lambda') = 0
\end{equation*}
by \Lem{desint.fct} and the Radon-Nikodym derivative.
\end{proof}

We are now able to state our main result on the growth of generalized
eigenfunction (cf.~\cite[Thm.~C.5.4]{simon:82}):
\begin{theorem}
  \label{thm:bd.gen.ef}
  Suppose $H$ is an operator with spectral measure $\rho$ satisfying
  \Ass{op} and $\Phi$ is a weight function satisfying \Ass{weight}.
  Then there exist a measurable disjoint decomposition
  \begin{equation*}
    \spec  H = \bigdcup_{n \in \N \cup \{\infty\}} \Sigma_n
  \end{equation*}
  (up to sets of $\rho$-measure $0$), and for each $n \in \N \cup
  \{\infty\}$ and $j=1,\dots, n$ there exists a measurable function
  $\map{\phi_j}{\Sigma_n \times X} \C$ such that
  $\{\phi_{\lambda,j}:=\phi_j(\lambda, \cdot)\}_{1\le j \le n} \subset
  \dom H_-$ are linearly independent,
  \begin{equation*}
    \sum_{j=1}^n \normsqr[-] {\phi_{\lambda,j}}=
    \sum_{j=1}^n \normsqr {\Phi \phi_{\lambda,j}} = 1
  \end{equation*}
  and $(H_- - \lambda) \phi_{\lambda,j} = 0$ for $\rho$-almost all
  $\lambda \in \Sigma_n$.

  In addition,
  \begin{equation}
    \label{eq:supp.sp.meas}
    \Sigma:= \set {\lambda \in \R}
        { \exists \phi \in \dom H_-\setminus \{0\} :
              \quad H_- \phi = \lambda \phi}
  \end{equation}
  is a support for the measure $\rho$, i.e, $\rho(\compl \Sigma) = 0$.
\end{theorem}
\begin{proof}
  Let $\Sigma_n := \set{\lambda \in \spec H} {\dim \ran
    E(\lambda)=n}$. By \Cor{desint}, there exists a orthonormal system
  of eigenfunctions $(\wt \psi_{\lambda,j})_j$ of $E(\lambda)$ with
  nonnegative eigenvalues such that
  \begin{equation*}
    E(\lambda) \wt \psi_{\lambda,j} =
   \eps_{\lambda,j} \wt \psi_{\lambda,j}.
  \end{equation*}
  We set $\psi_{\lambda,j} := \sqrt \eps_{\lambda,j} \wt
  \psi_{\lambda,j}$. Then we have $e_\lambda(x,y) = \sum_j \conj
  {\psi_{\lambda,j}(x)} \psi_{\lambda,j}(y)$ for the kernel in the
  weak sense, i.e.,
  \begin{equation}
    \label{eq:weak.sum}
    \iprod {\conj f \otimes g} {e_\lambda} =
    \sum_j \iprod f {\psi_{\lambda,j}} \iprod {\psi_{\lambda,j}} g
  \end{equation}
  for $f,g \in \HS$.
  Now set $\phi_{\lambda,j} := \Phi^{-1} \psi_{\lambda,j} \in
  \HS_-$. Then
  \begin{equation*}
    \sum_j \normsqr[-]{\phi_{\lambda,j}} =
    \sum_j \normsqr{\psi_{\lambda,j}} =
    \sum_j \eps_{\lambda,j} =
    \norm[\TRsymb] {E(\lambda)} = 1.
  \end{equation*}
  Finally, from~\eqref{eq:gen.ef} we get
  \begin{align*}
    0 & =
    \int_X \int_X \wt e_\lambda(x,y) \conj{f(x)}\, (H_+ - \lambda)g(y)
        \dd x \dd y \\ &=
    \int_X \int_X e_\lambda(x,y) \Phi(x)^{-1}\conj{f(x)}\,
             \Phi(y)^{-1}(H_+ - \lambda)g(y) \dd x \dd y
        \\ &\stackrel{\text{\eqref{eq:weak.sum}}}=
    \sum_j \iprod {\Phi^{-1} f} {\psi_{\lambda,j}}
           \iprod {\psi_{\lambda,j}} {\Phi^{-1}(H_+-\lambda)g}
  \end{align*}
  for $f \in \HS_+$ and $g \in \dom H_+$. Setting $f =
  \phi_{\lambda,j_0}$ we obtain
  \begin{equation*}
    0 =  \bigiprod {\psi_{\lambda,j_0}} {\Phi^{-1}(H_+-\lambda)g} =
    \bigl( \phi_{\lambda,j_0}, (H_+-\lambda)g \bigr)
  \end{equation*}
  for all $g \in \dom H_+$ and $\lambda \in \spec H \setminus N_g$
  where $\rho(N_g)=0$.

  We have to show that we can choose a set of measure $0$
  \emph{independent} of $g$ (cf.~\cite[Proof
  of~Thm.~2.2~(b)]{poerschke-stolz:93}): This can be done since $H_+$
  is closed in $\HS_+$, i.e., $\dom H_+$ with its graph norm is a
  Hilbert space, and $\HS_+$ is separable ($\Lsqr{X,\dd\mu_+}$ is
  $\sigma$-finite!). It follows that $\dom H_+$ with its graph norm is
  separable (this is true for a self-adjoint operator; for a
  general operator, note that $H_+$ and $|H_+|$ define the same graph
  norm).

  Therefore, we can choose the union $N=\bigcup N_g$ for countable
  many $g$ and $N$ still has measure $0$. Therefore, $\phi_{\lambda,j}
  \in \HS_-$ and $(H_--\lambda) \phi_{\lambda,j} = 0$ for
  $\rho$-almost all $\lambda \in \R$. We have therefore shown that
  $\spec H$ is included in $\Sigma$ up to a set of $\rho$-measure $0$,
  and therefore, $\Sigma$ is a support for $\rho$.
\end{proof}
Dealing with one-dimensional problems, we easily get more information
on the eigenfunction expansion as a by-product of the previous
theorem:
\begin{lemma}
  \label{lem:ef.exp}
  Suppose that the vector space of generalized eigenfunctions in the
  sense of \Def{gen.ef.weight} is generated by compactly supported
  functions and a finite number of functions with infinite
  support. Then the weak eigenfunction expansion~\eqref{eq:weak.sum}
  holds pointwise almost everywhere, i.e.,
  \begin{equation}
    \label{eq:ptw.sum}
    \wt e_\lambda(x,y) =
    \sum_j \conj {\phi_{\lambda,j}(x)} \phi_{\lambda,j}(y)
  \end{equation}
  for $\mu$-almost all $x,y \in X$ and $\rho$-almost all $\lambda \in \R$.
\end{lemma}
\begin{proof}
  We can choose the orthogonal basis $\wt \psi_{\lambda.j}$ to have
  compact support except than a finite number of vectors. Then the
  weak sum~\eqref{eq:weak.sum} is indeed a locally finite sum, and
  therefore exists also pointwise.
\end{proof}

%----------------------------------------------------------------------
%
\section{Line-like graphs and bounds on generalized eigenfunctions}
\label{app:ll.gr.gen.ef}
%
%----------------------------------------------------------------------

In this section we specify the analysis done in the previous section
to line-like graphs. We will show that the assumptions made in the
previous sections are fulfilled. In particular, we get integral bounds
on generalized eigenfunctions. In the concrete situation here, we can
also prove pointwise estimates on generalized eigenfunctions
(\Sec{ptw.bd}) and a spectral resolution of the spectral projector
(\Sec{gen.ef}).

%----------------------------------------------------------------------
%
\subsection{Quadratic forms and operators on line-like graphs}
\label{sec:ll.ops}
%
%----------------------------------------------------------------------
In this section, we determine the operator domain of the reduced
Hamiltonian on a line-like graph $L$ and show that the operator is
essentially self-adjoined on compactly supported functions. This has
been shown for graphs with a global lower bound on \emph{all} length,
i.e., $\ell_e \ge \ell_->0$ for all $e \in E(L)$, and global bounds on
the boundary conditions, i.e., $1 \le b_n \le b_+$ and $q_- \le q_n
\le q_+$ for example in \cite{kuchment:04} (see also \cite{carlson:97}
for the case of tree graphs). Although we assume that the lengths of
the edges connecting the vertices $(n+1)_+$ and $n_-$ have a global
lower bound (and some other conditions, see~\eqref{eq:ass.graphs}), we
want to allow edges of arbitrary small size inside the decoration
graph.

For simplicity, we only consider the line-like graph $L=L_0$. Clearly,
all statements hold similarly for $L_n$.  We begin with some
Sobolev-type estimates which follow from our assumptions on the
decoration graphs in~\eqref{eq:ass.graphs}:
\begin{lemma}
  \label{lem:sob}
  Suppose that the decoration graph $G_*$
  satisfies~\eqref{eq:ass.graphs}. Then there exists $C_1'>0$ such
  that
  \begin{align}
    \label{eq:sob}
    |f(x)|^2 &\le \eps \normsqr[G_*]{f'} + \frac 4 \eps
    \normsqr[G_*] f,&
          & 0<\eps \le \ell_-,\, x \in G_*,\, f \in \Sob{G_*}\\
    \label{eq:sob'}
    |f^\dag(o_0)|^2 &\le C_1'\bigl ( \normsqr[G_*]{f''} + \normsqr[G_*]
    {f'} \bigr),&
          & f \in \Sob[2]{G_*}.
  \end{align}
\end{lemma}
\begin{proof}
  Let $\eps>0$. For the first estimate, note that due
  to~\eqref{eq:lower.len}, every point $x \in G_*$ has a path $\gamma$
  of length larger than $\ell_-/2$ either to $o_0$ or to $o_1$, in
  particular, there is a nonclosed path $\gamma$ of length
  $|\gamma|=\eps/2$ starting at $x$. Denote the sequence of segments
  between the vertices on $\gamma$ by $e_i$, $i=1, \dots, n$ joining
  the vertices $x_i$, $1 \le i \le n-1$ and $x_0=x$ and $x_n$. In
  addition, let $\chi$ be the affine linear function on $\gamma$ with
  $\chi(x_0)=1$ and $\chi(x_n)=0$ where $x_1$ is the endpoint of
  $\gamma$. In particular, $|\chi'(x)|=1/|\gamma|$ along $\gamma$. Due to
  the continuity of $f$ at the vertices, we have
  \begin{equation*}
    f(x)=
    (\chi f)(x) =
    \sum_{i=1}^n
    \bigl((\chi f_{e_i})(\bd_+ e_i) - (\chi f_{e_i})(\bd_- e_i) \bigr)=
    \sum_{i=1}^n \int_{e_i} (\chi f_{e_i})'(s) \dd s.
  \end{equation*}
  A simple estimate using Cauchy-Schwartz yields
  \begin{equation*}
    |f(x)|^2 \le
    2 |\gamma| \normsqr[\gamma] {f'} + \frac 2 {|\gamma|} \normsqr[\gamma]
f.
  \end{equation*}
  The second estimate follows from
  \begin{equation*}
    |f_e'(o_0)|^2 \le
    2 \kappa \ell_- \normsqr[e]{f''} +
       \frac 2 {\kappa \ell_-} \normsqr[e]{f'}
  \end{equation*}
  using the previous estimate for a path $\gamma$ lying completely in
  the \emph{single} edge $e_0$ emanating $o_0$
  (cf.~\eqref{eq:start.vx2}).  Assumption~\eqref{eq:start.vx} assures,
  that $\gamma$ can be chosen to have length $\kappa \ell_-$.
  Therefore,
  \begin{equation}
    \label{eq:f.dag.0}
    \bigl| f^\dag(o_0) \bigr|^2 =
    \bigl| f_{e_0}'(o_0) \bigr|^2 \le
    2\kappa \ell_- \normsqr[e_0]{f''} +
    \frac 2 {\kappa \ell_-} \normsqr[e_0]{f'}
  \end{equation}
  and in particular, the estimate follows with $C_1':= 2 \max\{\kappa
  \ell_-,1/(\kappa \ell_-) \}$.
\end{proof}

In order to compare several Sobolev spaces and operator domains, we
need to define a cut-off function $\chi$ on a quantum graph leaving
the vertex boundary conditions invariant:
\begin{lemma}
  \label{lem:cut.off}
  There exists a nonnegative function $\chi \in \Sob[2] {G_*}$,
  smooth on each edge, constant near each vertex such that
  $\chi(o_0)=1$, $\chi(O_1)=0$ and
  \begin{equation*}
    \norm[\infty]{\chi^{(m)}} \le
    \Bigl(\frac 2 {\ell_-} \Bigr)^m
  \end{equation*}
  for $m=0, \dots, 3$.
\end{lemma}
We call such functions $\chi$ \emph{smooth cut-off functions}. Note
that $f \in \Sob[2] {G_*}$ iff $\chi f \in \Sob[2] {G_*}$ since $\chi$
is constant near each vertex.
\begin{proof}
  Choose a function $\hat \chi$ affine linear on each vertex with
  $\hat \chi(o_0)=0$ and $\hat \chi(o_1)=1$ and $0 \le \hat \chi(v)
  \le 1$ in such a way, that no slope exceeds the minimal needed slope
  $1/\ell_-$ on the shortest path from $o_0$ to $o_1$ (due
  to~\eqref{eq:lower.len}). Let $\chi$ be a slight modification of
  $\hat \chi$ such that $\chi$ is constant near each vertex, smooth on
  each edge and such that the $m$th derivative is bounded by
  $(2/\ell_-)^m$.
\end{proof}

Associated to the branching number sequence $\{b_n\}_n$ and the vertex
potential strength $\{q_n\}_n$ we define several Sobolev spaces on a
line-like graph $L$:
\begin{align}
  \label{eq:sob1.ll}
  \Sob L &:=
  \Bigset {f \in \bigoplus_{n \ge  1}\Sob{G_n}}
     {\forall n \ge 1\colon \,\,\,
     f_n(n_-)= b_n^{-1/2} \,  f_{n+1}(n_+)}\\
  \label{eq:sob2.ll}
  \Sob[2] L &:=
  \Bigset{f \in \bigoplus_{n \ge  1}\Sob[2]{G_n}}
     {\forall n \ge 1\colon
       \begin{array}[c]{l}
         f_n(n_-)= b_n^{-1/2} \,  f_{n+1}(n_+),\\
         f_n^\dag(n_-)= b_n^{1/2} \,f_{n+1}^\dag(n_+)
       \end{array}
     }
\end{align}
and
\begin{multline}
  \label{eq:sob.max}
  \dom H^{\max} :=
  \Bigset{f \in \Lsqr L}
     {f ''=\{f_e''\}_e \in \Lsqr L, \quad \forall n \ge 1\colon
           f_n '' \in \Sob[2] {G_n}\\
       \begin{array}[c]{l}
         f_n(n_-)= b_n^{-1/2} \,  f_{n+1}(n_+),\\
         f_n^\dag(n_-)= b_n^{1/2} \,f_{n+1}^\dag(n_+)
       \end{array}
     }
\end{multline}
where $f_n:= f \restr {G_n}$. The corresponding norms are given by
\begin{multline}
  \label{eq:norm.sob}
  \normsqr[\Sob L] f := \normsqr f + \normsqr {f'}, \qquad
  \normsqr[{\Sob[2] L}] f := \normsqr f + \normsqr {f'} +
         \normsqr{f''}\\ \text{and} \qquad
  \normsqr[H^{\max}] f := \normsqr f + \normsqr{f''}.
\end{multline}
We denote by $H^{\max}=H_L^{\max}$ the \emph{maximal} Hamiltonian with
domain $\dom H^{\max}$ acting as $(H^{\max} f)_e = -f_e''$ on each
edge.

Our aim is to show that $\dom H^{\max}=\Sob[2] L$ and that their norms
are equivalent.  The following lemma is a useful tool to get rid of
the first derivative:
\begin{lemma}
  \label{lem:res.trick}
  Suppose that $\chi$ is a function smooth on each edge and constant
  near each vertex. Suppose in addition that $H$ is a self-adjoint
  operator in $\Lsqr L$ such that $\chi f \in \dom H$ if $f \in \dom
  H^{\max}$, $H \ge \lambda_0$ and such that $Hf=-f''$ for functions
  with support away from the vertices. Then
  \begin{multline}
    \label{eq:res.trick}
    \norm{(\chi f)'} \le
     \norm{d R^{1/2}} \Bigl(
      \norm[\infty] \chi \norm{f''-\lambda_0 f} \\+
      \bigl(  \norm[\infty]{\chi''} +
                  2 \norm{d R^{1/2}}\norm[\infty] {\chi'} +
                  \norm[\infty]\chi \bigr)  \norm f
    \Bigr)
  \end{multline}
  for all $f \in \dom H^{\max}$ where $R:=(H-\lambda_0+1)^{-1}$ and
  $df:=f'$.
\end{lemma}
\begin{proof}
  The assumptions on $H$ imply that $\dom H \subset \dom H^{\max}$ and
  that $H(\chi f) = -(\chi f)''= -\chi f'' - 2(\chi'f)'+\chi''f$. Then
  we can write
  \begin{equation*}
    (\chi f)' =
    d R (H - \lambda_0 + 1)(\chi f) =
    d R \bigl(\chi (-f''+(-\lambda_0+1)f) - 2 d(\chi'f) + \chi'' f \bigr).
  \end{equation*}
  Since $\chi'$ has support away from the vertices,
  $(\chi'f)'=-d^*(\chi'f)$ where $d^*$ is the adjoint of $d$. Using
  $\norm{dRd^*}=\normsqr{dR^{1/2}}$ we obtain the desired estimate.
\end{proof}

\begin{lemma}
  \label{lem:sob''}
  The space $\dom H^{\max}_\mathrm{c}$ of compactly supported
  functions (not necessarily disjoint from the root vertex) in $\dom
  H^{\max}$ is dense. In addition, there is a constant $C_1''>0$ such
  that
  \begin{equation}
    \label{eq:sob''}
    \normsqr[L] {f'} \le
    C_1'' \bigl( \normsqr[L] {f''} + \normsqr[L] f\bigr)
  \end{equation}
  holds for all $f \in \dom H^{\max}$.
\end{lemma}
\begin{proof}
  For a function $f \in \dom H^{\max}$ let $f_n := \chi_n f$ where
  $\chi_n$ is the smooth cut-off function with $\chi_n(n-1)=0$ and
  $\chi_n(n)=1$ on $G_n$ as constructed in \Lem{cut.off} extended on
  $G_k$ by $0$ for $k<n$, respectively, by $1$, for $k>n$. Now
  \begin{equation}
    \label{eq:norm.est1}
    \norm{f-f_n} \le \normsqr[L_n] f \to 0
  \end{equation}
  as $n \to \infty$ since $f \in \Lsqr L$. Furthermore,
  \begin{multline}
    \label{eq:norm.est2}
    \norm{(f-f_n)''} \le
    \norm{((1-\chi_n)f''} + \norm{\chi_n''f}
                         + 2 \norm{(\chi_n' f)'} \\\le
    \norm[L_n]{f''} + \frac 4 {\ell_-^2} \norm[L_n] f
                         + 2 \norm{(\chi_n' f)'}.
  \end{multline}
  Now, the latter term can be estimated by
  \begin{equation*}
    \norm[L_n]{(\chi_n' f)'} \le
    \frac 2 {\ell_-} \norm[L_n] {f''} +
    \Bigl( \frac 8 {\ell_-^3} + \frac 8{\ell_-^2} + \frac 2 {\ell_-} \Bigr)
       \norm[L_n] f
  \end{equation*}
  applying the previous lemma with $H:=\laplacianD {L_n}$, the
  Dirichlet Laplacian on $L_n$ defined via the quadratic form $\qf
  d(f)=\normsqr{f'}$ with domain $\Sobn {L_n}$. Note that the estimate
  $\norm{d R^{1/2}} \le 1$ is equivalent to $\norm{df} \le \qf
  d(f)+\normsqr f$ which is obviously true. Since $f$ and
  $f''=\{f_e''\}_e \in \Lsqr L$, the left hand side
  of~\eqref{eq:norm.est2} tends
  to $0$ as $n \to \infty$.  We have therefore shown that compactly
  supported functions are dense in $\dom H^{\max}$.

  To show~\eqref{eq:sob''}, we can restrict ourselves to compactly
  supported functions.  Partial integration taking the inner boundary
  conditions into account yields
  \begin{equation*}
    \normsqr[L_n] {f'} \le
    \normsqr[L_n]{f''} + \normsqr[L_n] f + |f^\dag(0) f(0)|.
  \end{equation*}
  The last term can be estimated using~\eqref{eq:sob}--\eqref{eq:sob'}
  and $ab \le \eta a^2/2 + b^2/(2\eta)$ as
  \begin{multline*}
    |f^\dag(0) f(0)| \le
    \sqrt {C_1'}\bigl( \norm {f'} + \norm{f''} \bigr)
       \bigl( \sqrt \eps \norm {f'} +
                \frac 2 {\sqrt \eps} \norm f \bigr) \\ \le
      \sqrt {C_1'\eps} \normsqr {f'} + \frac 14 \normsqr {f'} +
         \Bigl(\frac {8 C_1'}\eps + \frac 1{\sqrt \eps} \Bigr)
              \normsqr f +
         \Bigl(4 \eps  + \frac 1{\sqrt \eps} \Bigr)
              \normsqr {f''}
  \end{multline*}
  provided $0 < \eps \le \ell_-$ where all the norms are
  $\Lsymb_2$-norms on $G_1$. Choosing $\eps=\min\{1/(16C_1'),\ell_-\}$
  we obtain
  \begin{equation*}
    |f^\dag(0) f(0)| \le \frac 12 \normsqr{f'} +
    \Bigl(\frac {8 C_1'} \eps  +
       \frac 1 {\sqrt \eps} \Bigr) \normsqr f +
    \Bigl(4 \eps + \frac 1 {\sqrt \eps} \Bigr) \normsqr{f''}.
  \end{equation*}
  Subtracting the contribution of $\normsqr{f'}$ on the right hand side
  we finally
  see that there is a constant $C_1''$ depending only on $C_1'$ and
  $\ell_-$ such that~\eqref{eq:sob''} holds.
\end{proof}
\begin{remark}
   \label{rem:sob}
   Note that for a fixed decoration graph, we can prove the
   estimate
   \begin{equation}
     \label{eq:sob''1}
     \normsqr[G_*] {f'} \le
     \wt C_1'' \bigl( \normsqr[G_*] {f''} + \normsqr[G_*] f \bigr)
   \end{equation}
   for all $f \in \Sob[2]{G_*}$ similar as in the above proof.  But to
   do so, we need an estimate on $f^\dag(o_1)$ as in~\eqref{eq:sob'}
   assuming that there is no vertex potential.  Therefore, the
   constant $\wt C_1''$ depend on the minimal length of all edges
   adjacent to $o_1$ and the vertex degree of $o_1$ similar
   to~\eqref{eq:f.dag.0}, which does \emph{not} admit a global lower
   bound in our family of decoration graphs.
\end{remark}

The next lemma deals with the Sobolev spaces $\Sob[m] L$ and their
relation with $\dom H^{\max}$:
\begin{lemma}
  \label{lem:sob.max}
  The spaces $\Sob L$ and $\Sob[2] L$ with their natural norms given
  in~\eqref{eq:norm.sob} are Hilbert spaces. The spaces $\Sob[2] L$
  and $\dom H^{\max}$ are equal and have equivalent norms.  In
  addition, the subspaces $\Sobc L$, respectively, $\Sobc[2] L$, of 
functions
  in $\Sob L$ resp.\ $\Sob[2] L$ with compact support (not necessarily
  away from the root vertex $0$) are dense.
\end{lemma}
\begin{proof}
  The completeness of $\Sob L$ and $\Sob[2] L$ follows from the fact
  that $\Sob L$, respectively, $\Sob[2] L$, are closed subspaces in the 
Hilbert
  space $\bigoplus_n \Sob[m] {G_n}$: Note that~\eqref{eq:sob}
  and~\eqref{eq:sob'} imply the continuity of $f \mapsto f(n_\pm)$,
  respectively, $f \mapsto f^\dag(n_\pm)$.

  From~\eqref{eq:sob''} we see that $\dom H^{\max} \subset \Sob[2]L$ and
  that the inclusion is continuous. The opposite inclusion is trivial.
  The density of the space of compactly functions in $\Sob[2] L$ now
  follows from \Lem{sob''}. The similar assertion for $\Sob L$ follows
  in the same way.
\end{proof}

To summarize, we can characterize the domain of the maximal
Hamiltonian as follows:
\begin{lemma}
  \label{lem:graph.norm}
  The maximal Hamiltonian $H^{\max}=H^{\max}_L$ on $L$ with domain
  $\dom H^{\max}=\Sob[2] L$ is a closed operator. In addition, $f \in
  \dom H^{\max} = \Sob[2] L$ iff
  \begin{enumerate}
  \item $f, f''=\{f_e''\}_e \in \Lsqr L$,
  \item
    \label{in.bc} $f$ satisfies all vertex boundary
    conditions at inner vertices $V_0(G_n)$, i.e.,
    $f_{e_1}(v)=f_{e_2}(v)$ for all $e_1, e_2 \in E_v(G_n)$, $v \in
    V(G_n)$, and $f_{G_n}'(v)=0$ for all $v \in V_0(G_n)$ and $n \ge 1$
    (cf.~\eqref{eq:def.der} for the notation);
  \item $f$ satisfies all vertex boundary conditions~\eqref{eq:ll.bc2}
    at the connecting vertices $n_\pm$, $n \ge 1$.
  \end{enumerate}
\end{lemma}
\begin{proof}
  The domain $\dom H^{\max}$ with its graph norm is a complete space
  by the previous lemma. The characterization of $\dom H^{\max}$ is
  just a reformulation of~\eqref{eq:sob.max}.
\end{proof}

We will see in \Lem{ess.sa} that the maximal operator $H^{\max}$ is
maximal in the sense that only a boundary condition at the root vertex
$0$ is missing in order to have a self-adjoint operator. We will
impose a \emph{Dirichlet boundary condition}.

Let $\qf h_{G_n}$ be the quadratic form on $\Sob{G_n}$ defined by
\begin{equation*}
  \qf h_{G_n}(f) =
  \sum_{e \in E(G_n)} \normsqr[e] {f'} +
                  q_n |f(n_-)|^2
\end{equation*}
where $\{q_n\}$ is the strength of the vertex potential
satisfying~\eqref{eq:q.bdd}. We define the quadratic form on the
line-like graph $L=L_0$ as
\begin{equation}
  \label{eq:qf.line}
  \qf h_L ( f) :=
       \sum_{n \ge 1} \qf h_{G_n}( f_n)
\end{equation}
with domain
\begin{equation}
  \label{eq:dom.qf.line}
  \Sobn L := \dom \qf h_L :=
  \bigset {f \in \Sob L} { f(0)=0}
\end{equation}
where $\Sob L$ has been defined in~\eqref{eq:sob1.ll}.

Let $\qf d_L$ be the ``free'' quadratic form, i.e., the form without
vertex potential, namely
\begin{equation*}
  \qf d_L(f) := \sum_{e \in E(L)} \normsqr[e]{f'}
\end{equation*}
for $f \in \Sobn L$. Remember that the vertex potential in $\qf h$ has
support only at the ending vertices $n_-$ of $G_n$.
\begin{lemma}
  \label{lem:form.bdd}
  The quadratic form $\qf d=\qf d_L$ is closed. Furthermore, $\qf
  h=\qf h_L$ is relatively form-bounded with respect to the form $\qf
  d$ with relative bound $0$. In particular, $\qf h$ is a closed form
  on $\Sobn L$.
\end{lemma}

\begin{proof}
  Since $\Sobn L$ is a closed subspace of the Hilbert space $\Sob L$
  with norm given by $\normsqr f + \qf d(f)$, $\qf d$ is a closed form
  on $\Sob L$.  Furthermore, we have
  \begin{multline*}
    |\qf h(f)- \qf d(f)| \le
    q_0 \sum_{n \in \N} |f(n_-)|^2 \\\le
    q_0 \sum_{n \in \N}
      \Bigl( \eps \normsqr[G_n] {f'} +
                \frac 4 \eps \normsqr[G_n] f \Bigr) \le
    q_0 \Bigl( \eps \normsqr[L] {f'} +
                \frac 4 \eps \normsqr[L] f \Bigr)
  \end{multline*}
  for any $0<\eps<\ell_-$ using~\eqref{eq:sob} and~\eqref{eq:q.bdd}
  where $q_0:=\max\{|q_-|,|q_+|\}$, with shows the assertion.
\end{proof}
\begin{corollary}
  \label{cor:form.bdd}
  We have $0 \le \qf d \le 2 (\qf h - \lambda_0)$ where $\lambda_0:= -
  \max \{ 4 q_0/\ell_-, 8 q_0^2\} \le 0$. We might choose
  $\lambda_0:=0$ if the strength of the vertex potential is
  nonnegative, i.e., $q \ge 0$.
\end{corollary}
\begin{proof}
  A simple application of the last estimate shows that
  \begin{equation*}
    (1- q_0 \eps) \qf d(f) \le
    \qf h(f) + \frac {4q_0} \eps \normsqr f.
  \end{equation*}
  Choose $\eps=\min\{\ell_-, 1/(2q_0)\}$. If $q \ge 0$ then clearly
  $\qf d \le \qf h$.
\end{proof}

Denote the self-adjoint operator associated to the quadratic form $\qf
h=\qf h_L$ by $H=H_L$, and similarly $\laplacianD L$ the operator
associated to $\qf d$.
\begin{lemma}
  \label{lem:ess.sa}
  A function $f$ is in the domain of $H_L$ iff~(i)--(iii) of
  \Lem{graph.norm} are fulfilled and if
  \begin{enumerate}
  \setcounter{enumi}{3}
  \item $f(0)=0$.
  \end{enumerate}
  Furthermore, $H_L$ is essentially self-adjoint on all functions $f
  \in \dom H_L$ with compact support.
\end{lemma}
\begin{proof}
  A function $f \in \dom \qf h$ in the domain of the operator
  associated to $\qf h$ satisfies $\qf h (f,g)= \iprod {Hf} g$ where
  $Hf$ denotes an element in $\Lsqr L$. Choosing only functions $g$
  with support inside an edge $e$, partial integration shows that
  $(Hf)_e=-f''_e$ in a distributional sense; and therefore $f_e \in
  \Sob[2] e$.  Taking general $g \in \dom \qf h$, it is an easy
  exercise to see that the boundary terms from partial integration
  vanish iff $f$ satisfies the conditions of
  \Lemenum{graph.norm}{in.bc} and~\eqref{eq:ll.bc2} for all inner,
  respectively, connecting, vertices and $f(0)=0$. Therefore all
  conditions~(i)--(iv) are fulfilled.

  If a function $f$ satisfies the condition~(i), we know from
  \eqref{eq:sob''} that also $f' \in \Lsqr L$. Together
  with~(ii)--(iv) we have $f \in \dom \qf h$. Furthermore, the
  same argument as before shows that for each $f$ there is
  $Hf:=\{-f_e''\}_e$ such that $\qf h(f,g)=\iprod {Hf} g$, i.e., $f$
  is in the domain of the associated operator.

  The essential self adjointness follows from~\Lem{sob.max}.
\end{proof}

Finally, we also need the following estimate in \Sec{weyl}:
\begin{lemma}
  \label{lem:f'.int}
  If $f \in \Sob[2] L$ then $|(f f^\dag)(n_+)| \to 0$ as $n \to \infty$.
\end{lemma}
\begin{proof}
  Suppose first that $f$ has compact support. Partial integration on
  the line-like graph $L_n$ and the boundary
  condition~\eqref{eq:ll.bc2} yield
  \begin{equation*}
    |(f f^\dag)(n_+)|=|(f f^\dag)(n_-)| \le
    |\iprod[L_n]{Hf } f| + \normsqr[L_n]{f'} \le
    \normsqr[{\Sob[2] {L_n}}] f.
  \end{equation*}
  This inequality extends to all functions $f \in \Sob[2] L$ due
  to \Lem{sob.max}. Now, if $f \in \Sob[2] L$, then
  $\normsqr[{\Sob[2]{L_n}}] f = \normsqr[L_n] f + \normsqr[L_n] {f'} +
  \normsqr[L_n] {f''} \to 0$ and the result follows.
\end{proof}

%----------------------------------------------------------------------
%
\subsection{Generalized eigenfunctions and integral kernels}
\label{sec:gen.ef}
%----------------------------------------------------------------------

In this section we first provide the necessary
estimate~\eqref{eq:carleman} in order to show that the results of
\App{bd.ef} apply.
\begin{lemma}
  \label{lem:carleman}
  Suppose that $H=H_L$ is the self-adjoint operator on the line-like
  graph $L=L_0$ with Dirichlet boundary condition at $0$ constructed
  as below. Then
  \begin{equation*}
    \normsqr[\Linfty L] f \le
    4\Bigl(\ell_- + \frac 1 {\ell_-} \Bigr) 
               \normsqr[\Lsqr L]{(H - \lambda_0 + 1)^{1/2} f}
  \end{equation*}
  for all $f \in \dom \qf h=\dom(H-\lambda_0)^{1/2}$ where $\lambda_0$
  is given in \Cor{form.bdd} and $\ell_->0$ is defined
  in~\eqref{eq:lower.len}. In particular, \Ass{op} is fulfilled with
  $m=1$.
\end{lemma}
\begin{proof}
  We have
  \begin{multline*}
    |f(x)|^2 \le
    \ell_- \normsqr[G_n] {f'} + \frac 4 {\ell_-} \normsqr[G_n] f \le
    2 \ell_- \bigl(\qf h(f) - \lambda_0 \normsqr[L] f\bigr) +
           \frac 4 {\ell_-} \normsqr[L] f \\\le
    4\Bigl(\ell_- + \frac 1 {\ell_-} \Bigr)
          \normsqr[\Lsqr L]{(H - \lambda_0 + 1)^{1/2} f}
  \end{multline*}
  using~\eqref{eq:sob} if $x \in G_n$ and $f \in \dom \qf h$. In
  addition, $H$ is a local operator, so that \Ass{op} is fulfilled.
\end{proof}

On a quantum graph, we can define the notion of a generalized
eigenfunction as follows:
\begin{definition}
  \label{def:gen.ef}
  We say that $\phi$ is a \emph{generalized eigenfunction} of the
  graph Hamiltonian $H$ on the line-like graph $L$ with eigenvalue
  $\lambda \in \R$ if $\phi \restr e \in \Sob[2] e$ and
  $-\phi''=\lambda \phi$ on each edge $e$ and if $\phi$ satisfies the
  vertex boundary conditions of \Lemenum{graph.norm}{in.bc} at each
  inner vertex $v \in V_0(G_n)$ and the boundary
  conditions~\eqref{eq:ll.bc2} at the connecting vertices $n_\pm$ and
  $f(0)=0$.
\end{definition}
Note that automatically, a generalized eigenfunction is smooth on each
edge since it must have the form~\eqref{edgesoln1} on each edge.

The next lemma assures that the notion of generalized eigenfunction of
\Def{gen.ef} and \Def{gen.ef.weight} agree up to an integrability
condition:
\begin{lemma}
  \label{lem:eq.gen.ef}
  Suppose that $\phi \in \dom H_-$ and $H_-\phi = \lambda \phi$ then
  $\phi$ is a generalized eigenfunction in the sense of \Def{gen.ef}.
  On the other hand, suppose that $\phi$ is a generalized
  eigenfunction in the sense of \Def{gen.ef} and that $\Phi \phi \in
  \Lsqr L$, then $\phi \in \dom H_- \subset \HS_-$ and $H_-\phi =
  \lambda \phi$.
\end{lemma}
\begin{proof}
   For the first assertion, $H_- \phi = \lambda \phi$ is
   equivalent to
   \begin{equation}
     \label{eq:gen.ef2}
     (\phi, (H_+ - \lambda)f)=0
   \end{equation}
   for all $f \in \dom H_+$. Using $f \in \Cci e$ one sees that
   $-\phi''=\lambda \phi$ in the distributional sense, and from
   regularity theory, we obtain $\phi \in \Ci e$. It follows that for
   the boundary terms we have
   \begin{equation*}
     \sum_{v \in V} \sum_{e \in E_v}
        \bigl( \conj {\phi_e(v)} \, f_e'(v) -
               \conj{\phi_e'(v)} \, f(v) \bigr) = 0
   \end{equation*}
   for all $f \in \dom H_+$. Using the argument of
   \cite[Lem.~2.2]{kostrykin-schrader:99} we see that $\phi$ has to
   satisfy the same boundary conditions as $f$ at each vertex $v$.

   For the converse we have $\phi \in \HS_-$ and one easily sees
   that~\eqref{eq:gen.ef2} is fulfilled for all $f \in \dom H_+$
   using partial integration and the boundary conditions for $f$ and
   $\phi$.
\end{proof}

We prove next a representation of the integral kernel of the resolution
of unity:
\begin{lemma}
  \label{lem:kernel.gen.ef}
  Let $\lambda \in \R$.
  Then the integral kernel of $\wt E(\lambda)$ associated to the
  operator $H$ has the representation
  \begin{equation}
    \label{eq:kernel.gen.ef}
    \wt e_\lambda(x,y)=
    \sum_j \conj{\phi_{\lambda,j}(x)} \phi_{\lambda,j}(y)
  \end{equation}
  where $\{\phi_{\lambda,j}\}_j$ forms a basis of generalized
  eigenfunctions. Even if the family is infinite, the sum is locally
  finite and defined everywhere. In particular, $\wt e_\lambda(x,y)$
  is continuous on each edge and satisfies the boundary condition at
  each vertex in $x$ and $y$.

  In particular, if $\lambda$ is not an exceptional energy
  (cf.~\Def{non.sep.l}), i.e., $\lambda \notin E(L)$, then the sum
  reduces to
  \begin{equation}
    \label{eq:kernel.gen.ef2}
    \wt e_\lambda(x,y)=c_\lambda \conj{\phi_\lambda(x)} \phi_\lambda(y)
  \end{equation}
  where $c_\lambda = 1/\normsqr[L]{\Phi \phi_\lambda} \in \R$ and
  $\phi_\lambda$ is the generalized eigenfunction with
  $\phi_\lambda(0)=0$ and $\phi_\lambda'(0)=1$.  All statements
  hold for almost all $\lambda$ w.r.t.~a spectral measure of $H$.
\end{lemma}
\begin{proof}
  A generalized eigenfunction is $\Contsymb^\infty$ on each edge and
  satisfies the boundary conditions due to \Lem{eq.gen.ef}.  Since the
  space of generalized eigenfunctions (without conditions at $0$ and
  $\infty$) is generated by compactly supported functions and at most
  two noncompactly supported functions (cf.~\Lem{dim.ef}), we can
  apply \Lem{ef.exp} and obtain~\eqref{eq:kernel.gen.ef}.

  For the second assertion, note that for nonexceptional energies, the
  space of generalized eigenfunctions without conditions at $0$ and
  $\infty$ is two-dimensional (cf.~\Lem{dim.ef}). In addition, there
  is only one function satisfying the boundary condition
  $\phi_\lambda(0)=0$ and $\phi_\lambda^\dag(0)=1$.  The value of the
  normalization constant $c_\lambda$ follows from $1=\norm[\TRsymb]
  {E(\lambda)} = \int_L \Phi(x)^2 \wt e_\lambda(x,x) \dd x = c_\lambda
  \normsqr[L]{\Phi \phi_\lambda}$. In addition, we have $c_\lambda \in
  (0,\infty)$ for almost all $\lambda$.
\end{proof}

We finally need an integral representation of the Green's
functions:
\begin{corollary}
   \label{cor:green}
   The Green's function (i.e., the kernel of $(H-z)^{-1}$) can be
   written as
   \begin{equation}
     \label{eq:green}
     G_z(x,y)= \int_\R \frac 1 {\lambda - z} \wt e_\lambda(x,y)
              \dd \rho(\lambda)
   \end{equation}
   for \emph{all} $x,y \in L$.  In particular, $G_z$ is continuous
   (even $\Contsymb^\infty$) outside the vertices and satisfies the
   boundary conditions of $H$ at each vertex in each variable.
\end{corollary}
\begin{proof}
   We obtain the kernel representation from \Lem{desint.fct} a priori
   only for \emph{almost} all $x,y \in L$, but the
   representation~\eqref{eq:kernel.gen.ef} assures that $G_z(x,y)$ is
   smooth outside the edges and satisfies the boundary conditions
   (since $\wt e_\lambda$ does).
\end{proof}

%----------------------------------------------------------------------
%
\subsection{Polynomial bounds on generalized eigenfunctions}
\label{sec:ptw.bd}
%----------------------------------------------------------------------

In this section we show weighted $\Lsymb_2$-bounds on generalized
eigenfunctions on line-like graphs. In addition, we prove pointwise
bounds on the eigenfunctions. To do so, we fix the weight function
$\Phi$ needed in order to apply the results of \App{bd.ef}. The metric
measure space $(X,\mu)$ will be the metric graph $L$ with its natural
measure. For example, if $\Phi(x) = 1/n$ for $x \in G_n$ then we have
\begin{equation*}
  \normsqr[L] \Phi = \sum_n \frac 1 {n^2} \ell(G_n) < \infty
\end{equation*}
by~\eqref{eq:vol.bd}.  Therefore, \Ass{weight} is also fulfilled.
{}From \Thm{bd.gen.ef} we obtain that a generalized eigenfunction
$\phi_\lambda$ satisfies $\norm[L]{\Phi \phi_\lambda} < \infty$ for
almost all $\lambda$ with respect to a spectral measure of $H$:
\begin{theorem}
  \label{thm:sp.meas.ef}
  Suppose that the assumptions~\eqref{eq:ass.graphs} on the decoration
  graphs $\{G_n\}_n$ are fulfilled. Then the spectral measure is
  supported by those $\lambda$ for which there is a generalized
  eigenfunction $\phi$ of polynomial growth rate (in the sense that
  $\norm[L]{\Phi \phi} < \infty$).
\end{theorem}
Our second  main result is:
\begin{theorem}
  \label{thm:ptw.bd.ef}
  Suppose the assumptions~\eqref{eq:ass.graphs} are fulfilled. Let
  $\phi=\phi_\lambda$ be a generalized eigenfunction of $H$ in the
  sense of \Def{gen.ef.weight} and $\Phi(x)>0$ for all $x \in L$. Then
  there exist $C_3, C_3'>0$ such that
  \begin{gather*}
    |\Phi(x)\phi(x)| \le C_3 w_n \norm[\Lsqr G]{\Phi \phi}\\
    |\Phi(n_+) \phi^\dag(n_+)| \le C_3' w_n \norm[\Lsqr G]{\Phi \phi}
  \end{gather*}
  for $x \in G_n \subset L$ and $n \in \N$ for almost all $\lambda \in
  \R$ with respect to a spectral measure of $H$ where
  \begin{equation*}
    w_n := \frac{\Phi_+(n)}{\Phi_-(n)} :=
    \frac{\sup \set{\Phi(x)}{x \in G_n}} {\inf \set{\Phi(x)}{x \in G_n^+}}
  \end{equation*}
  and $G_n^+ := L_{n-2,n+1}$, the concatenation of $G_k$, $k=n-1,
  \dots, n+1$.  In particular, if $\Phi(x)=1/n$ for $x \in G_n$ then
  $w_n \le 2$ and $|\phi(x)|$ ($x \in G_n$) and $|\phi^\dag(n_+)|$ are
  polynomially bounded in $n$.
\end{theorem}
\begin{remark}
  Note that we state the result only with respect to a spectral
  measure! Generally this is of course false, take for example
  $b_n=2$, $H=\laplacian L$ and a function constant on each decoration
  graph $G_n$ satisfying the boundary condition $\phi_n(n_-)=2^{-1/2}
  \phi_{n+1}(n_+)$. Then $H_+ \phi=0$, but $\phi$ has exponential
  growth since $\phi_n(x)=2^{(n-1)/2} \phi(0)$ for $x \in G_n$. The
  important point here is that $0 \notin \spec H$.
\end{remark}
\begin{proof}
  Let $x \in G_n$ and $\chi$ a function such that $\chi=1$ on $G_n$,
  $\chi=0$ on $G_k$, $|n-k|\ge 1$ and $\chi(n-1)=0$, $\chi(n)=1$,
  $\chi(n+1)=1$ and $\chi(n+2)=$ on $G_{n-1}$ and $G_{n+1}$ as
  constructed in \Lem{cut.off}. Note that $\chi$ has support in
  $G_n^+$.  Now,
  \begin{equation*}
    |\Phi(x)\phi(x)|^2 \le
    \Phi_+(n)^2
     \bigl( \ell_- \normsqr[G_n]{\phi'} +
             \frac 4 {\ell_-} \normsqr[G_n] \phi \bigr)
  \end{equation*}
  due to~\eqref{eq:sob}.  Furthermore, $\normsqr {df} \le 2
  \normsqr{(H-\lambda_0+1)^{1/2}f}$ due to \Cor{form.bdd} and
  therefore $\normsqr {dR^{1/2}} \le 2$. From \Lem{res.trick} we
  conclude that
  \begin{multline}
    \label{eq:est.der}
    \norm[G_n]{\phi'} \le
    \norm[G_n^+]{(\chi \phi)'} \\\le
    \sqrt 2 \Bigl(
       (\lambda-\lambda_0) + \frac 4 {\ell_-^2} +
       \frac {4\sqrt 2} {\ell_-} + 1 \Bigr) \norm[G_n^+] \phi =:
       C_2 \norm[G_n^+] \phi.
  \end{multline}
  Finally,
  \begin{multline*}
    |\Phi(x)\phi(x)|^2 \le
    \Phi_+(n)^2 \Bigr(\frac 4 {\ell_-} +
           \ell_- C_2^2 \Bigl) \normsqr[G_n^+] \phi \\\le
    \Bigl(\frac{\Phi_+(n)}{\Phi_-(n)} \Bigr)^2
        \Bigr(\frac 4 {\ell_-} +
           \ell_- C_2^2 \Bigl) \normsqr[L] {\Phi \phi} \le
    w_n^2 \Bigr(\frac 4 {\ell_-} +
               \ell_- C_2^2 \Bigl) \normsqr[L] {\Phi \phi} =:
    w_n^2 C_3^2 \normsqr[L] {\Phi \phi}.
  \end{multline*}

  The second assertion follows similarly:
  \begin{multline*}
    |\Phi(n_+)\phi^\dag((n-1)_+)|^2 \le
    \Phi_+(n)^2
      C_1'\bigl( \lambda^2 \normsqr[G_n] \phi +
            \normsqr[G_n] {\phi'} \bigr) \\\le
    w_n^2 C_1'(\lambda^2 + C_2^2) \normsqr[G_n^+] \phi \le
    w_n^2 C_1'(\lambda^2 + C_2^2) \normsqr[L] {\Phi \phi}
    =: C_3'{}^2 \normsqr[L] {\Phi \phi}.
  \end{multline*}
\end{proof}

%----------------------------------------------------------------------
%
\section{Transfer matrices and Weyl-Titchmarsh functions}
\label{app:tm.weyl}
%
%----------------------------------------------------------------------

%----------------------------------------------------------------------
%
\subsection{Transfer matrix for generalized eigenfunctions}
\label{sec:trans.mat}
%----------------------------------------------------------------------

We want to prove in this section that the transfer matrix is defined
up to an exceptional set. For a fixed sequence of graphs $\{G_n\}$,
the exceptional set is countable. The main ingredient is the
\emph{Dirichlet-to-Neumann map} (see for example \cite{ong:pre05}
or~\cite{fop:pre04}).  Let $G_*$ be one of the decoration graphs
replacing an edge in the tree graph with two boundary vertices $o_0$
and $o_1$.

Denote by $\laplacianD{G_*}$ the Dirichlet operator on $G_*$, i.e.,
the self-adjoint operator on functions $u \in \Sob[2]{G_*}$ satisfying
$u(o_0)=0$ and $u(o_1)=0$.  Denote its spectrum repeated according to
multiplicity by $\{\lambda_k\}_k$ and the corresponding orthonormal
basis of \emph{real} eigenfunctions by $\{\phi_k\}_k$. We first state
some results on the solution map
\begin{equation}
  \label{eq:dir.sol.map}
  \map {H_z} {\C^2} {\Sob[2]{G_*}}
\end{equation}
of the Dirichlet problem, i.e., $f=H_z(F_0,F_1)$ solves the equation
$(\laplacian[\max]{G_*}-z)f=0$ with initial data $f(o_0)=F_0$ and
$f(1)=F_1$ for $z \notin \spec{\laplacianD{G_*}}$. We need the
following extension map
\begin{equation}
  \label{eq:ext.op}
  \map E {\C^2} {\Sob[2]{G_*}},
\end{equation}
i.e., $\wt f=E(F_0,F_1) \in \Sob[2]{G_*}$ such that $f(o_0)=F_0$,
$f(o_1)=F_1$ and $\wt f^\dag(o_i)=0$ for $i=0,1$. For example, $\wt
f:= E(F_0,F_1) := F_0 \chi + F_1 (1-\chi)$ is a possible choice where
$\chi$ is the smooth cut-off function constructed in \Lem{cut.off}.
In particular, the derivatives of $\chi$ up to order $2$ enter in
$\norm E$ so that $\norm E$ can be bounded by a universal polynomial
of degree $2$ in $1/\ell_-$.

In addition, denote by $\laplacian[\max]{G_*}$ the differential
operator $\laplacian{G_*}$ with \emph{maximal} domain, i.e., $f \in
\dom {\laplacian[\max]{G_*}}$ iff $f, \laplacian{G_*}f \in
\Lsqr{G_*}$.

We can now give expressions for the Dirichlet solution map:
\begin{lemma}
  \label{lem:dir.sol.map}
  For $z \notin \spec{\laplacianD{G_*}}$ the solution map $H_z$
  in~\eqref{eq:dir.sol.map} is given by
  \begin{equation}
    \label{eq:dir.sm1}
    H_z =
    \bigl(\1 - (\laplacianD {G_*} - z)^{-1}
               (\laplacian[\max] {G_*} - z) \bigr) E
  \end{equation}
  and is bounded as map $\map {H_z}{\C^2}{\Lsqr{G_*}}$ with norm
  estimated by
  \begin{equation}
    \label{eq:dir.sm.norm}
    \norm {H_z} \le
    \Bigl( 1+ \frac {1+ |z|} {d(z,
         \spec{\laplacianD{G_*}})}\Bigr)
       \norm E.
  \end{equation}
  In addition, $H_z$ is also bounded as map $\map
  {H_z}{\C^2}{\Sob[2]{G_*}}$.  Furthermore, $z \mapsto H_z$ is
  norm-analytic with the series representation
  \begin{subequations}
  \label{eq:dir.sm.series}
  \begin{align}
    \label{eq:dir.sm.a}
    H_z(F_0,F_1)
   &= -\sum_k \frac1{\lambda_k-z}
     \bigl(  \phi_k^\dag(o_1)F_1 - \phi_k^\dag(o_0)F_0
             \bigr)\phi_k\\
    \label{eq:dir.sm.b}
   &= H_0(F_0,F_1) -\sum_k \frac z{\lambda_k(\lambda_k-z)}
     \bigl(  \phi_k^\dag(o_1)F_1 - \phi_k^\dag(o_0)F_0
             \bigr)\phi_k
  \end{align}
  \end{subequations}
  with $\wt f=E(F_0,F_1)$ where the first series converges in
  $\Lsqr{G_*}$ and the second in $\Sob[2]{G_*}$.
\end{lemma}
\begin{proof}
  First, $H_z$ is well defined as map from $\C^2$ into $\Lsqr{G_*}$.
  Next, it follows from $\dom \laplacianD{G_*} \subset \dom
  \laplacian[\max]{G_*}$ that $f=H_z(F_0,F_1)$ solves
  $(\laplacian[\max]{G_*}-z)f=0$. In addition, $f(o_0)=\wt f(o_0)=F_0$
  and similarly in $o_1$, since functions in the range of the
  resolvent vanish at the boundary. Furthermore, $H_z$ is bounded as
  map into $\Lsqr{G_*}$ or as map into $\dom \laplacian[\max]{G_*}$
  with the graph norm given by $\normsqr[{\laplacian[\max]{G_*}}] f :=
  \normsqr {\laplacian[\max]{G_*}f} + \normsqr f$.  Now, due
  to~\eqref{eq:sob''1}, we have
  \begin{equation*}
    \normsqr[{\Sob[2]{G_*}}] f \le
    (\wt C_1''+1) \normsqr[{\laplacian[\max]{G_*}}] f
  \end{equation*}
  but the constant $\wt C_1''$ is not uniform in the sense of
  \Ass{deco.gr}. Nevertheless, $H_z$ is continuous as map into
  $\Sob[2]{G_*}$.

  Expanding the resolvent into a series of eigenvector we obtain
  \begin{equation*}
    H_z(F_0,F_1) =
    \wt f - \sum_k \frac1{\lambda_k-z}
       \iprod {\phi_k}
       {(\laplacian[\max]{G_*}-z) \wt f} \phi_k.
  \end{equation*}
  Since $\wt f \in \dom \laplacian[\max]{G_*}$ the
  coefficients
  \begin{equation*}
    a_k :=
      \frac1{\lambda_k-z}
      \iprod {\phi_k} {(\laplacian[\max]{G_*}-z) \wt f}
      \qquad\text{and}\qquad
    \lambda_k a_k
  \end{equation*}
  form sequences in $\lsqr \N$. It follows that the series converge in
  $\Lsqr{G_*}$ and in $\dom \laplacian[\max]{G_*}$ with the graph
  norm, and therefore the series also converges in $\Sob[2]{G_*}$.
  The first series representation follows from partial integration.
  Note that
  \begin{equation*}
    b_k =
    \iprod{\phi_k}{\wt f} -       \frac1{\lambda_k-z}
      \iprod {\phi_k} {(\laplacian[\max]{G_*}-z) \wt f}=
    -\frac1{\lambda_k-z}
      \bigl( \phi_k^\dag(o_1)F_1 - \phi_k^\dag(o_0)F_0 \bigr)
  \end{equation*}
  is in $\lsqr \N$ since $\wt f$ and $\laplacian[\max]{G_*} \wt f$ are
  both in $\Lsqr {G_*}$.
\end{proof}

For $z \in \C \setminus \spec {\laplacianD{G_*}}$, we can define the
Dirichlet-to-Neumann map:
\begin{definition}
  \label{def:d2n}
  The \emph{Dirichlet-to-Neumann map} $\Lambda(G_*,z) =
  \Lambda(z)$ is the $2\times2$-matrix
  defined by\footnote{In this section, we assume that there is no
    vertex potential, i.e., $q(o_1)=0$ (cf.~\eqref{eq:gen.der}).}
  \begin{equation}
      \label{eq:def.d2n}
      \Lambda_{ij}(z) = H_z (\vec e_j)^\dag (o_i)
  \end{equation}
  for $i,j=0,1$, where $\vec e_0=(1,0)$ and $\vec e_1=(0,1)$.  Let
  \begin{equation}
    \label{eq:d2n.coeff}
    \Psi_i(\lambda)=
        \set{\phi^\dag_k(o_i)}{\text{$k$ with $\lambda=\lambda_k$}}
  \end{equation}
  be the vector of boundary derivatives with dimension equal to the
  multiplicity of the eigenvalue $\lambda$. We denote
  \begin{equation}
    \label{eq:red.dir}
    \spec[red]{\laplacianD{G_*}} :=
    \bigset{\lambda \in \spec{\laplacianD{G_*}}}
        {\Psi_0(\lambda) \ne 0 \text{ or } \Psi_1(\lambda) \ne 0}
  \end{equation}
  the \emph{reduced Dirichlet spectrum of $G_*$}.
\end{definition}
The next lemma explains the reason for introducing the reduced spectrum:
\begin{lemma}
  \label{lem:d2n}
  The Dirichlet-to-Neumann map $\Lambda(z)$ is meromorphic in $z$ with
  poles of order $1$ in the reduced spectrum of $\laplacianD{G_*}$ and
  has the absolutely convergent series
  \begin{equation}
    \label{eq:d2n.series}
    \Lambda_{ij}(z) = \Lambda_{ij}(0)+
     (-1)^j\sum_{ \lambda \in \spec[red]{\laplacianD{G_*}}}
            \frac {z \Psi_{ij}(\lambda)}
                   {\lambda(\lambda-z)}
  \end{equation}
  where $\Psi_{ij}(\lambda) := \Psi_i(\lambda) \cdot \Psi_j(\lambda) =
  \sum_{k,\lambda_k=\lambda} \phi_k^\dag(o_i)\phi_k^\dag(o_j)$.  In
  addition, we have $\Lambda_{10}(z)=-\Lambda_{01}(z)$.
\end{lemma}
\begin{proof}
  The series representation~\eqref{eq:d2n.series} and the absolutely
  convergence follows from~\eqref{eq:dir.sm.b} and the fact that
  $d_0=(\cdot)^\dag(o_0)$ is continuous on $\Sob[2]{G_*}$
  by~\eqref{eq:sob'} (a similar \emph{nonuniform} estimate holds for
  $d_1$). Note that if $\lambda \in \spec{\laplacianD{G_*}} \setminus
  \spec[red]{\laplacianD{G_*}}$, then $\Psi_0(\lambda)=0$ and
  $\Psi_1(\lambda)=0$, i.e., the pole $\lambda$ does not appear in the
  series. The last statement follows from~\eqref{eq:d2n.series} once
  we have $\Lambda_{01}(0)=-\Lambda_{10}(0)$: To see the last
  equality, note that if $f=H_0(F_0,F_1)$ is a harmonic function with
  boundary values $F_0$ and $F_1$, then
  \begin{equation*}
    0 =
    \iprod {\laplacian[\max]{G_*} f} f -
             \iprod f {\laplacian[\max]{G_*} f} =
    \iprod[\C^2] {\Lambda(0)J \vec F} {\vec F} -
    \iprod[\C^2] {\vec F}{\Lambda(0)J \vec F}
  \end{equation*}
  where $J= \begin{pmatrix} -1 & 0 \\ 0 & 1 \end{pmatrix}$. In
  particular, $\Lambda(0)^* = J\Lambda(0)J$ and since $\Lambda(0)$ is
  a real matrix, we obtain the claim.
\end{proof}
\begin{remark}
  \begin{enumerate}
  \item Note that $\Lambda(z) \ne \Lambda(z)^{\tr}$ due to our
    definition of $f^\dag$, with has a different orientation at $o_0$
    and $o_1$.
  \item We have seen the phenomena that the set of poles of
    $\Lambda(z)$ is smaller than the Dirichlet spectrum already in the
    necklace decoration model (see~\Footnote{red.dir}). For the $p-1$
    linearly independent eigenfunctions $\phi_i$ with eigenvalue
    $\lambda=(\pi k/\ell)^2$ living on the $p$ loop edges we have
    $\phi_i=0$ near $o_0$ and $\phi_i^\dag(o_1)=0$ since
    $(\cdot)^\dag(o_1)$ is the sum over all edges meeting in $o_1$. In
    particular, the boundary derivative vectors $\Psi_i(\lambda)$
    vanish for $i=0,1$.
\end{enumerate}
\end{remark}

In order to define the transfer matrix, we need the following notion:
\begin{definition}
  \label{def:non.sep}
  We say that $\lambda \in \C$ is a \emph{separating energy for $G_*$}
  if there exists a nontrivial solution $\phi$ of the equation
  $(\laplacian[\max]{G_*}-\lambda)\phi=0$ such that $\vec \Phi_0=\vec
  0$ or $\vec \Phi_1=\vec 0$ where $\vec
  \Phi_i=(\phi,\phi^\dag)(o_i)$.  Denote
  \begin{equation}
    \label{eq:sep.en}
    \wt E(G_*):= \bigset {\lambda \in \C}
         {\text{$\lambda$ is a separating energy for $G_*$}}
  \end{equation}
  the set of separating energies for $G_*$.

  We call $\lambda$ an \emph{exceptional energy $G_*$} iff $\lambda$
  is an element of
  \begin{equation}
    \label{eq:exc.g}
    E(G_*):= \bigset {\lambda \in \C \setminus \spec {\laplacianD{G_*}}}
         {\Lambda_{01}(\lambda)=0 } \cup \spec{\laplacianD{G_*}}
  \end{equation}
  the set of exceptional energies.
\end{definition}

We will see in the next two lemmas, that for nonexceptional energies, we
can uniquely define the transfer matrix.

\begin{lemma}
  \label{lem:exc.set}
  \begin{enumerate}
  \item \label{sep-en} $E(G_*) \setminus \spec{\laplacianD{G_*}}
    \subset \wt E(G_*) \subset \R$.
  \item \label{tm} Let $z \notin E(G_*)$, i.e., $z$ is not in the
    Dirichlet spectrum and $\Lambda_{01}(z) \ne 0$.  Then for each
    $\vec F_0 \in \C$ there exists a unique solution of the equation
    $\laplacian[\max]{G_*} f = z f$ with $f \in \Sob[2]{G_*}$ and
    $\vec F_0=(f(o_0),f^\dag(o_0))$. We set $\vec F_1=(f(o_1),
    f^\dag(o_1)) \in \C^2$ and denote the solution by $T_z(x) \vec
    F_0:=T_z(x,G_*) \vec F_0 = f(x)$ for $x \in G_*$.
    
    The \emph{transfer} or \emph{monodromy} matrix $T_z=T_z(G_*)$ is
    uniquely defined by $\vec F_1=T(z) \vec F_0$.  The transfer matrix
    is unimodular, i.e., $\det T_z(G_*)=1$, and satisfies
  \begin{equation}
    \label{eq:def.tm}
    T_z = T_z(G_*) = \frac 1 {\Lambda_{01}(z)}
    \begin{pmatrix}
      -\Lambda_{00}(z) & 1\\
      -\det \Lambda(z) & \Lambda_{11}(z)
    \end{pmatrix}.
  \end{equation}
  The transfer matrix is still uniquely defined for $\lambda \in \spec
  {\laplacianD{G_*}} \setminus \spec[red] {\laplacianD{G_*}}$ and
  $\Lambda_{01}(\lambda)\ne 0$, although the solution ``map''
  $T_\lambda(\cdot)$ is no longer uniquely defined.

\item \label{tm.cont} Suppose that $\lambda_k \in
  \spec[red]{\laplacianD{G_*}}$ is a simple eigenvalue such that
  $\phi_k^\dag(o_0) \ne 0$ and $\phi_k^\dag(o_1) \ne 0$. Then the
  transfer matrix $T_z$ has an analytic continuation into $\lambda_k$
  given by
  \begin{equation}
    \label{eq:def.tm2}
    T(\lambda_k) =
    \begin{pmatrix}
      \dfrac{\phi_k^\dag(o_0)} {\phi_k^\dag(o_1)} & 0\\
      T_{21} (\lambda_k)  & \dfrac{\phi_k^\dag(o_1)} {\phi_k^\dag(o_0)}
    \end{pmatrix}.
  \end{equation}
   with
   \begin{multline}
     \label{eq:def.tm21}
     T_{21}(\lambda_k)=
     \frac 1 {\phi_k^\dag (o_0)\phi_k^\dag (o_1)}
       \Bigl(-z\sum_{n\ne k}
         \frac{ \bigl( \phi_k^\dag (o_0)\phi_n^\dag (o_1) -
                \phi_k^\dag (o_1)\phi_n^\dag (o_0) \bigr)^2}
              {\lambda_k(\lambda_n-\lambda_k)} + \\
      \phi_k^\dag(o_0)^2 \Lambda_{11}(0) -
      \phi_k^\dag(o_1)^2 \Lambda_{00}(0) -
      2 \phi_k^\dag(o_0) \phi_k^\dag(o_1) \Lambda_{01}(0)\Bigr).
   \end{multline}
\end{enumerate}
\end{lemma}
\begin{proof}
  \eqref{sep-en}~A separating energy is an eigenvalue for a
  self-adjoint operator, i.e., $\wt E(G_*) \subset \R$.  Note that the
  boundary condition depends on the fixed solution $\phi$. Let
  $\lambda \in E(G_*)$ and $\lambda \notin \spec{\laplacianD{G_*}}$ %% . If
$\lambda \in \spec[red] {\laplacianD{G_*}}$
  then $\Lambda_{01}(\lambda)=0$ and therefore also
  $\Lambda_{10}(\lambda)=0$, so that $\Lambda(\lambda)$ is a diagonal
  matrix.  In this case, there exist two linearly independent
  solutions $\phi^{(0)}$, $\phi^{(1)}$ of the eigenvalue equation such
  that
  \begin{align*}
    \vec \Phi_0^{(0)} =
      \begin{pmatrix} 1\\ \Lambda_{00}(\lambda) \end{pmatrix}, \quad
    \vec \Phi_1^{(0)} = \begin{pmatrix} 0\\ 0 \end{pmatrix}, \quad
    \vec \Phi_0^{(1)} = \begin{pmatrix} 0\\ 0 \end{pmatrix}, \quad
    \vec \Phi_1^{(1)} =
      \begin{pmatrix} 1\\ \Lambda_{11}(\lambda) \end{pmatrix}
  \end{align*}
  where $\vec \Phi_i^{(j)}=(\phi^{(j)},\phi^{(j)}{}^\dag)(o_i)$.
  
  \eqref{tm}~If $\Lambda_{01}(z) \ne 0$ then a simple calculation
  shows that the transfer matrix $T_z(G_*)$ is given
  by~\eqref{eq:def.tm}.  Furthermore, $\det T_z(G_*)=1$ since
  $\Lambda_{10}(z)=-\Lambda_{01}(z)$. Note that if $\lambda \in \spec
  {\laplacianD{G_*}} \setminus \spec[red] {\laplacianD{G_*}}$ then the
  derivatives of all Dirichlet solutions with eigenvalue $\lambda$
  vanish at both boundary points $o_0$ and $o_1$ and can therefore be
  added to a solution $f$ without infecting the boundary vectors $\vec
  F_0$ and $\vec F_1$ and in particular, the transfer matrix
  $T_\lambda$.
  
  \eqref{tm.cont}~The last assertion follows by a straightforward
  calculation.
\end{proof}

\begin{remark}
   \label{rem:exc.set}
   \begin{enumerate}
   \item \label{exc.set.real} We do not show in general that $E(G_*)$
     is discrete, but this is always fulfilled in our examples: The
     only point to check for the discreteness is that
     $\Lambda_{01}(z)$ is not constant.
   \item
     \label{nonseparating} The name \emph{separating energy}
     comes from the fact that if e.g.~$\lambda \in E(G_*) \setminus
     \spec{\laplacianD{G_*}}$ then $\Lambda_{01}(\lambda)=0$; we have
     seen in the proof that there exist two independent separating
     solutions. In particular, the recursion equation is
     ``interrupted'' or ``separated'' at such a decoration graph.
   \item We do not give the possible extension of the solution map
     $T_z(\cdot)$ into (parts) of the Dirichlet spectrum, although the
     norm estimate~\eqref{eq:norm.sol.map} in the next lemma is quite
     rough. But in our applications, it does not matter if our
     exceptional set is larger then necessary.
   \end{enumerate}
\end{remark}
Next, we give an expression for the solution of the eigenvalue
equation on $G_*$ in terms of $\vec F(0)$. Its proof follows
immediately from \Lem{dir.sol.map} and a simple calculation.
\begin{lemma}
  \label{lem:l2.L2}
  Let $z \notin E(G_*)$ then the solution map
  $\map{T_z(\cdot)}{\C^2}{\Sob[2]{G_*}}$ defined in
  \Lemenum{exc.set}{tm} is given by
  \begin{equation*}
    T_z(\cdot)
    \begin{pmatrix}
      F_0 \\ F'_0
    \end{pmatrix} = H_z
    \begin{pmatrix}
      F_0 \\ F_1
    \end{pmatrix}
    \quad\text{with}\quad
    \begin{pmatrix}
      F_0 \\ F_1
    \end{pmatrix}:=
    \begin{pmatrix}
      1 & 0 \\
      -\dfrac{\Lambda_{00}(z)}{\Lambda_{01}(z)} &
        \dfrac 1 {\Lambda_{01}(z)}
    \end{pmatrix}
    \begin{pmatrix}
      F_0 \\ F'_0
    \end{pmatrix}
  \end{equation*}
  and defines a continuous map from $\C^2$ into $\Lsqr {G_*}$,
  respectively, $\Sob[2]{G_*}$.  Its norm as map into $\Lsqr{G_*}$ is
  bounded by
  \begin{equation}
    \label{eq:norm.sol.map}
    \norm{T_z(\cdot)} \le
    \Bigl( 1+ \frac {1+ |z|} {d(z, \spec{\laplacianD{G_*}})}\Bigr)
    \norm E
    \Bigl( 1 + \frac {|\Lambda_{00}(z)|+1}{|\Lambda_{01}(z)|}\Bigr)
  \end{equation}
  where the norm $\norm E$ of the extension operator~\eqref{eq:ext.op}
  is bounded by a universal polynomial of degree $2$ in $1/\ell_-$.
\end{lemma}

We now consider a sequence of decoration graphs $G_n$, attached to a
line-like graph $L$:

\begin{definition}
  \label{def:non.sep.l}
  For a line-like graph $L$ consisting of the concatenations of
  $\{G_n\}_n$, we say that $\lambda \in \R$ is an \emph{exceptional
    energy for $L$} if $\lambda$ is an exceptional energy for at least
  one decoration graph $G_n$.  Denote
  \begin{equation}
    \label{eq:exc.l}
    E(L):= \bigcup_n E(G_n)
  \end{equation}
  the set of all exceptional energies for $L$.
\end{definition}

\begin{definition}
  \label{def:ess.nc}
  Let $f$ be a generalized eigenfunction associated to the eigenvalue
  $z$ in the sense of \Def{gen.ef}, but without condition at the
  vertex $0$ and no decay condition at $\infty$. We set $\vec F(v):=
  (f(v),f^\dag(v))$ and $\vec F(n):= \vec F(n_+)$, where $n_+$ is the
  starting vertex of
  $G_{n+1}$.

  A generalized eigenfunction is called \emph{essentially noncompactly
    supported} if there is $n_0 \in \N$ such that $\set{n \in \N}{\vec
    F(n) \ne \vec 0} = \set {n \in \N}{n \ge n_0}$ and any linear
  combination of $f$ does not contain a compactly supported
  eigenfunction.
\end{definition}
  The next lemma makes an assertion about the dimension of the space
  of generalized eigenfunctions:
\begin{lemma}
  \label{lem:dim.ef}
  If $z \notin E(L)$, i.e., if $z$ is nonexceptional for the line-like
  graph $L$, then the the space of generalized eigenfunctions $f$
  (without condition at $0$ and $\infty$) with eigenvalue $z$ is
  completely determined by the solution space of the recursion
  equation
  \begin{equation*}
    \vec F(n) = D(b_n) T_z(G_n) \vec F(n).
  \end{equation*}
  In addition, the solution space is two-dimensional.  Finally, if
  $\lambda \in E(L)$, then the space of essentially not-compactly
  supported generalized eigenfunctions has dimension at most $2$.
\end{lemma}
\begin{proof}
  The first assertion is a simple consequence of \Lem{exc.set}. The
  second assertion follows from the fact that an essentially noncompactly
  supported eigenfunction $f$ is completely determined by its start
  vector (the first nonvanishing vector $\vec F(n_0)$.
\end{proof}

The next lemma is a simple consequence of the definition of the
transfer matrix:
\begin{lemma}
  \label{lem:wronskian}
  Suppose that $f$ and $g$ are two generalized eigenfunctions in the
  sense of \Def{gen.ef} associated to the same eigenvalue $\lambda$,
  but without condition at the vertex $0$. Then the so-called
  \emph{Wronskian}
  \begin{equation}
    \label{eq:wronskian}
    W(f,g)(n):= f^\dag(n_+) g(n_+) - f(n_+) g^\dag(n_+)
  \end{equation}
  is independent of $n$.  In addition, if $\lambda \notin E(L)$,
  $W(f,g)=0$, $f \ne 0$ and $f(0)=0$ then also $g(0)=0$.
\end{lemma}
\begin{proof}
  The generalized eigenfunctions can be constructed via
  \begin{equation*}
    \vec F_\lambda(n) = U_\lambda(n) \vec F(0)
      \qquad \text{and} \qquad
    \vec G_\lambda(n) = U_\lambda(n) \vec G(0).
  \end{equation*}
  In particular, the Wronskian is given by
  \begin{multline*}
    W(f,g)(n) =
    \det (\vec F_\lambda(n),\vec G_\lambda(n)) =
    \det U_\lambda(n) (\vec F_\lambda(0),\vec G_\lambda(0)) \\ =
    \det U_\lambda(n) W(f,g)(0)
  \end{multline*}
  and the result follows from the fact that the transfer matrices
  $U_\lambda(n)=T_\lambda(G_n) \cdot \ldots \cdot T_\lambda(G_1)$ are
  unimodular (cf.~\eqref{eq:def.tm}).

  The last assertion follows from the fact, that a generalized
  eigenfunction $f$ is uniquely determined by $\vec
  F=(f(0),f^\dag(0))$ if $\lambda \notin E(L)$ (see \Lem{dim.ef}), so
  $f \ne 0$ implies $\vec F(n_+) \ne 0$ for some $n$ (\Lem{exc.set})
  and therefore $\vec F(0) \ne 0$. Finally, $f(0)=0$ implies
  $f^\dag(0) \ne 0$ and therefore also $g(0)=0$ since $W(f,g)=0$.
\end{proof}

%----------------------------------------------------------------------
\subsection{Weyl-Titchmarsh functions}
\label{sec:weyl}
%----------------------------------------------------------------------

In this section we define the Weyl-Titchmarsh function associated to
an operator on a line-like graph $L=L_0$. It will encode the
corresponding spectral measure. This function will be used in
\App{sp.av} in order to show that the corresponding \emph{averaged}
measure is absolutely continuous with respect to the Lebesgue measure.

The Weyl-Titchmarsh function is defined as follows: Let $\psi=\psi_z$ be a
generalized eigenfunction solving
\begin{equation}
  \label{eq:gen.ef3}
  H \psi = z \psi
\end{equation}in the sense of \Def{gen.ef}, but without
condition at the vertex $0$. Here, $z \in \C_+:=\set {z \in \C}{\Im z
  > 0}$ is in the upper half-plane. The aim of the next lemma is to
show that there is exactly one such function $\psi$ satisfying $\psi
\in \Lsqr L$ and $\psi(0)=1$.

To give~\eqref{eq:gen.ef3} a proper meaning, we let $H$ be the maximal
operator $H^{\max}$ as defined in~\eqref{eq:sob.max} with domain
$\dom H^{\max} = \Sob[2] L$. We derived an equivalent characterization
of $\Sob[2] L$ in \Lem{graph.norm}.
\begin{lemma}
  \label{lem:def.weyl}
  For each $z \in \C_+$ there exists exactly one function $\psi=\psi_z
  \in \dom H^{\max}$ such that $H^{\max} \psi=z\psi$, $\psi(0)=1$ and
  $\psi \in \Lsqr L$.
\end{lemma}
\begin{proof}
  We use an argument similar to the definition of the
  solution map in~\eqref{eq:dir.sm1}. Let $\wt \psi$ be a
  function in $\Sob[2] L$ with compact support such that $\psi(0)=1$.
  We set
  \begin{equation*}
    \psi_z := \wt \psi - (H^\Dir - z)^{-1} (H^{\max} - z) \wt \psi.
  \end{equation*}
  Here, $H^\Dir$ is the \emph{Dirichlet} Hamiltonian as defined in
  \Def{ll.graph.qg}. The function $\psi_z$ is well-defined, in $\Lsqr L$
  and satisfies the eigenvalue equation. Furthermore, since functions
  in the domain of the Dirichlet Hamiltonian vanish at $0$, we also
  have $\psi_z(0)=\wt \psi(0)=1$. This proves the existence of
  $\psi_z$.

  Uniqueness follows from the fact that $H^\Dir$ is self-adjoint:
  Suppose there is another function $\hat \psi_z \in \Sob[2] L$ solving
  the eigenvalue equation. Then $u:=\hat \psi_z - \psi_z \in \Lsqr L$
  is also a nontrivial solution of~\eqref{eq:gen.ef3} and $u(0)=0$.
  In particular, $u \in \dom H^\Dir$. Since a self-adjoint operator
  cannot have a nonreal eigenvalue, we have $u=0$, i.e, $\hat
  \psi_z=\psi_z$.
\end{proof}

We can now define the \emph{Weyl-Titchmarsh} function as\footnote{For
  the notation $(\cdot)^\dag$ see~\eqref{eq:gen.der}.}
\begin{equation}
  \label{eq:def.weyl}
  m(z):= \frac {\psi_z^\dag(0)}{\psi_z(0)} = \psi_z^\dag(0),
\end{equation}
due to normalization.  Note that $m$ is an analytic function on
$\C_+$. The next lemmas will show that $m$ maps $\C_+$ into $\C_+$,
i.e., that $m$ is a \emph{Herglotz} function.\footnote{A
  \emph{Herglotz} or \emph{Nevanlinna} function $m$ is an analytic
  function on the upper half-plane $\C_+$ such that $\Im m(z)>0$ for
  all $z \in \C_+$, or, equivalently, an analytic function $\map m
  {\C_+} {\C_+}$.}
\begin{lemma}
  \label{lem:herglotz}
  We have $m(z) = \Im z \normsqr[L]{\psi_z}$. In particular, $m$ is a
  Herglotz function.
\end{lemma}
\begin{proof}
  We have
  \begin{equation*}
    \iprod[L_{0,n}] \psi {H^{\max} \psi} -
    \iprod[L_{0,n}]  {H^{\max} \psi} \psi =
    W(\conj \psi, \psi)(n_-) - W(\conj \psi, \psi)(0)
  \end{equation*}
  since all other boundary terms vanish due to the inner vertex
  boundary conditions. Now, the left hand side equals $2\im \, \Im z
  \normsqr[L_{0,n}] \psi$, and $-W(\conj \psi,\psi)(0)= 2 \im \, m(z)$
  and
  \begin{equation*}
   |W(\conj \psi,\psi)(n_-)| \le 2 |(\psi^\dag \psi)(n_-)| =
    2 |(\psi^\dag \psi)(n_+)| \to 0
  \end{equation*}
  using the boundary condition at $n$ and \Lem{f'.int}. The result
  follows as $n \to \infty$.
\end{proof}

We now want to relate the Weyl-Titchmarsh function with (a component
of) the spectral measure $\rho$ associated to $H:=H^\Dir$. To do so we
need the Green's function near the connecting vertices $n$. Let
\begin{equation}
  \label{eq:split}
  L_\Lsplit :=
  \set{x \in L} { \text{$L \setminus \{x\}$ has two disjoint components
      $L_{0,x}$ and $L_{x,\infty}$ } }.
\end{equation}
In particular, points on a loop of the graph do not lie in
$L_\Lsplit$. {}From Assumption~\eqref{eq:start.vx2} it follows that $n \in
L_\Lsplit$ is \emph{not} an isolated point. In particular, $n_+$ is
always succeeded by an interval contained in $L_\Lsplit$.
\begin{lemma}
  \label{lem:green}
  For \emph{nonisolated} points $x,y$ in $L_\Lsplit$ we have
  \begin{equation}
    \label{eq:green2}
    G_z(x,y) = (s_z \wedge \psi_z)(x,y)
  \end{equation}
  where
  \begin{equation*}
    (f\wedge g)(x,y):=
    \begin{cases}
      f(x) g(y) & \text {if $y \in L_{x,\infty}$,}\\
      f(y) g(x) & \text {if $y \in L_{0,x}$,}
    \end{cases}
  \end{equation*}
  $s=s_z$ is the (unique)\footnote{Note that $z \in \C_+$ is always a
    \emph{nonexceptional energy (cf.~\Def{non.sep}).}}  generalized
  eigenfunction with $s_z(0)=0$ and $s_z^\dag(0)=1$, and $G_z(x,y)$ is
  the Green's function for $z \in \C_+$, i.e., the kernel of
  $(H-z)^{-1}$.
\end{lemma}
\begin{proof}
  Let $f \in \Lsqr L$ and
  \begin{equation*}
    g(x):= \int_L (s \wedge \psi)(x,y) f(y) \dd y =
    \int_{L_{0,x}} s(y) f(y) \dd y \, \psi(x) +
    \int_{L_{x,\infty}} \psi(y) f(y) \dd y \, s(x).
  \end{equation*}
  It is easy to see that $g$ is smooth on the interior of $L_\Lsplit$,
  that $g$ satisfies the boundary conditions at those vertices in
  $L_\Lsplit$ (here, we need the fact that $x$ is not isolated in
  $L_\Lsplit$ in order to apply a limit argument).  Furthermore, a
  simple calculation shows that $-g''(x) - z g(x) = W(s,\psi)(x) f(x)
  = f(x)$ since the Wronskian is constant on $L_{\Lsplit}$ (see
  \Lem{wronskian}) and equals $1$ due to our boundary condition at
  $x=0$.  The Green's function is pointwise defined, smooth away from
  the vertices and satisfies the boundary conditions at each vertex in
  $x$ and $y$ separately (cf.~\Cor{green}). In particular, $g(x)=
  \int_L G_z(x,y) f(x) \dd y$ for all $x \in L$. Since a continuous
  kernel is uniquely defined,~\eqref{eq:green2} follows for
  nonisolated points in $L_\Lsplit$.
\end{proof}

In the general graph-decorated case, it may happen that $\delta_0'$ is
not a cyclic vector. In this case, one has to assure that the spectral
measure on the complement can only be pure point:

\begin{lemma}
  \label{lem:m-fct}
  We have
  \begin{equation}
    \label{eq:green.der}
    m(z) =
    \partial_{xy} G_z(0,0) =
    \int_\R \frac 1 {\lambda-z} \dd \hat \rho(\lambda)
  \end{equation}
  where $\dd \hat \rho(\lambda):= \sum_j|\phi_{\lambda,j}'(0)|^2 \dd
  \rho(\lambda)$.  In addition, there is a countable set $E_{\mathrm
    {pp}} \subset \R$ such that the measure $\hat \rho +
  \rho_{\mathrm{pp}}$ is a spectral measure for $H$ and
  $\rho_{\mathrm{pp}}$ is pure point and a spectral measure for $H
  \1_{E_{\mathrm{pp}}}(H)$. In particular, on a tree graph,
  $E_{\mathrm{pp}}= \emptyset$, i.e., $\hat \rho=\rho$ itself is a
  spectral measure for $H$.
\end{lemma}
\begin{proof}
  The first equality follows from~\eqref{eq:green2}, the fact that $0$
  is not isolated in $L_\Lsplit$ (cf.~\eqref{eq:start.vx2}) and the
  definition of $m(z)$ in ~\eqref{eq:def.weyl}.

  For the second equality, we use the pointwise representation of
  \Cor{green} and~\eqref{eq:kernel.gen.ef} for $x,y=0$. We set
  $E_{\mathrm{pp}} := \set{\lambda}{\partial_{xy}\wt
    e_\lambda(0,0)=\sum_j |\phi_{\lambda_j}^\dag(0)|^2=0}$. In
  \Lem{kernel.gen.ef} we have seen that $E_{\mathrm{pp}} \subset
  E(L)$, where $E(L)$ is the \emph{countable} set of exceptional
  energy values (see \Lem{exc.set}).

  Since the derivative of the kernel in~\eqref{eq:kernel.gen.ef} is
  also defined at $x,y=0$, we have
  \begin{equation}
    \label{eq:sp.meas}
    \hat \rho(I) =
    (\delta_0', \1_I(H) \delta_0') =
    \int_I \partial_{xy} \wt e_\lambda(0,0) \dd \rho(\lambda)
  \end{equation}
  which shows that $\hat \rho$ is a spectral measure for the part of
  the operator $H$ on the complement of $E_{\mathrm{pp}}$. Clearly,
  the countable set $E_{\mathrm{pp}}$ can only support a pure point
  measure, i.e., $\rho_{\mathrm {pp}}(I) := \rho(I \cap
  E_{\mathrm{pp}})$ is a pure point measure and $\hat \rho +
  \rho_{\mathrm {pp}}$ is a spectral measure for the whole operator
  $H$. Here, $(\cdot, \cdot)$ is the dual pairing of the Hilbert
  scales $\HS_{-2}$ and $\HS_2 := \dom H$. In addition, we set
  $\delta_0' f := f^\dag(0)$.  Note that $\delta_0' \in \HS_{-2}$ due
  to \eqref{eq:sob'} and \eqref{eq:sob''}.
\end{proof}

%----------------------------------------------------------------------
%
\section{Spectral averaging}
\label{app:sp.av}
%
%----------------------------------------------------------------------

Using the Weyl-Titchmarsh function, we want to prove a spectral
averaging formula in the sense that integrating the spectral measure
of $H=H(\omega)$ on $L_0$ with respect to the first random variable
$\omega_1 \in \Omega_1$ yields in a measure \emph{absolutely
  continuous} with respect to the Lebesgue measure.  The
Weyl-Titchmarsh function associated to $H=H(\omega)$ (see \Sec{weyl})
has the advantage that there is a formula (cf.~\eqref{eq:weyl.split1})
separating the first random variable $\omega_1 \in \Omega_1$ from the
other ones $\hat \omega =(\omega_2,\dots) \in \hat \Omega$ where
$\omega=(\omega_1,\hat \omega) \in \Omega=\Omega_1 \times \hat
\Omega$.

Let $L=L(\omega)$ be a random line-like graph as defined in
\Sec{reduct.gen}.  For a unimodular matrix $A \in \SL_2 (\C)$ we denote
by $\map {\proj A} \C \C$ the corresponding M\"obius transformation
associated to $A$, i.e., if
\begin{equation*}
   A :=
   \begin{pmatrix}
     a & b \\ c & d
   \end{pmatrix}\in \SL_2(\C)
   \qquad \text{then} \qquad
   \proj A m := \frac {c + dm} {a + bm}
\end{equation*}
for $m \ne -a/b$. Our definition of $\proj A$ differs from the
standard one due to the fact that the projection of $\vec
\Psi=(\psi,\psi') \in \C^2$ ($\vec \Psi \ne 0$) onto the complex
projective line is $\proj{\vec \Psi}:=\psi'/\psi$.  We use this
convention since the transfer matrix $A$ acts on $\vec
\Psi:=(\psi,\psi^\dag)$ and the Weyl-Titchmarsh function associated to
$H=H(\omega)$ on the line-like graph $L=L(\omega)$ is given by
$m(z)=\proj {\vec \Psi_z} = (\psi_z^\dag/\psi_z)(0)$.

We denote by $T_z(\omega_1)$ the transfer matrix of a single graph
decoration (see \Lem{exc.set}). Note that the transfer matrix is
defined for all $z \in \C \setminus \R$.

Our main tool in this section will be the following estimate. Let
$\lambda_\pm \in \R$.
\begin{definition}
  We say that spectral averaging holds in the compact set $\Sigma_0
  \subset [\lambda_-,\lambda_+]$ if for $C_4>0$ and $\eps_0>0$ there
  exists a constant $C_3=C_3(\lambda_\pm, \eps_0,\Omega_1,C_4)$ such
  that
  \begin{equation}
    \label{eq:weyl.bdd}
    \int_{\Omega_1} \Im \bigl( \bigproj{T_z(\omega_1)^{-1}} m \bigr)
           \dd \Prob_1(\omega_1) \le C_3
  \end{equation}
  for all $z=\lambda + \im \eps \in \Sigma_0 \times \im
  (0,\eps_0]$ and all $m \in \C_+$ such that $\eps |m| \le C_4$ and
  $\Im (\proj{T_z(\omega_1)^{-1}} m) >0$.
\end{definition}

We will see in~\eqref{eq:moeb1}--\eqref{eq:moeb2} that the M\"obius
transformation in~\eqref{eq:weyl.bdd} has no poles in $\C_+$.
Furthermore, we will relate this estimate to the Weyl-Titchmarsh
function $m(z)$ for the line-like graph $L=L_0$. Here and in the
sequel, $\psi=\psi_z$ is the unique eigenfunction $H\psi=z\psi$ with
$\psi(0)=1$ and $\psi \in \Lsqr L$.  If $L=L(\omega)$, then $\psi_z$
also depends on $\omega$. More generally, we define the
Weyl-Titchmarsh function for the line-like \emph{subgraph} $L_n$ (see
\Sec{reduct.gen}) as
\begin{equation}
  \label{eq:weyl.sub.gr}
  m_n(z) := (\psi_z^\dag / \psi_z)(n_+).
\end{equation}
\begin{lemma}
  The function $m_n$ is a Herglotz function, i.e., $m_n$ is the
  Weyl-Titchmarsh function for $L_n$. In addition, $m_n(z)$ only
  depends on the random variables $\hat \omega_n :=
  (\omega_{n+1},\omega_{n+2},\dots)$, and is given explicitly by
  \begin{equation}
    \label{eq:weyl.split}
    m_n(z,\hat \omega_n) =
    \proj {T_z(\omega_n)} m_{n-1}(z,\hat \omega_{n-1})
  \end{equation}
\end{lemma}
\begin{proof}
  The proof that the solution space of $Hu=zu$ on $L_n$ is
  one-dimensional is the same as the proof of~\Lem{def.weyl}. In
  particular, the solution is determined by its value at $\psi_z(n_+)$
  and $m_n(z)$ only depends on the data of $L_n$, i.e., $m_n(z)$ only
  depends on $\hat \omega_n$. {}From \Lem{herglotz} applied to $L_n$, we
  see that also $\Im m_n(z)>0$ for $z \in \C_+$. The last equality
  follows from the definition of the transfer matrix
  (cf.~\Sec{trans.mat}).
\end{proof}

In particular, we have
\begin{equation}
  \label{eq:weyl.split1}
    m_0(\omega,z) = \proj{T_z(\omega_1)^{-1}} m_1(\hat \omega,z)
\end{equation}
where $m_0$ is the Weyl-Titchmarsh function on $L=L_0(\omega)$. In
addition, $m_1$ is the Weyl-Titchmarsh function on $L_1=L_1(\hat
\omega)$ and $\hat \omega:= \hat \omega_1$, i.e.,
$\omega=(\omega_1,\hat \omega) \in \Omega = \Omega_1 \times \hat
\Omega$.
\begin{remark}
  \label{rem:tm.compl}
  Note that $\Im m_0(\omega,z)>0$ if $\Im m_1(\hat \omega,z)>0$
  although $T_z(\omega)$ generally has \emph{complex} entries. It is
  the nontrivial dependence of $z$ entering in $m_1$ \emph{and} the
  transfer matrix $T_z(\omega_1)$ which makes the quantum graph
  problem different from spectral averaging methods considered for
  other models before (see e.g.~\cite{gesztesy-makarov:03}) where
  usually only \emph{real} entries are considered.
\end{remark}

We can now prove the main result of this section:
\begin{theorem}
  \label{thm:sp.av}
  The spectral measure $\rho=\rho_\omega$ of $H=H(\omega)$ on the
  line-like graph $L=L(\omega)$ splits into two measures $\rho=\hat
  \rho + \rho_{\mathrm{pp}}$ where $\rho_{\mathrm {pp}}$ is pure point.

  In addition, if~\eqref{eq:weyl.bdd} holds in $\Sigma_0 \subset
  [\lambda_-,\lambda_+]$, then the measure $\hat \rho=\hat
  \rho_{\omega}$ averaged over the first random variable $\omega_1$ is
  absolutely continuous w.r.t.\ the Lebesgue measure, i.e., there is a
  constant $C_5=C_5(\lambda_\pm,\Omega_1)>0$ uniform in $\hat \omega$
  such that
  \begin{equation}
    \label{eq:sp.av}
    \int_{\Omega_1} \hat \rho_{(\omega_1,\hat \omega)}(I)
         \dd \Prob_1(\omega_1) \le C_5 \leb(I)
  \end{equation}
  for all measurable sets $I \subset \Sigma_0$, where
  $\leb$ denotes Lebesgue measure.
\end{theorem}
\begin{proof}
  {}From~\eqref{eq:green.der} and the theory of Herglotz functions
  (see e.g.~\cite[App.~A]{pastur-figotin:92})
  we have
  \begin{equation*}
    \hat \rho_\omega (I) =
    \lim_{\eps \to 0} \frac 1 \pi
        \int_I \Im m_\omega(\lambda + \im \eps) \dd \lambda
  \end{equation*}
  provided $\bd I$ does not contain eigenvalues of $H=H(\omega)$.
  Note that $\int_\R \frac 1 {1+|\lambda|} \dd \rho(\lambda)< \infty$
  by \Cor{desint} and \Lem{carleman}.  Now,
  \begin{multline*}
    \pi \int_{\Omega_1}
          \hat \rho_{\omega_1,\hat \omega}(I) \dd \Prob_1(\omega_1) =
    \int_{\Omega_1}
       \Bigl(\lim_{\eps \to 0}
       \int_I \Im m_0 \bigr(\lambda+\im \eps, (\omega_1,\hat \omega) \bigr)
           \dd \lambda
       \Bigr) \dd \Prob_1(\omega_1) \\ =
    \int_{\Omega_1}
        \Bigl( \lim_{\eps \to 0} \int_I
         \Im \proj{ T_z(\omega_1)^{-1}}
                m_1(\lambda+\im \eps, \hat \omega)
         \dd \lambda \Bigr)
            \dd \Prob_1(\omega_1).
  \end{multline*}
  Now $m_1(z,\hat \omega)$ is a Herglotz function and all components
  of $\hat \omega$ are iid\ random variables. In particular, there
  exists a constant $C_4=C_4(\lambda_\pm, \eps_0,\Omega_1)$ such that
  $\eps |m_1(\lambda+\im \eps,\hat \omega)| \le C_4$ for all $z \in
  \Sigma_0 + \im (0,\eps_0) \subset \C_+$.  Interchanging the first
  integral and the limit by Fatou's lemma, we use Fubini's theorem to
  exchange the order of integration and obtain
  from~\eqref{eq:weyl.bdd},
  \begin{equation*}
    \pi \int_{\Omega_1} \hat \rho_{\omega_1,\hat \omega}(I)
            \dd \Prob_1(\omega_1)
    \le C_3 \leb(I)
  \end{equation*}
  i.e., $C_5=C_3/\pi$.
\end{proof}

In the rest of this section we provide some criteria
guaranteeing~\eqref{eq:weyl.bdd}. In our applications, it will be more
convenient to use $w=\sqrt z$ as spectral parameter where we choose
the branch with $\Im \sqrt z > 0$, i.e., cut along $\R_+$. We write
the transfer matrix as
\begin{equation}
  \label{eq:tm.sp.av}
  T_z(t)= D(b) \hat T_z(t), \qquad
  \hat T_z(t) =
  \begin{pmatrix}
    t_{11}(t,w) & t_{12}(t,w)\\
    t_{22}(t,w) & t_{22}(t,w)
  \end{pmatrix}
\end{equation}
where $\hat T_z(t)=T_z(G_*(t))$ or $\hat T_z(t)=S(t) T_z(G_*)$ in a
random length, respectively, Kirchhoff model, denotes the transfer matrix of
the decoration graph $G_*(t)$ as defined in~\eqref{eq:def.tm}. We also
assume that $\Omega_1 =[t_-,t_+]$ and often write $t=\omega_1$ for the
integration parameter.

Let $\Ln$ be the complex logarithm on the infinite sheeted Riemann
surface $\wt \C^*$ with branching points at $0$ and $\infty$. For a
map $t \to a(t) \in \C^*:=\C \setminus \{0\}$ we denote by $t \to \wt
a(t)$ the lift of $t \mapsto a(t)$ onto $\wt \C^*$ such that $\wt
a(0)$ lies in the first sheet (given by the argument $0 \le \phi <
2\pi$). Note that if $a(t)=r(t)\e^{\im \phi(t)}$ for \emph{continuous}
functions $r(\cdot)$ and $\phi(\cdot)$, then $\Ln \wt a(t)= \ln r(t) +
\im \phi(t)$.  In particular, we decompose the denominator of the
M\"obius transformation $\proj{\hat T_z(t)^{-1}} m$, namely
\begin{equation}
  \label{eq:def.f}
    f_{w,m}(t):=
    t_{22}(t,w) - t_{12}(t,w) m =
    r_{w,m}(t) \e^{\im \phi_{w,m}(t)},
\end{equation}
into its polar decomposition with continuous functions $r_{w,m}$ and
$\phi_{w,m}$.

\begin{lemma}
  \label{lem:sp.av.int}
  Suppose that $\Prob_1$ has a bounded density on
  $\Omega_1:=[t_-,t_+]$ with respect to the Lebesgue measure, i.e.,
  $\dd \Prob_1(t)= \eta(t) \dd t$ and $0 \le \eta(t) \le \norm[\infty]
  \eta$ for almost all $t$. Suppose in addition, that there are
  complex constants $A_w$, $B_w \in \C$
  such that
  \begin{equation}
    \label{eq:integrand}
    \proj{\hat T_z(t)^{-1}} m =
    -A_w \frac {f'_{w,m}(t)}{f_{w,m}(t)} + B_w
  \end{equation}
  and measurable subsets $\Sigma_j \subset [\lambda_-,\lambda_+]$ with
  $\bigcup \Sigma_j = [\lambda_-,\lambda_+]$ up to a discrete set such
  that for all $w=\sqrt{\lambda+\im \eps}$ with $\lambda \in
  \Sigma_j$, $0<\eps<\eps_0$ we have
  \begin{enumerate}
  \item for each $j \in \N$, there is a constant
    $C_6=C_6(j,\lambda_\pm,\eps_0)>0$ such that
    \begin{equation*}
      |\Re A_w| \le C_6, \quad |B_w| \le C_6 \quad \text{and} \quad
      |\Im A_w| \le C_6 \, \eps;
    \end{equation*}
  \item the winding number is bounded, i.e., there exists $N>0$ such
    that $|\phi_{w,m}(t_+)-\phi_{w,m}(t_-)| \le N$ for all $m \in
    \C_+$.
  \end{enumerate}
  All constants and error estimates are supposed to depend only on
  $\Sigma_j$ and $\eps_0$. It suffices to choose $m \in \C_+$ such
  that $\eps |m| \le C_4$. Here, $C_6$ and $N$ may depend on $C_4$ but
  \emph{not} on $m$ directly.  Then there exist $\Sigma_j' \subset
  \Sigma_j$ such that $\bigcup_j \Sigma_j' = \Sigma_0$ up to a
  discrete set such that~\eqref{eq:weyl.bdd} is fulfilled in
  $\Sigma_j$ with
  \begin{equation*}
    C_3= C_3(j) =
    C_6 \norm[\infty] \eta
       \bigl(\eps O(|\ln \eps|)+ N  + (t_+-t_-) \bigr).
  \end{equation*}
  If $\Im A_w=0$ we can choose $\Sigma_j'=\Sigma_j$.
\end{lemma}
\begin{proof}
  From~\eqref{eq:integrand} we obtain
  \begin{multline*}
   \Im \int_{t_-}^{t_+} \proj{\hat T_z(t)^{-1}} m \, \eta(t) \dd t \le
   \norm[\infty] \eta
     \Im \bigl[-A_w \Ln \wt f_{w,m}(t) + B_w t \bigr]_{t_-}^{t_+} \\=
   \norm[\infty] \eta
  \Bigl[ -\Im A_w \ln \frac {r_{w,m}(t_+)}{r_{w,m}(t_-)} -
          \Re A_w \bigl(\phi_{w,m}(t_+)- \phi_{w,m}(t_-) \bigr) +
          B_w(t_+-t_-)
     \Bigr] \\ \le
   \norm[\infty] \eta
   \Bigl[ |\Im A_w| \Bigl| \ln \frac {r_{w,m}(t_+)}{r_{w,m}(t_-)} \Bigr| +
           |\Re A_w| \bigl|\phi_{w,m}(t_+)- \phi_{w,m}(t_-) \bigr| +
          |B_w| |t_+-t_-|
   \Bigr]
\end{multline*} so that the estimate follows from the assumptions
once we have shown that $r_{w,m}(t_+)$ is bounded from above by a
polynomial in $\eps^{-1}$ and that $r_{w,m}(t_-)$ is bounded from
below by a polynomial in $\eps$; uniformly for all $w =\sqrt
{\lambda+\im \eps}$, $\lambda \in \Sigma_j$, $0 < \eps\le \eps_0$
and for all $m \in \C_+$ such that $|m| \le C_4/\eps$ (the
polynomials may depend on $C_4$, but not on $m$ itself). Note that if
$\Im A_w=0$, we can skip the estimate on $r_{w,m}(t_\pm)$ and we are
done.

To estimate $r_{w,m}(t_\pm)$, we write
  \begin{equation}
    \label{eq:moeb1}
    f_{w,m}(t) =
    t_{12}(t,w) \Bigl( \frac {t_{22}(t,w)}{t_{12}(t,w)} -m \Bigr)=
    \frac 1 {\Lambda_{01}(t,z)} \bigl( \Lambda_{11}(t,z) - m \bigr)
  \end{equation}
  for $z \notin E(G_*(t))$ using~\eqref{eq:def.tm}. Note that
  $t_{12}(t,w)=1/\Lambda_{01}(t,z) \ne 0$ due to~\eqref{eq:def.tm}.
  The series representation~\eqref{eq:d2n.series} of the
  Dirichlet-to-Neumann map $\Lambda(t,z)$ of the decoration graph
  $G_*(t)$ yields
  \begin{equation}
    \label{eq:moeb2}
    \Im \frac{t_{22}(t,w)}{t_{12}(t,w)} =
    \Im \Lambda_{11}(t,z) =
    - \eps \sum_k \frac{|\phi_k^\dag(o_1)|^2}{|\lambda_k - z|^2} =:
    - \eps \, C_7(z).
  \end{equation}
  Now let $k_0$ be an index for which $\lambda_k > \lambda_+$ and
  $z_0:=\lambda_-+\im \eps_0$. Then
  \begin{equation}
    \label{eq:def.c8}
    C_7(z) \ge
    \sum_{k \ge k_0} \frac{|\phi_k^\dag(o_1)|^2}{|\lambda_k - z|^2} \ge
    \sum_{k \ge k_0} \frac{|\phi_k^\dag(o_1)|^2}{|\lambda_k - z_0|^2}
    =: C_8.
  \end{equation}
  We also need a lower bound on the module of $t_{12}=1/\Lambda_{01}$, i.e.,
  an upper bound on $|\Lambda_{01}(t,z)|$, namely
  \begin{equation*}
    |\Lambda_{01}(t_-,z)| \le
    \sum_k
       \frac{|z||\phi_k^\dag(o_0) \phi_k^\dag(o_1)|}
            {\lambda_k|\lambda_k - z|} + |\Lambda_{01}(t_-,0)|.
  \end{equation*}
  We restrict the values of $\lambda$ to the subset
  \begin{equation*}
      \Sigma_j':= \bigset{\lambda \in \Sigma_j}
        {|\Lambda_{01}(t,\lambda)| \ge 1/j \text{ and } |\lambda -
          \lambda_k|\ge 1/j \text{ for all $k$}, t=t_\pm}
  \end{equation*}
  and assume that $z=\lambda+\im \eps$ with $\lambda \in \Sigma_j'$.
  A compactness argument yields the existence of a constant $C_9>0$
  depending only on $j$, $t_-$ $\lambda_\pm$ and $\eps_0$ such that
  $|\Lambda_{01}(t_-,z)| \le C_9$.  Since $m \in \C_+$ and $\Im
  t_{22}/t_{11} \le -\eps \, C_8$, we deduce $r_{w,m}(t_-) \ge \eps
  C_8/C_9$.
  
  The upper bound can be obtained similarly: Here, we need an upper
  bound on $|t_{12}|$ and $|t_{22}/t_{12}|$, i.e., a lower bound on
  $|\Lambda_{01}|$ and an upper bound on $|\Lambda_{11}|$.  The upper
  bound $|\Lambda_{11}(t_+,z)| \le C_{10}$ for $z \in \Sigma_j' + \im
  (0,\eps_0]$ can be obtained as above for $\Lambda_{01}$. For the
  global lower bound on $\Lambda_{01}(z)=\Lambda_{01}(t_+,z)$ we have
  \begin{equation*}
    |\Lambda_{01}(z)| \ge
    |\Lambda_{01}(\lambda)| -
         \eps |\Lambda_{01}'(\lambda+\im \tau \eps)| \ge
    \frac 1 j - \eps C_{11}
  \end{equation*}
  where $z=\lambda+\im \eps$, $\tau \in (0,1)$ and $C_{11}$ is the
  maximum of $\Lambda_{01}'(z)$ in a compact set avoiding the poles of
  $\Lambda_{01}$ (where $\Lambda_{01}(z)$ is large).  Therefore, there
  exists $C_{12}=C_{12}(j)$ such that $|\Lambda_{01}(z)| \ge C_{12}$
  for all $z \in \Sigma_j \times (0,\eps_0]$ and for
  $\eps_0=\eps_0(j)$ small enough. Note that still $\bigcup \Sigma_j'
  = [\lambda_-,\lambda_+]$ up to a discrete set since by
  \Assenum{ran.graph}{exc.set},
  $\set{\lambda}{\Lambda_{01}(t_\pm,\lambda)=0}$ is discrete.
  Finally, we have shown $r_{w,m}(t_+) \le (C_{12})^{-1} (C_{10} +
  C_4/\eps) =O(\eps^{-1})$.
\end{proof}
\begin{remark}
  \label{rem:why.sp.av}
  Note that the constants defined in the proof below (for example
  $C_8$ in~\eqref{eq:def.c8}) still depends on the decoration graph
  $G_*(t_\pm)$ via the eigenvalues and eigenvectors of $G_*(t_\pm)$.  But
  here we see the advantage of the spectral averaging: After
  integrating, we only have to control the behavior at the points
  $t_\pm$ of the random space, \emph{not} a uniform estimate over all
  $t=\omega_1 \in \Omega_1$ (which is in general not possible). In
  fact, even if we would have global lower bounds on the denominator
  of the M\"obius transformation, we are usually not done, since the
  estimates are of order $\eps^{-1}$ and therefore unbounded as in the
  proof above.
\end{remark}

We will give two particular examples in which the spectral averaging
estimate can be deduced from \Lem{sp.av.int}.
%----------------------------------------------------------------------
\subsection*{Random length models}
%\label{sec:rlm.sp.av}
%----------------------------------------------------------------------
There is a particular simple form of the transfer matrix in certain
random length models: Suppose that $T_z(\ell)=D(b) \hat T_z \wt
T_z(\ell)$ where $\ell=\omega_1 \in \R$, $\hat T_z=(\hat t_{ij}(z))
\in \SL_2(\C)$ and $\ell \mapsto \wt T_z(\ell)=\e^{-\ell X_z}$ is a
one-parameter group in $\SL_2(\C)$ with $X_z \in \mathrm{sl}_2(\C)$,
the Lie algebra of $\SL_2(\C)$.  We assume that
\begin{equation}
  \label{eq:one-par}
  \wt T_z(\ell) =
  \begin{pmatrix}
    \wt t_{11}(\ell,w) & \wt t_{12}(\ell,w)\\
    \wt t_{21}(\ell,w) & \wt t_{22}(\ell,w)
  \end{pmatrix}
  \qquad \text{and} \qquad
  X_z =
  \begin{pmatrix}
    \beta_z & \alpha_z \\ \gamma_z & -\beta_z
  \end{pmatrix}.
\end{equation}
Using $\frac \dd {\dd \ell} T_z(\ell)=\wt T_z(\ell) X_z$ we obtain
(denoting $(\cdot)'$ the derivative w.r.t.\ $\ell$)
\begin{equation*}
  \wt t_{12}' = \alpha \wt t_{11} - \beta \wt t_{12} \qquad \text{and}
\qquad
  \wt t_{22}' = \alpha \wt t_{21} - \beta \wt t_{22}.
\end{equation*}
If $\alpha\ne0$, we can decompose
\begin{multline*}
  \proj{\wt T_z(-\ell) \hat T_z^{-1}} m =
  -\frac{\wt t_{21}(\hat t_{22} - \hat t_{21} m) +
        \wt t_{11}(\hat t_{21} - \hat t_{11} m) }
       {\wt t_{22}(\hat t_{22} - \hat t_{21} m) +
        \wt t_{12}(\hat t_{21} - \hat t_{11} m) } \\=
  -\frac{(\wt t_{22}' + \beta \wt t_{22})(\hat t_{22} - \hat t_{21} m) +
        (\wt t_{12}' + \beta \wt t_{12})(\hat t_{21} - \hat t_{11} m) }
       {\alpha \bigl(\wt t_{22}(\hat t_{22} - \hat t_{21} m) +
        \wt t_{12}(\hat t_{21} - \hat t_{11} m)\bigr) } =
      - \frac {f_{w,m}'(\ell)}{\alpha f_{w,m}(\ell)} -
      \frac \beta \alpha
\end{multline*}
where $f_{w,m}(\ell)$ denotes the denominator of the M\"obius
transformation so that $A_w:= 1/\alpha$ and $B_w:= -\beta/\alpha$ in
the notation of~\eqref{eq:integrand}.

Typically, $\ell$ denotes the length and $\wt T_z(\ell)=R_{pw}(w\ell)$
where $R_w(\phi)$ is defined in~\eqref{eq:def.el.mat} and $p>0$ is a
fixed parameter, so in particular, $\beta=0$, $\alpha=1/p \in \R$ and
$\gamma_z=p z \in \C$.  Then
\begin{equation}
  \label{eq:sol.1.par}
  \wt T_z(\ell)=
  \begin{pmatrix}
    \cos w \ell &  \dfrac {\sin w\ell} {pw} \\
    -pw \sin w\ell & \cos w \ell
  \end{pmatrix}
\end{equation}
where $w=\sqrt z$ and $\Re z, \Im z > 0$ (we choose the branch with
$\Im \sqrt z > 0$).  In this case, we obtain from the previous lemma:
\begin{corollary}
  \label{cor:sp.av.rlm}
  Assume that the single transformation matrix has the form
  \begin{equation*}
    T_z(\ell) = D(b) \hat T_z R_{pw}(w \ell).
  \end{equation*}
  Suppose in addition that $\Omega_1 =[\ell_-,\ell_+]$ and that
  $\Prob_1$ has a bounded density with respect to the Lebesgue
  measure, i.e., $\dd \Prob_1(\ell)= \eta(\ell) \dd \ell$ and $0 \le
  \eta(\ell) \le \norm[\infty] \eta$ for almost all $\ell$.
  Then~\eqref{eq:weyl.bdd} is fulfilled for all $\lambda \in
  [\lambda_-,\lambda_+]$.
\end{corollary}
\begin{proof}
  In our case, we have $A_w=1/\alpha=p$ and $B_w=0$. Furthermore, the
  winding number of $\wt f_{w,m}$ can be estimated by a fixed number
  depending only on $\lambda_\pm$ and $\ell_\pm$.  In particular, the
  assumptions of \Lem{sp.av.int} are fulfilled. Since $\Im A_w=0$, we
  can skip the estimate on the real part of the logarithm and do not
  need the exceptional sets $\Sigma_j$.
\end{proof}

In general, changing the length of a subgraph $G_n$ does not yield a
one-parameter group. For such general random length model the
integrand is in general a very complicated rational function in $w,
\sin w\ell$ and $\cos w \ell$.

%----------------------------------------------------------------------
\subsection*{Random Kirchhoff models}
%\label{sec:rkm.sp.av}
%----------------------------------------------------------------------
Suppose that $T_z(q)=D(b) S(q) \hat T_z$ where $S(q)$ is the shearing
matrix as in~\eqref{eq:def.el.mat}, where $q=\omega_1 \in
\Omega_1=[q_-,q_+]$ and where $\hat T_z$ is the transition matrix for
a (fixed) decoration graph, i.e., we assume a Kirchhoff model where
the vertex potential is at the end point of the decoration graph. A
simple calculation shows that
\begin{multline*}
  \proj{T_z(q)^{-1}} m =
  \proj {\hat T_z^{-1}} (-q + bm) =
  \frac {-t_{21} + t_{11} (-q + bm)} {t_{22} - t_{12} (-q+bm)} \\ =
  \frac 1 {t_{12}} \Bigl(
      \frac 1 {t_{22} - t_{12} (-q+bm)} - t_{11} \Bigr) =
  \frac 1 {(t_{12})^2} \, \frac {f_{w,m}'(q)}{f_{w,m}(q)} -
          \frac {t_{11}}{t_{12}}
\end{multline*}
with the notation of~\eqref{eq:tm.sp.av} ($t_{ij}=t_{ij}(q,w)$)
and~\eqref{eq:def.f}.

\begin{corollary}
  \label{cor:sp.av.rkm}
  Suppose that $\Omega_1 =[q_-,q_+]$ and that $\Prob_1$ has a bounded
  density with respect to the Lebesgue measure, i.e., $\dd \Prob_1(q)=
  \eta(q) \dd q$ and $0 \le \eta(q) \le \norm[\infty] \eta$ for almost
  all $q$.  Then there is a sequence $\Sigma_j' \subset
  [\lambda_-,\lambda_+]$ with $\bigcup_j \Sigma_j =
  [\lambda_-,\lambda_+]$ up to a discrete set such that
  ~\eqref{eq:weyl.bdd} is fulfilled for all $\lambda \in
  \Sigma_j$ with a constant $C_3$ depending on $j$.
\end{corollary}
\begin{proof}
  Again, we use \Lem{sp.av.int}. We have seen in the calculation above
  that $A_w:=1/(t_{12})^2=(\Lambda_{01}(z))^2$ and
  $B_w:=-t_{11}/t_{12}=\Lambda_{00}(z)$ (see~\eqref{eq:def.tm}). The
  upper bounds on $|\Re A_w|$ and $|B_w|$ can be found as in the proof
  of \Lem{sp.av.int}. Note in addition that $\Im A_w = 2 \Re
  \Lambda_{01}(z) \Im \Lambda_{01}(z) = O(\eps)$ using again the
  series representation of the Dirichlet-to-Neumann
  map~\eqref{eq:d2n.series}. The winding number is bounded by $\pi$
  since $q \mapsto f_{w,m}(q)$ describes a line in the complex plane.
\end{proof}

%----------------------------------------------------------------------
%\bibliographystyle{amsalpha}
%\bibliography{/home/post/Aktuell/BibTeX/literatur}
%----------------------------------------------------------------------

\providecommand{\bysame}{\leavevmode\hbox to3em{\hrulefill}\thinspace}
\renewcommand{\MR}[1]{%
%\relax\ifhmode\unskip\space\fi MR }
}
\renewcommand{\MRhref}[2]{%
%  \href{http://www.ams.org/mathscinet-getitem?mr=#1}{#2}
}
\providecommand{\href}[2]{#2}

\end{document}